\documentclass[nofootinbib,a4paper]{revtex4-2}
\usepackage{ulem}
\usepackage{comment}
\usepackage{braket}
\usepackage{mathrsfs}
\usepackage{tensor}

\usepackage{amsfonts}

\usepackage[utf8]{inputenc} 
\usepackage[T1]{fontenc} 
\usepackage[english]{babel}

\usepackage{ulem}
\usepackage{xr}

\usepackage{amsmath,dsfont,url,amssymb}
\usepackage{slashed,cancel}
\usepackage[usenames,dvipsnames]{xcolor}
\usepackage[colorlinks=true, linkcolor=black, citecolor=Forest Green]{hyperref}

\usepackage{color}
\usepackage{threeparttable}
\usepackage{cancel}
\usepackage{tikz}
\usetikzlibrary{decorations.pathmorphing}
\usepackage{feynmf}
\usepackage[compat=1.1.0]{tikz-feynman}
\tikzfeynmanset{compat=1.1.0}
\usepackage{verbatim}
\usepackage{xcolor}

\begin{document}

\title{Relativistic Hydrodynamics under Rotation: Prospects \& Limitations from a Holographic Perspective }

\author{Markus A. G. Amano} 
\email{markus@henu.edu.cn}
\affiliation{Institute of Contemporary Mathematics, School of Mathematics and Statistics, Henan University, Kaifeng, Henan 475004, P. R. China}

\author{Casey Cartwright}
\email{c.c.cartwright@uu.nl}
\affiliation{Institute for Theoretical Physics, Utrecht University, Princetonplein 5, 3584 CC Utrecht, The Netherlands}

\author{Matthias Kaminski}
\email{mski@ua.edu}
\author{Jackson Wu}
\email{jknwgm13@gmail.com}
\affiliation{Department of Physics and Astronomy, University of Alabama, Tuscaloosa, AL 35487, USA}

\date{\today}



\begin{abstract}
{The AdS/CFT correspondence, or holography, has provided numerous important insights into the behavior of strongly-coupled many-body systems. Crucially, it has provided a testing ground for the construction of new effective field theories, especially those in the low frequency, long wavelength limit known as hydrodynamics. 
In this article we continue the study of strongly-coupled rotating fluids using holography and hydrodynamics. We provide an overview of recent developments arising from the study of simply spinning Myers-Perry black holes. We review techniques to obtain hydrodynamic and non-hydrodynamic modes, describe how branch point singularities in the complex frequency and momentum space limit the radius of convergence of the hydrodynamic gradient expansion, and outline the relation of pole-skipping in the linear response functions to early time chaotic behavior.  }
\end{abstract}


 


\newcommand{\exd}{\mathrm{d}}
\newcommand{\fpd}[1]{\frac \partial {\partial #1}}
\newcommand{\ffact}[1]{%
  ^{\underline{#1}}%
} 

\newcommand{\lsigma}{\overline{\sigma}}

\maketitle

\tableofcontents

\section{Introduction}  \label{sec:intro}
In this article we review recent results about rotating holographic fluids and add previously unpublished results.  
The rotating holographic fluid which we study here may be viewed as a model of real-world rotating fluids, with our results aiding the construction of effective (e.g. hydrodynamic) descriptions of such fluids. These investigations were motivated by the recent report of large vorticity in the quark-gluon-plasma generated at RHIC~\cite{STAR:2017ckg}. 

\subsection{Overview}
Rotating fluids occur in great abundance in Nature, from eddies in rivers to rotating binary systems of merging black holes (BHs) or neutron stars creating gravitational waves which are measured with increasing precision, requiring a more precise theoretical understanding and modeling~\cite{LIGOScientific:2016aoc,Bailes:2021tot}. Thus the effective description of rotating systems is of high interest. Hydrodynamics provides such a universal effective description for low energy and long wavelength excitations around an equilibrium state. In this article, we will see that the rotating holographic fluids we study present such equilibrium states allowing a hydrodynamic description. 

In 2017, the hyperon polarization measurement by STAR~\cite{STAR:2017ckg} indicated that the quark-gluon-plasma generated in the heavy-ion-collisions at RHIC could possess a large vorticity. Since then, there has been an increased interest in rotating (in part strongly coupled and confined) systems~\cite{Karpenko:2016jyx,Becattini:2017gcx,Florkowski:2017ruc,Montenegro:2020paq,Bhadury:2020cop,Weickgenannt:2020aaf,Fukushima:2020ucl,Li:2020eon,Becattini:2021iol,Gallegos:2021bzp,Hongo:2021ona,Gallegos:2022jow,Wang:2021wqq,Weickgenannt:2022zxs,Weickgenannt:2022qvh}. 
Rotating systems, specifically rotating BHs, have also been investigated in the context of AdS/CFT correspondence, or holography~\cite{Myers:1986un,Carter:1968ks,Gibbons:2004ai,Gibbons:2004js,Hawking:1999dp,Murata:2008xr,Murata:2008yx,Murata:2007gv,Ishii:2021xmn,Ishii:2019wfs,Cardoso:2013pza,Gregory:2013xca}. 
The eternal rotating black holes (BHs) we consider here are in thermodynamic equilibrium, as their observable properties, namely temperature and angular momentum, do not change over time. 
In both rotating fluids and rotating BHs, the rotation introduces nontrivial inertial effects into the dispersion relations of eigenmodes (with respect to an inertial observer in relation to the fluid rest frame). Holography relates rotating fluids to rotating BHs~\cite{Hawking:1998kw,Hawking:1999dp}, which allows to study rotating BHs and derive properties of the holographically dual rotating fluids. 

In this article, we present the hydrodynamic regime and parts of the non-hydrodynamic regime of a rotating holographic plasma. 
In particular, we find hydrodynamic behavior in large simply spinning five-dimensional Myers-Perry (MP) Anti-de Sitter (AdS) BHs which are solutions of type IIB Supergravity. These are dual to rotating quantum fluids within strongly coupled $\mathcal{N} = 4$ Super-Yang-Mills theory (SYM) through the AdS/CFT correspondence.  
Therefore, it is possible to compute metric fluctuations around such rotating BHs and then translate the results into properties of rotating strongly coupled SYM plasma. In particular, we focus on the computation of the dispersion relations of hydrodynamic and non-hydrodynamic modes. 
%

Several limitations of the hydrodynamic description of the rotating holographic fluid are discussed, among them the explicit calculation of the radius of convergence of the gradient expansion facilitated by the study of so-called {\it critical points in complex momentum space}. In addition, cases are considered in which the low energy regime is dominated by non-hydrodynamic modes, leading to a breakdown of the hydrodynamic description.     
%
We compute the lowest few non-hydrodynamic modes and study their dynamics describing the fluctuations around the rotating holographic fluid, reaching far beyond the standard regime of hydrodynamics (thus we also include fluctuations with frequencies and momenta on the same order or much larger than the fluid temperature). 
Prospects are explored, pushing the hydrodynamic description beyond its limitations and exploring fluid properties outside the hydrodynamic regime, for example with the so-called {\it pole-skipping points} relating chaotic dynamics to the hydrodynamic sound dispersion relation. 

\subsection{Summary of Fluid Results}
\label{sec:introSummary}
The rotating holographic fluid which we study here may be viewed as a model of real-world rotating fluids, aiding the construction of effective (e.g. hydrodynamic) descriptions of such fluids. For convenience, we begin with a quick guide to the main results, which will be discussed and explained further below.
\begin{itemize}
    \item At any temperature (for any size BHs):
    \begin{itemize} 
        \item Hydrodynamic dispersion relations with modified transport coefficients are a good fit already at small to intermediate temperatures (corresponding to a horizon radius $r_+\approx 10^4$, and possibly even smaller), see Fig.~\ref{fig:horizonRadiusDependence}.
        \item There is a remarkable window of horizon values in the range of $1000 \lesssim r_+ \lesssim 10^7$, in which these holographic rotating fluids display hydrodynamic behavior that is distinct from a boosted fluid, see Sec.~\ref{sec:HorizonDependence}.
        \item Multiple {\it level crossings} occur between distinct non-hydrodynamic modes, as illustrated in Fig.~\ref{fig:vary_A_small_scale}.
        \item At small temperature (horizon radius $r_+=10$), the lowest few hydrodynamic and non-hydrodynamic modes are computed for the tensor (Fig.~\ref{fig:tensor_modes_over_a}), vector (Fig.~\ref{fig:vector_modes_over_a}), and scalar (Fig.~\ref{fig:scalar_modes_over_a}) sectors. 
        \item At {\it extremality}, holographic fluids display a similarity with rotating charged fluids. Consider a state of fixed energy with just rotational and thermal energy. When all the energy is rotational and the temperature vanishes, the zero momentum eigenmodes appear to coalesce, forming a branch cut along the imaginary axis similar to what is observed in charged holographic fluids~\cite{Edalati:2009bi,Edalati:2010hk,Janiszewski:2015ura}, see bottom graph ($a/\ell=0.9$, angular momentum parameter $a$ normalized to the AdS radius $\ell$) in Fig.~\ref{fig:tensor_modes_over_a}, \ref{fig:vector_modes_over_a}), and~\ref{fig:scalar_modes_over_a}. 
        \item At small temperatures, corresponding to horizon radius $r_+=\ell$ (with the AdS radius $\ell$) the fluid becomes unstable at smaller angular momentum, as superradiance occurs below extremality (i.e.~below rotational parameter $a=\ell$) at $a/a_{\text{ext}}\sim 0.916$, see Fig.~\ref{fig:tensor_unstable_im_part}.
        \item Radius of convergence of the hydrodynamic expansion in momentum space was computed for $100\le r_+\le 10^6$, see Fig.~\ref{fig:convergenceRadius}; the convergence is affected very little by the value of the horizon radius, see Fig.~\ref{fig:Horizon_dep}. 
    \end{itemize}
    \item At large temperature (for large BHs):
    \begin{itemize}
        \item Fluid is nontrivially rotating, low-energy (hydrodynamic) excitations {\it see} it as a fluid boosted with velocity $a$, see Sec.~\ref{sec:previousResults}.
        \item Hydrodynamic dispersion relations with boosted fluid transport coefficients are given by eqs.~\eqref{eq:etaPerpSum} to~\eqref{eq:GammaSum}.
        \item Eikonal limit (large momentum limit): scaling as expected, see Fig.~\ref{fig:vary_A_small_scale}.
        \item Superradiance instabilities occur only at extremality, $a=\ell$, while the fluid is stable (according to our numerical checks) against all fluctuations at $a<\ell$. 
        \item The pole skipping point which is believed to be related to quantum chaotic dynamics, encoding the quantum butterfly velocity and Lyapunov exponent, is shown to persist under rotation. This may indicate a relation between chaotic dynamics and hydrodynamics persisting under rotation.  Pole skipping frequencies (and momenta) are determined in a rotating fluid (these are points where the numerator and the denominator of an energy-energy Green's function vanish simultaneously, and consequently the pole is "skipped''~\cite{Blake:2017ris,Blake:2018leo}). Fig.~\ref{fig:dispersion} shows that the pole skipping point lies on the hydrodynamic sound dispersion relation; the associated quantum chaos butterfly velocity and Lyapunov exponent are displayed in Fig.~\ref{fig:butterfly}.
    \end{itemize}
\end{itemize}
%
Most strikingly, we find hydrodynamic behavior in the rotating fluids already at moderate temperatures, as noted before~\cite{Garbiso:2020puw}. We holographically compute fluctuations of the hydrodynamic fields (temperature, fluid velocity) and determine their spectrum of eigenmodes. Already at fairly low temperatures, the spectrum of this fluid contains modes which have vanishing frequency in the limit of vanishing momentum eigenvalue, a defining property of hydrodynamic modes. Furthermore, the frequencies of these hydrodynamic mode candidates scale with the (angular) momentum of the fluctuations linearly or quadratically, closely resembling the linear scaling of sound modes with the speed of sound and the quadratic damping of diffusion modes. 
Notably, this momentum of the fluctuations is now a discrete angular momentum eigenvalue because the rotating fluid enforces an expansion into eigenmodes which resemble spherical harmonics, being labelled by discrete angular momentum eigenvalues $\mathcal{J}$.  

Our analysis here goes beyond~\cite{Garbiso:2020puw} in that we present non-hydrodynamic modes along with the hydrodynamic modes. Extending~\cite{Garbiso:2020puw} even further, we present results at moderate to low temperatures (corresponding to medium to small BH horizon radius). At such low temperatures, even the lowest energy eigenmodes may not behave hydrodynamically, in that their dispersion relations deviate from hydrodynamic scaling.  These results demonstrate how the hydrodynamic scaling of the lowest lying eigenfrequencies with angular momentum emerges as the temperature increases. Furthermore, non-hydrodynamic modes are responding strongly to the coupling of modes in the scalar and vector sectors. This is seen in the level crossing in Fig.~\ref{fig:vary_A_small_scale}.  
Furthermore, we study large momentum values $\mathcal{J}$ and large frequencies $\nu$, beyond the naive regime of validity of hydrodynamics, and find that the hydrodynamic momentum diffusion mode dominates the low energy dynamics in the whole regime of momentum $j=0, ..., 10$ for all values of $a$ which we scanned, see Fig.~\ref{fig:vector_modes_over_a} from which it is obvious that the diffusion mode originating at zero frequency $\nu=0$ with momentum $\mathcal{J}=0$ (marked by triangle) always has the smallest imaginary part, i.e. the smallest damping, thus it is the longest lived mode dominating the late time low energy dynamics.
In the case of the two sound modes, the behavior is more complicated, as displayed in Fig.~\ref{fig:scalar_modes_over_a}. However, at least one of the two sound modes is dominating the low energy late time dynamics for all values of $a$ and $\mathcal{J}$ which we scanned (very similar to the diffusion mode described above). 
The other sound mode, at a specific momentum value $\mathcal{J}$, becomes more damped than the lowest-lying non-hydrodynamic modes, leading to a change in the late-time transport properties of the fluid, as seen also at vanishing rotation. 
There is a third hydrodynamic mode appearing in Fig.~\ref{fig:scalar_modes_over_a} at the triangle marking $\mathcal{J}=0$, $\nu=0$, namely the momentum diffusion mode, which mixes with the sound modes at finite values of the horizon radius $r_+$ (moderate to small temperatures).  
At large values of the momentum, we observe the expected~\cite{Festuccia:2008zx, Fuini:2016qsc} linear scaling of the eigenfrequencies with momentum, see Fig.~\ref{fig:hpplargej}. 

One obvious limitation of hydrodynamics when viewed as an expansion into gradients of the hydrodynamic quantities (temperature, fluid velocities) is the radius of convergence of this series expansion. 
While it is currently impossible to determine the radius of convergence of the hydrodynamic derivative expansion when the microscopic theory is QCD, this task is feasible when the microscopic theory is SYM~\cite{Grozdanov:2019uhi,Heller:2020hnq,Heller:2020uuy}. Thus, we here review the radius of convergence of the linearized relativistic hydrodynamic expansion around a non-trivially rotating strongly coupled SYM plasma~\cite{Cartwright:2021qpp}. Our results show that the validity of hydrodynamics is sustained and can even get enhanced in a rapidly rotating plasma, see Fig.~\ref{fig:convergenceRadius}, and this result holds true for intermediate down to small temperatures, see Fig.~\ref{fig:Horizon_dep}. 

Furthermore, we study the relationship between many-body quantum chaos and energy dynamics in holographic quantum field theory states dual to the simply-spinning MP-AdS5 BH~\cite{Amano:2022mlu}. In the large BH limit we are able to obtain a simple analytic expression for the out-of-time-order-correlator (OTOC) for operator configurations on Hopf circles, and demonstrate that the associated Lyapunov exponent and butterfly velocity are robustly related to the locations of a family of pole-skipping points in the energy response. 
We demonstrate that the dispersion relations of sound modes in the energy response explicitly pass through our pole-skipping locations. 

{\bf Large black hole (BH) limit:} 
One truly remarkable result is:
{\it Hydrodynamic fluctuations at large temperatures {\it perceive} the rotating holographic fluid as if it was a boosted fluid.}  

On the gravity side, the statement above is verified as follows. 
In the case of large rotating BHs, the metric simplifies slightly. If one now chooses to probe this slightly simplified rotating background metric only with metric fluctuations which have {\it small frequency and momentum compared to the horizon radius}, then these fluctuations effectively {\it see} a boosted black brane metric.\footnote{We stress, however, that the background metric itself is not that of a boosted black brane.} 
This is explicitly seen in the equations of motion obeyed by the metric fluctuations around the rotating background metric, as will be explained in Sec.~\ref{sec:results}. 

Previously, in a particular limit (the large BH limit),
analytic results for the hydrodynamic dispersion relations and transport coefficients of this holographic rotating plasma as a function of their values in a plasma at rest were given~\cite{Garbiso:2020puw}.    
Such analytic expressions are provided for the two shear viscosities, the longitudinal momentum diffusion coefficient, two speeds of sound, and two sound attenuation coefficients. The shear viscosity to entropy density ratio varies between zero and $1/(4\pi)$ depending on the direction of the shear. 
At vanishing rotation, $a=0$, the longitudinal shear viscosity $\eta_{\perp,||}$ (along the rotation axis) and the longitudinal momentum diffusion coefficient $D_{||}$, and the two sound attenuations $\Gamma_\pm$ satisfy the known (Einstein) relations, $D_{||}\propto \eta_{||}$ and $\Gamma_{||}\propto \eta_{||}$. At nonzero rotation, $a\neq 0$, known relations between these coefficients are generalized to include dependence on angular momentum. 
These previously discussed relations are reviewed here in Sec.~\ref{sec:previousResults}. 

Now this seems to indicate that the hydrodynamic description of this rotating holographic fluid studied here may be trivial in that it may be equivalent to a boosted fluid. It turns out that this suspicion is {\it not} true. 
In the present article, we analyze to what degree this hydrodynamic behavior of a boosted fluid persists to lower temperatures and especially away from the large black hole limit predominantly considered in~\cite{Garbiso:2020puw}. We show that this holographic fluid behaves hydrodynamically, with transport coefficients deviating significantly from those of a boosted fluid, see section~\ref{sec:HorizonDependence}.

\subsection{Gravity Results}
\label{sec:gravityResults}
With the help of the AdS/CFT correspondence\cite{Maldacena:1997re}  
one can study the dissipative dynamics of (confined) systems with a nontrivial amount of angular momentum. 
This is made possible by the equivalence of the spectrum of quasinormal modes (QNMs) around asymptotically AdS black holes and the poles of retarded Green's functions in the dual field theory~\cite{Kovtun:2005ev}. 
QNMs are characteristic linearized metric solutions of Einstein's field equations. 
So, while we report on dispersion relations of hydrodynamic and nonhydrodynamic modes, we simultaneously obtain results for QNMs. 

In particular, we consider (4+1)-dimensional rotating asymptotically AdS Myers-Perry black holes (MP BHs), where the horizon induces a temperature and the angular momenta introduce angular momenta in the dual field theory~\cite{Hawking:1998kw,Hawking:1999dp}.  
Previous work modeled a rotating quark gluon plasma  holographically~\cite{McInnes:2014haa, McInnes:2016dwk, McInnes:2017rxu,Garbiso:2020puw,Chen:2020ath}. 

We provide an overview of the recent progress in the analysis of the linearized field equations of metric fluctuations around (4+1)-dimensional MP BHs, see section~\ref{sec:model}. 
In the large black hole limit, \eqref{eq:LBH1} and~\eqref{eq:LBH2}, these black holes are perturbatively stable against all metric fluctuations~\cite{Garbiso:2020puw}.  
However, near extremal BHs become unstable. This so-called superradiant instability has been known~\cite{Murata:2008xr, Cardoso:2013pza, Aliev:2008yk, Monteiro:2009tc}. We report some details at small horizon radius $r_+=\ell$ (with the AdS radius $\ell$): the BH becomes unstable at small angular momentum, as superradiance occurs below extremality (i.e.~below rotational parameter $a=\ell$) at $a/a_{\text{ext}}\sim 0.916$, see Fig.~\ref{fig:tensor_unstable_im_part}. 

Reference values for all QNMs in all three channels of fluctuations which we studied for various values of the angular momentum parameter $a$ of the rotating black hole are recorded in tables in appendix~\ref{sec:QNMdata}. Examples of QNMs are visualized in Fig.~\ref{fig:tensor_modes_over_a}, \ref{fig:vector_modes_over_a}, \ref{fig:scalar_modes_over_a}, and shown together for comparison in Fig.~\ref{fig:all_sectors_compared_a_0_not_0}. 

Throughout this article, w point out various similarities between the QNM spectra of rotating MP BHs studied in the present article and the QNM spectra charged Reissner-Nordstr\"om black branes studied in~\cite{Janiszewski:2015ura}. Most striking is the occurence of a set of ``purely imaginary QNMs'' at nonzero rotation or charge parameter, respectively. These modes at large rotation or charge parameter coalesce and appear to form a branch cut along the negative imaginary frequency axis as the black holes approach extremality, as studied in the lower dimensional AdS4 Reissner-Nordstr\"om case~\cite{Edalati:2009bi,Edalati:2010hk}.  

Pole skipping points as well as critical points in the QNM spectra of these MP BHs have been computed and are reported in sections~\ref{sec:critical_Points} and~\ref{sec:pole_skipping}, where the significance of both critical and pole skipping points is discussed in detail.

\section{The Holographic Setup 
}
\label{sec:model}
In this section we describe the theory which we study, the background metric our work centers on, and the separability of the equations of motion of the metric fluctuations around the background metric. Our discussion of the separability of the field equations follows closely the original work~\cite{Murata:2007gv}. 

\subsection{Rotating background metric and BH thermodynamics} \label{sec:thermo}
We are interested in the metric fluctuations around stationary (4+1)-dimensional, asymptotically locally AdS (AlAdS) metrics, which are solutions to the equations of motion obtained from the Einstein-Hilbert action
\begin{equation}
    S_{EH}=\frac{1}{16\pi G_5}\int \exd^5 x \sqrt{-g}\left(R-2\Lambda\right) +S_{ct} \, .
\end{equation}
Here, $\Lambda=-6/\ell^2$ denotes the cosmological constant given in terms of the AdS radius, $G_5$ is the five-dimensional Newton's constant, and $S_{ct}$ is the counterterm action containing the Gibbons-Hawking-York boundary term required for a well-defined variation and other  counterterms required to render the boundary action finite. Variation of $S_{EH}$ leads to the standard source free Einstein equations with cosmological constant
\begin{equation}
    R_{\mu\nu}-\frac{1}{2}(R-2\Lambda)g_{\mu\nu}=0\, .
\end{equation}

Given our focus on the finite temperature rotating states of the dual field theory, we are interested in the most general rotating BH solution with boundary topology $S^3\times \mathbb{R}$ 
to the Einstein equations.\footnote{For higher dimensional BHs we note that other topologies are possible besides $S^3\times \mathbb{R}^{1,1}$, e.g. black saturns~\cite{Elvang:2007rd} and black rings~\cite{Emparan:2001wn}. We do not consider these more exotic topologies here.} 
The solution is given by the following metric~\cite{Hawking:1998kw,Hawking:1999dp}
\begin{align}\label{eq:MP_Black_Hole}
ds^2&=\frac{\left(1+r_H^2\ell^{-2}\right)}{\rho ^2 r_H^2} \left(a b \exd t_H-\frac{b  \left(a^2+r_H^2\right) \sin ^2(\theta_H )}{\Xi_a}\exd \phi_H-\frac{a  \left(b^2+r_H^2\right) \cos ^2(\theta_H )}{\Xi_b}\exd\psi_H\right)^2 \nonumber\\
    &-\frac{\Delta_r}{\rho ^2} \left(\exd t_H-\frac{ a \sin ^2(\theta_H )}{\Xi_a}\exd \phi_H-\frac{b  \cos ^2(\theta_H )}{\Xi_b}\exd \psi_H\right)^2 +\frac{ \rho ^2}{\Delta_\theta}\exd \theta_H^2+\frac{ \rho ^2}{\Delta_r}\exd r_H^2\nonumber\\
    &+\frac{\Delta_\theta \sin ^2(\theta_H )}{\rho ^2} \left(a \exd t_H-\frac{ a^2+r_H^2}{\Xi_a}\exd \phi_H\right)^2+\frac{\Delta_\theta \cos ^2(\theta_H )}{\rho ^2}  \left(b \exd t_H-\frac{ b^2+r_H^2}{\Xi_b}\exd \psi_H\right)^2 \, ,
\end{align}
with coordinates $(r_H,t_H,\theta_H,\phi_H,\psi_H)$ residing in ranges
\begin{subequations}
\begin{align}
    -\infty &< t_H <\infty\, , \\
    0 &< r_H <\infty \, ,\\
    0 &< \theta_H < \pi/2\, , \\ 
    0 &< \phi_H < 2\pi \, , \\
    0 &<\psi_H <2\pi\, ,
\end{align} 
\end{subequations}
and we define
\begin{align}
    \Delta_r&=\frac{1}{r_H^2}(r_H^2+a^2) (r_H^2+b^2)(\frac{r_H^2}{\ell^2}+1)-2M \, ,\\
     \Delta_\theta&=1-\frac{a^2}{\ell^2}\cos(\theta_H)^2-\frac{b^2}{\ell^2}\sin(\theta_H)^2\, , \\
    \rho^2&=r_H^2+a^2\cos(\theta_H)^2+b^2\sin(\theta_H)^2 \, ,\\
    \Xi_a&=1-\frac{a^2}{\ell^2}\, , \quad \Xi_b=1-\frac{b^2}{\ell^2} \,.
\end{align}

Although originally discovered in~\cite{Hawking:1998kw}, this solution is referred to as the Myers-Perry BH due to the pioneering work of Malcolm Perry and Robert Myers to extend Kerr BHs to higher dimensions~\cite{Myers:1986un}~\footnote{It should be noted that their generalized Kerr solutions were computed without a cosmological constant. The generalization of the Kerr solution in (3+1)-dimensions to include AdS-dS boundary conditions was found by Carter~\cite{Carter:1968ks} and subsequently generalized to all higher dimensions by Gibbons, Lu, Page and Pope~\cite{Gibbons:2004js}.}. The solution features two rotational parameters, $(a,b)$, related to the two angular momenta, $(J_\phi,J_\psi)$. This can be anticipated from the fact that in (4+1)-dimensions the rotation group is $SO(4)$ whose double cover is $SU(2)\times SU(2)$; irreducible representations of the rotation group are thus labeled by two parameters. It is however important to note that the full $SO(4)$ is not present in the solution, eq.~\eqref{eq:MP_Black_Hole}, which is only invariant under time translations and the two $U(1)$ transformations associated with independent rotations in $(\phi_H,\psi_H)$. 

The two angular momenta, $(J_\phi,J_\psi)$, are given by~\cite{Papadimitriou:2005ii}
\begin{equation}\label{eq:ang_mmt}
    J_\phi=\frac{4\pi^2 M}{\kappa^2\Xi_a^2\Xi_b}a \,, \qquad  
    J_\psi=\frac{4\pi^2 M}{\kappa^2\Xi_a\Xi_b^2}b\,,
\end{equation}
where $\kappa^2=8\pi G_5$. Associated with $(J_\phi,J_\psi)$ are the two angular velocities, $(\Omega_a,\Omega_b)$, which are non-vanishing throughout the entire spacetime. In particular, at the horizon and the conformal infinity, i.e. the AdS boundary, they taken on values
\begin{subequations}\label{eq:horizon_boundary_ang_vel}
\begin{align}
    \Omega_a\vert_{hor} &= \frac{a-a^3/\ell^2}{a^2+r_H^{+2}} \,, &
    \Omega_b\vert_{hor} &= \frac{b-b^3/\ell^2}{b^2+r_H^{+2}} \,, \\
    \Omega_a\vert_{bdy} &= -a/\ell^2 \,, &
    \Omega_b\vert_{bdy} &= -b/\ell^2 \,,  
\end{align}
\end{subequations}
where $r_H^+$ denotes the radius of the outer horizon defined as the largest real root of the equation $\Delta_r=0$. Defining the relative angular velocities, $(\widetilde\Omega_a,\widetilde\Omega_b)$, by the respective differences~\cite{Papadimitriou:2005ii}:
\begin{subequations}\label{eq:thermodynamic_ang_vel}
\begin{align}
    \widetilde{\Omega}_a &= \Omega_a\vert_{hor} - \Omega_a\vert_{bdy} 
    = \frac{(1+r_H^{+2}/\ell^2)}{(a^2 + r_H^{+2})}a \,, \\
    \widetilde{\Omega}_b &= \Omega_b\vert_{hor} - \Omega_b\vert_{bdy} 
    = \frac{(1+r_H^{+2}/\ell^2)}{(b^2 + r_H^{+2})}b \,,
\end{align} 
\end{subequations}
we have the quantities thermodynamically conjugate to $(J_\phi,J_\psi)$.

As the metric is independent of the coordinates $(t,\phi,\psi)$, $(\partial_t,\partial_\phi,\partial_\psi)$ are Killing vectors. The combination $\partial_t+\Omega_a\,\partial_\phi+\Omega_b\,\partial_\psi$ is tangent to the null generators of the horizon from which one can obtain the Hawking temperature of the horizon (and thus the temperature of the dual field theory) from the surface gravity of the horizon:
\begin{equation}
     T=\frac{r_H^{+4} \left(a^2/\ell^2+b^2 /\ell^2+1\right)-a^2 b^2+2 r_H^{+6}/\ell^2}{2 \pi  r_H^+ \left(a^2+r_H^{+2}\right) \left(b^2+r_H^{+2}\right)} \, .
\end{equation}
One can notice that there exist extremal solutions for which the temperature vanishes, an explicit relation between the horizon radius and the rotational parameters where this occurs will be given in section~\ref{sec:seperability}. 

As stated in the beginning of this section, the geometry described by the metric in eq.~(\ref{eq:MP_Black_Hole}) is AlAdS, 
which is a geometry whose solutions does not approach AdS exactly but rather a general conformal structure. The first law of BH thermodynamics for AlAdS and the definition of conserved holographic charges can be found in~\cite{Papadimitriou:2005ii} along with the analysis for MP BHs in (4+1)-dimensions. The entropy which appears in this first law is given by the Bekenstein-Hawking formula $S=A/(4G_5)$, where $A$ is area of the event horizon (as given by the integral over the square root of the spatial determinant of the metric evaluated at the event horizon):
\begin{equation}
    A=\frac{2 \pi ^2 \left(a^2+r_H^{+2}\right) \left(b^2+r_H^{+2}\right)}{\Xi_a \Xi_b r_H^+} \, .
\end{equation}
As usual, the entropy density is obtained from the entropy by dividing out the volume. The relevant volume here is the spatial volume of the boundary metric, $V_{bdy} = 2\pi^2\ell^3$, and thus the entropy density of the dual field theory is
\begin{eqnarray}
    s = \frac{S}{V_{bdy}} = \frac{A/(4G_5)}{V_{bdy}} 
      = \frac{l\left(a^2+r_H^{+2}\right)\left(b^2+r_H^{+2}\right)}{4G_5\,r_H^+(\ell^2-a^2)(\ell^2-b^2)} \,.
\end{eqnarray}

\subsection{Linearized metric fluctuation equations and separability}
\label{sec:seperability}
As stated in the introduction, our goal is to provide an overview of the recent progress in the analysis of the field equations in (4+1)-dimensional MP BH. To that end we consider perturbations of the metric in eq.~(\ref{eq:MP_Black_Hole}) as was done in~\cite{Murata:2007gv,Murata:2008xr,Garbiso:2020puw,Cartwright:2021qpp,Amano:2022mlu}:
\begin{align}\label{eq:pertgeneric}
        g^{p}_{\mu\nu} {dx}^\mu {dx}^\nu = \left(g_{\mu\nu}+\epsilon~h_{\mu\nu}+O(\epsilon^2)\right) {dx}^\mu {dx}^\nu\,,
\end{align}
where $g$ is the metric satisfying the Einstein equations (at order $\epsilon^0$). 
Expanding the Einstein equations to leading order in $\epsilon$, the perturbation $h_{\mu\nu}$, is required to satisfy~\cite{Wald:1984rg}
\begin{align}\label{eq:pertgenericeom}
       -\frac{1}{2}\nabla_\mu \nabla_\nu h-\frac{1}{2}\nabla^\lambda \nabla_\lambda h_{\mu\nu}+\nabla^\lambda \nabla_{(\mu}h_{\nu)\lambda} = \frac{2\Lambda}{D-2}h_{\mu\nu}\,,
\end{align}
where $h=h^{~~\mu}_{\mu}=h_{\nu \mu} g^{\mu \nu}$, and the covariant derivatives in eq.~(\ref{eq:pertgenericeom}) are constructed entirely from the background metric $g_{\mu\nu}$. Although the metric in eq.~(\ref{eq:MP_Black_Hole}) may not be the most complicated solution to the Einstein equations, it is not immediately clear if or when field equations in this geometry are separable. It is typically the case that separability is intimately tied to symmetry, for instance the field equations in a highly symmetric Schwarzschild geometry may be easily separated by considering an expansion in spherical harmonics. Furthermore, for a clever choice of field variables, fields carrying a spin under the little group may be decoupled into invariant sectors. While this may be expected in highly symmetric spacetimes, it is obviously not guaranteed in general. For the (4+1)-dimensional AlAdS MP BH we consider, separability of the field equations of a propagating spin 2 field was demonstrated in~\cite{Murata:2007gv} for the case of equal angular momenta.\footnote{Separability of the field equations for dimensions larger than six was demonstrated in~\cite{Kunduri:2006qa} for equal angular momenta.}  We will follow this technique of decomposition of the field equations in what follows. 

The metric in eq.~(\ref{eq:MP_Black_Hole}) simplifies in two special cases: 
\begin{equation}
  \text{(i)}\quad  J_\phi=J_\psi\,, \qquad \text{(ii)} \quad J_\psi =0 \, .
\end{equation}
In this work, we focus on case (i) for which the MP BH is referred to as simply spinning. Imposing  $J_\phi=J_\psi$ (and thus $a=b$) leads to a symmetry enhancement of the metric crucial to the following analysis. To see this, first transform to a new set of coordinates:
\begin{subequations}\label{eq:Murata_Form}
\begin{align}
    \theta &=2 \theta _H\, ,\hspace{.15cm}\phi =\psi _H-\phi _H\, ,\hspace{.15cm}\psi =2 a \ell^{-2} t_H+\psi _H+\phi _H, \\
    t&=-t_H\, , \hspace{.15cm}r=\sqrt{\frac{a^2+r_H^2}{1-a^2 \ell^{-2}}} \, ,\hspace{.15cm} M= \mu  \left(1-a^2 \ell^{-2}\right)^3\hspace{.15cm} \, ,
\end{align}
\end{subequations}
and we have defined the effective mass parameter $\mu$. It is important to note that the coordinate ranges are now given by
\begin{subequations}\label{eq:Simple_Ranges}
\begin{align}
    -\infty &< t <\infty\, , \\
    0 &< r <\infty \, ,\\
    0 &< \theta< \pi\, , \\ 
    0 &< \phi < 2\pi \, , \\
    0 &<\psi <4\pi\, .
\end{align}
\end{subequations}
Under the transformation of eq.~(\ref{eq:Murata_Form}) the metric is brought to the form first described in~\cite{Murata:2007gv,Murata:2008xr}
\begin{equation}\label{eq:Simple_Spin_MP_Metric}
    d s^2=\frac{{dr}^2}{G(r)}-dt^2 \left(\frac{r^2}{\ell^2}+1\right)+\frac{1}{4} r^2 \left((\sigma^1)^2+(\sigma^2)^2+(\sigma^3)^2\right) +\frac{2 \mu  \left(\frac{a \sigma^3}{2}+dt\right)^2}{r^2}\, ,
\end{equation}
with the blackening factor given by
\begin{equation}
G(r)=-\frac{2 \mu  \left(1-a^2\ell^{-2}\right)}{r^2}+\frac{2 a^2 \mu }{r^4}+\frac{r^2}{\ell^2}+1 \, ,
\end{equation}
and we have introduced the one-forms $\sigma^a$, $a=1,2,3$, defined by
\begin{align}\label{eq:form_basis}
    \sigma^1&=d\phi \sin (\theta ) \cos (\psi )-d\theta \sin (\psi ) \, ,\\
    \sigma^2&=d\theta  \cos (\psi )+d\phi \sin (\theta ) \sin (\psi ) \, , \\
    \sigma^3&=d\psi+d\phi  \cos (\theta )\, .
\end{align}

The location of the event horizon in this set of coordinates corresponds to the largest positive real root of the polynomial equation $G(r_+)=0$. Expanding this equation one can define a relation between the horizon radius $r_+$ and the effective mass $\mu$:
\begin{equation}
 \mu= \frac{r_+^4 \left(\ell^2+r_+^2\right)}{2 \ell^2 r_+^2-2 a^2 \left(\ell^2+r_+^2\right)}\, . 
 \end{equation}
In terms of the redefined parameters and coordinates~\footnote{The thermodynamic data displayed in previous sections are in the Boyer-Lindquist coordinate system used in~\cite{Hawking:1998kw,Hawking:1999dp}.} the Hawking temperature, in particular, is given by
\begin{equation}\label{eq:temperature_Murata}
  T=  \frac{r_+^2 \ell ^2 \left(2 r_+^2+\ell ^2\right)-2 a^2 \left(r_+^2+\ell ^2\right)^2}{2 \pi  r_+^2 \ell ^3 \sqrt{r_+^2 \ell ^2-a^2 \left(r_+^2+\ell ^2\right)}} \, ,
\end{equation}
and the entropy density of the dual field theory has the expression
\begin{equation}
    s=\frac{r_+^4}{4G_5\ell ^2 \sqrt{r_+^2 \ell ^2-a^2 \left(r_+^2+\ell ^2\right)}} \,.
\end{equation}
The extremal value of the rotational parameter, $a_{ext}$, at which the temperature vanishes is given by
\begin{equation}\label{eq:extremal}
    a_{\text{ext}} =\pm \frac {r_+^2} {1 + r_+^2/\ell^2} \sqrt{ \frac 1 {2 r_+^2} + \frac 1 {\ell^2} }\, .
\end{equation} 

For $a=b$, one can see from eq.~\eqref{eq:ang_mmt} the two angular momenta coincide and thus so does the two associated angular velocities, $\Omega_a = \Omega_b = \Omega$.
In particular, in the new coordinates we have
\begin{equation}\label{eq:horizon_boundary_ang_vel_murata}
    \Omega\vert_{hor} = \frac{a}{r_+^{2}} \,, \qquad
    \Omega\vert_{bdy} = -\frac{a}{\ell ^2} \,, 
\end{equation}
and the thermodynamically relevant relative angular velocity is given by~\footnote{The thermodynamic value of the angular velocity is mistakenly given in place of the horizon value on pg. 7 of~\cite{Garbiso:2020puw}. The same mistake is made in~\cite{Amano:2022mlu} on pg. 4, including $\Omega'=2\Omega$, as alluded to below eq.~(\ref{eq:angular_momentum_murata}).} 
\begin{equation}\label{eq:thermodynamic_ang_vel_murata}
    \widetilde{\Omega}= \Omega\vert_{hor} - \Omega\vert_{bdy} 
    = a \left(\frac{1}{r_+^2}+\frac{1}{\ell ^2}\right) \,.
\end{equation}
The angular momentum thermodynamically conjugate to $\widetilde\Omega$ has the expression
\begin{equation}\label{eq:angular_momentum_murata}
    J=\frac{4 a \pi^2 \mu}{\kappa^2} \, .
\end{equation}
Note that the form of the first law now contains identical factors related to the contribution from the rotation, viz. $\sum_i J_i\widetilde{\Omega}_i=2 J \widetilde{\Omega}$. Different conventions have been used regarding the factor of two: some leave it in place while others absorb it into $J'=2J$~\cite{Cartwright:2021qpp} or $\Omega'=2\Omega$ ~\cite{Amano:2022mlu}. We choose here to leave the factor of two alone.

With the thermodynamic information associated with the metric in these new coordinates recounted, we now turn to the symmetries of the metric. We begin our discussion with the form basis $\sigma^a$ introduced in eq.~(\ref{eq:form_basis}). These forms are the left-invariant 1-forms of the group $SU(2)$.\footnote{We recall here some basic results of the theory of Lie groups and their algebras (see for instance~\cite{nakahara2003} or the classic text~\cite{chevalley1946}, in addition~\cite{fecko_2006} covers explicitly the $SU(2)$ case).  Let $E^a$ ($E_a$) be a basis of the cotangent (tangent) space at the unit element of a group $G$, the left invariant 1-forms on $G$ are generated by $E^a$. Given a parameterization of a generic element of the group, $A$, the left-invariant 1-forms (also known as the Maurer-Cartan form) can be found from $A^{-1}\exd A= \sigma^a E_a$; the right-invariant 1-forms, $\xi^a$, are likewise given by $\exd A A^{-1}=\xi^aE_a$. In the case of $SU(2)$ of interest here, the basis we choose to work in is the standard one for the algebra $su(2)$: $E_a=-\frac{1}{2}i\tau_a$ where $\tau_a$ are the Pauli matrices. Parameterize $A\in SU(2)$ as
\begin{equation*}
A=\left(
\begin{array}{cc}
 z & -w^* \\
 w & z^* \\
\end{array}
\right) \,, \qquad
z=\cos(\theta/2)e^{-i(\psi+\phi)} \,, \quad w=\sin(\theta/2)e^{-i(\psi-\phi)}
\end{equation*}
where $(\theta,\psi,\phi)$ range as in eq.~\eqref{eq:Simple_Ranges}, in the standard basis $E_a=-\frac{1}{2}i\tau_a$ one would get $\sigma^1=-d\phi\sin(\theta )\cos(\psi )+d\theta\sin(\psi)$ and a corresponding opposite sign for $\sigma_1$. Since a constant multiple of a left-invariant field is still a left invariant field and $\sigma^1$ only appears in the metric as $(\sigma^1)^2$, we redefine these as displayed in the text.} Dual to the left-invariant 1-forms are the left-invariant vector fields $\sigma_a$ (which can be expanded in a basis $E_a$ of the tangent space):~\footnote{Note that Refs.~\cite{Murata:2007gv,Murata:2008xr,Murata:2008yx,Garbiso:2020puw,Cartwright:2021qpp,Amano:2022mlu} make the non-standard choice of labelling the vector fields dual to $\sigma^a$ with a different symbol, $e$, rather than $\sigma^a$ with the index lowered. In order to clear the notational mine-field surrounding the discussion of separability in MP BHs, we change to the standard notation in this work.}
\begin{align}\label{eq:Dual_Vectors_SU_2}
  \sigma_1 & = -\sin{\psi}\,\partial_\theta + \frac{\cos{\psi}}{\sin{\theta}}\,\partial_\phi - \cos{\theta} \sin{\phi}\,\partial_\psi\, ,\nonumber\\
    \sigma_2 & = \cos{\psi}\,\partial_\theta + \frac{\sin{\psi}}{\sin{\theta}}\,\partial_\phi - \cot{\theta} \sin{\psi}\,\partial_\psi\, ,\\
    \sigma_3 & = \partial_\psi\nonumber \, .
\end{align}
Expanding in the coordinate basis spanned by $Y=(\theta,\phi,\psi)$, the components of the left-invariant 1-forms are given by $\sigma^a=\sigma^a_i\exd Y^i$, while that of the left-invariant vector fields are given by $\sigma_a=\sigma^{i}_{a}\partial_i$, and by definition $\sigma^a_i\sigma^i_b=\delta^a_b$. Likewise, one can write down the right-invariant 1-forms and their dual right-invariant vector fields. While the right-invariant 1-forms are not used here, the right-invariant vector fields are, and they are given by~\footnote{Naively computing the right-invariant forms and vector fields as described in the previous footnote, one would find that $\xi_1$ and $\xi_2$ are interchanged along with the 1-forms $\xi^1$ and $\xi^2$. We rearrange them as displayed in the text.}
\begin{align}  \label{eq:kvfsu2}
   \xi_1 & = \cos{\phi}\,\partial_\theta + \frac{\sin{\phi}}{\sin{\theta}}\, \partial_\psi - \cot{\theta} \sin{\phi}\,\partial_\phi\, ,\nonumber\\
   \xi_2 & = -\sin{\phi}\,\partial_\theta + \frac{\cos{\phi}}{\sin{\theta}}\, \partial_\psi - \cot{\theta} \cos{\phi}\,\partial_\phi\, ,\\
   \xi_3 & = \partial_\phi\nonumber \, .
\end{align}
Naturally the left-invariant 1-forms are invariant under the action of a right translation generated by the curve $\xi_a$ which can be written as the Lie derivative of $\sigma^b$ along $\xi_a$, i.e. $\mathcal{L}_{\xi_a} \sigma^b=0$. Furthermore since these do not involve the temporal or radial coordinates it is then easy to see this can be extend to the whole metric $\mathcal{L}_{\xi_a} g_{\mu\nu}=0$ and hence the $\xi_a$ constitute a set of Killing vectors generating an $SU(2)$ symmetry of the spacetime. Furthermore, since the metric contains no explicit time dependence there exists a time translation symmetry generated by $\partial_t$. And, finally, while not obvious in the form notation it can be shown that \begin{equation}
    \mathcal{L}_{\sigma_3}\sigma^1=-\sigma^2\, , \quad \mathcal{L}_{\sigma_3}\sigma^2=\sigma^1\, , \quad \mathcal{L}_{\sigma_3}\sigma^3=0\, ,
\end{equation}
and therefore $\sigma_3=\partial_\psi$ generates an additional $U(1)$. Hence the full isometry group of the simply spinning MP BH metric given in eq.~(\ref{eq:Simple_Spin_MP_Metric}) is $\mathbb{R}_t \times (SU(2)\times U(1))/\mathbb{Z}_2$~\cite{Murata:2008xr}, generated by
$( \xi_a, \sigma_3, \partial_t)$. Given this set of isometries the separability of the field equations can be demonstrated, as in~\cite{Murata:2007gv}, by considering irreducible representations of $SU(2)\times U(1)$.

Defining the differential operators $W_a = i \sigma_a$ and $L_a = i \xi_a$ 
one can check that the following commutation relations are satisfied: 
\begin{equation}\label{eq:WLalgebra}
    [L_a,L_a] = i \epsilon_{abc} L_c \quad [W_a,W_b] = -i \epsilon_{abc} W_c \quad [W_a,L_b] = 0 \,  
 \end{equation}
and the symmetry group of interest, $SU(2)\times U(1)$, is generated by $L_a$ and $W_3$. In addition, one can show that the left- and right-invariant operators share the same quadratic Casimir $L_aL_a=W_aW_a=L^2$. As a result $L^2$, $L_3$, and $W_3$, are compatible (mutually commuting operators) which can be simultaneously diagonalized. The eigenfunctions associated with this set of compatible operators are the Wigner-D functions, $D^\mathcal{J}_{\mathcal{KM}}$~\cite{Murata:2007gv} which obey the following operator equations
\begin{align}\label{eq:wigeigenequations}
   L^2 D^\mathcal{J}_{\mathcal{KM}}&=\mathcal{J}(\mathcal{J}+1)D^\mathcal{J}_{\mathcal{KM}}\,\,\, , \\
    L_3D^\mathcal{J}_{\mathcal{KM}}&=\mathcal{M} D^\mathcal{J}_{\mathcal{KM}}\,\,\, ,\\
    W_3 D^\mathcal{J}_{\mathcal{KM}}&=\mathcal{K} D^\mathcal{J}_{\mathcal{KM}}\,\,\, .
\end{align}
The Wigner-D functions, expanded in coordinate basis $D^\mathcal{J}_{\mathcal{M}\mathcal{K}}(\psi,\phi,\theta)$, for different $\mathcal{J}$ are mutually orthogonal when integrated over the volume of $SU(2)$, defined in eq.~(\ref{eq:Simple_Ranges}) 
\begin{equation}
  \int_0^{4\pi}\exd\psi\int_0^\pi \exd\theta\sin(\theta) \int_0^{2\pi}\exd\phi D^{\mathcal{J}_2*}_{\mathcal{K}_2\mathcal{M}_2}(\psi,\phi,\theta) D^{\mathcal{J}_1}_{\mathcal{K}_1\mathcal{M}_1}(\psi,\phi,\theta)=\frac{16\pi^2}{2\mathcal{J}_1+1}\delta_{\mathcal{J}_1\mathcal{J}_2}\delta_{\mathcal{M}_1\mathcal{M}_2}\delta_{\mathcal{K}_1\mathcal{K}_2}
\end{equation}
and form a complete set of eigenfunctions 
\begin{align}
&\frac{1}{\sin(\theta_1)} \delta(\tilde{\psi} - \tilde{\psi}_1) \delta(\phi - \phi_1) \delta(\theta - \theta_1) \nonumber\\
&= \sum_{{\cal J}=0, 1/2, 1, \dots}^{\infty} \sum_{{\cal K}=-{\cal J}}^{\cal J}  \sum_{{\cal M}={-{\cal J}}}^{{\cal J}} \frac{2 {\cal J}+1}{16 \pi^2} D^{{\cal J}}_{{\cal K} {\cal M}}(\tilde{\psi}, \phi, \theta) D^{{\cal J}*}_{{\cal K} {\cal M}}(\tilde{\psi_1}, \phi_1, \theta_1) \, .
\label{eq:completeness}
\end{align}
on $S^3$ (see for instance~\cite{varshalovich_1988}). 

These commutation relations, eq.~(\ref{eq:WLalgebra}), and eigenvalue equations, eq.~(\ref{eq:wigeigenequations}), are immediately familiar since the Wigner-D functions are precisely the matrix elements of the rotation operator $\mathscr{D}(R)$ of ordinary, non-relativistic quantum mechanics~\cite{sakurai_napolitano_2017}
\begin{equation}
\mathscr{D}^\mathcal{J}_{\mathcal{K}\mathcal{M}}=\bra{\mathcal{J}\mathcal{K}}e^{\left(\frac{-i \mathbf{J}\cdot \hat{n}\phi}{\hbar}\right)}\ket{\mathcal{J}\mathcal{M}} \, .
\end{equation}
For this reason, we refer to $\mathcal{J}$ as the total angular momentum. 
Also, familiar from ordinary quantum mechanics is the ability to raise and lower the eigenvalue of $L_3$ and $W_3$ by the action of ladder operators which, for the eigenvalue of $W_3$, can be defined as $W_{\pm}:=W_1\pm i W_2$. 
Using (\ref{eq:wigeigenequations}), we can express the action of $W_\pm$ and $W_3$ on $D^\mathcal{J}_{\mathcal{KM}}$ as
\begin{align}
    W_+D^\mathcal{J}_{\mathcal{KM}}&=i\sqrt{(\mathcal{J}+\mathcal{K})(\mathcal{J}-\mathcal{K}+1)}D^\mathcal{J}_{\mathcal{K}-1\ \mathcal{M}} \, \\
    W_-D^\mathcal{J}_{\mathcal{KM}}&=-i\sqrt{(\mathcal{J}-\mathcal{K})(\mathcal{J}+\mathcal{K}+1)}D^\mathcal{J}_{\mathcal{K}+1\ \mathcal{M}} \, \\ 
    W_3 D^\mathcal{J}_{\mathcal{KM}}&=\mathcal{K} D^\mathcal{J}_{\mathcal{KM}}  \,
\end{align}
These ladder operators take a particularly useful form when rewritten in terms of the partial derivatives~\footnote{Note again here $a=1,2,3$ and $\partial_a=\partial/\partial Y^a$ where as before $Y^a$ is the angular set of coordinates  $Y^a=(\theta,\phi,\psi)$}, $\partial_\pm={\sigma_\pm}^a\partial_a$ and $\partial_3={\sigma_3}^a\partial_a$ 
\begin{align}
\partial_+D^\mathcal{J}_{\mathcal{KM}}&=\sqrt{(\mathcal{J}+\mathcal{K})(\mathcal{J}-\mathcal{K}+1)}D^\mathcal{J}_{\mathcal{K}-1\ \mathcal{M}} \, \\
    \partial_-D^\mathcal{J}_{\mathcal{KM}}&=-\sqrt{(\mathcal{J}-\mathcal{K})(\mathcal{J}+\mathcal{K}+1)}D^\mathcal{J}_{\mathcal{K}+1\ \mathcal{M}} \, \\ 
    \partial_3 D^\mathcal{J}_{\mathcal{KM}}&=-i\mathcal{K} D^\mathcal{J}_{\mathcal{KM}}  \, .
\end{align}
Finally, given that our goal is to use the expressions detailed above to separate the equations of motion of the metric fluctuations it will be advantageous to rephrase the metric in terms of the $\pm$ basis where $\sigma^\pm = \frac{1}{2} \left(\sigma^1 \mp i \sigma^2 \right)$ and  $\sigma_\pm  = \sigma_1 \pm i \sigma_2$. 
In this basis, the metric in eq.~(\ref{eq:Simple_Spin_MP_Metric}) now takes the form
\begin{align}\label{eq:metric5dmpadssigpm3}
    {ds}^2 = & -\left( 1 + \frac{r^2}{\ell^2} \right) {dt}^2 + \frac{{dr}^2}{G(r)} + \frac{r^2}{4} \left(4 \sigma^{+}\sigma^{-} + (\sigma^3)^2 \right) +  \frac{2 \mu}{r^2} \left(dt + \frac{a}{2} \sigma^3 \right)^2\, .
\end{align} 
\subsection{General ansatz for spin 2 fields}
We are now in the position to describe the general ansatz for the fluctuations. As originally discussed in~\cite{Murata:2007gv} the choice of ansatz is facilitated by two parts, a decomposition of coordinate dependence of $h_{\mu\nu}$ in terms of a suitable basis of eigenfunctions and a compensating choice of frame/basis, i.e. a decomposition of the fluctuation in the invariant basis $h_{\mu\nu}=h_{ij}\lsigma_\mu^i\lsigma_\nu^j$. Here we have introduced $\lsigma^a$ as an extension of the frame basis~\footnote{In these expressions Latin symbols near the beginning of the alphabet, $a,b,c$ take values $1,2,3$ over the angular coordinates, Latin symbols near the middle of the alphabet $i,j,k$ take the values $t,1,2,3,r$. } defined in eq. (\ref{eq:form_basis}) with $\lsigma^i=(\exd t, \sigma^+,\sigma^-,\sigma^3,\exd r)$.  However when constructing the decomposition it is important to note that $\sigma^b$ are charged under the operator $W_{3}$, 
\begin{equation}
\mathcal{L}_{W_3}\sigma^{\pm}=\pm\sigma^{\pm}\, , \qquad \mathcal{L}_{W_3}\sigma^{3}=0 \, .
\end{equation}
Hence in constructing the ansatz it is important to compensate for the additional charge of the basis under $W_3$. This can be written succinctly by introducing the function $Q$, defined as
\begin{equation}
    Q(\lsigma^i) = \left\{ \begin{array}{cc} 0 & i=r,t,3 \\1 & i=+ \\ -1 & i=- \end{array} \right. 
\end{equation}
This auxiliary function $Q$ gives the charge of the form basis under the operator $W_3$. With it, we write the ansatz for the fluctuations as
\begin{equation}\label{eq:pertsimplygeneric}
    h_{\mu\nu} = \int d\omega e^{-i\omega t} \sum_{\mathcal{J} = 0} \sum_{\mathcal{M}=\mathcal{J}}^{\mathcal{J}} \sum_{\mathcal{K'}=-(\mathcal{J}+2)}^{\mathcal{J}+2} h_{i j}(r,\omega, \mathcal{J},\mathcal{M},\mathcal{K}') \lsigma^i_{\mu} \lsigma^j_{\nu} D_{\mathcal{K'}-Q(\lsigma^{i})-Q(\lsigma^{j}) \mathcal{M}}^\mathcal{J}
\end{equation}
One can choose a particular mode since the modes of different $\left(\mathcal{J}, \mathcal{K'}\right)$ decouple. 

Note, we choose to work in the radial gauge $h_{r\mu} = 0$, in order to find solutions to \eqref{eq:pertgenericeom} in what follows.

\section{Results}
\label{sec:results}
In this section we collect and summarize the salient results from the recent studies of quasinormal modes (QNMs) of metric fluctuations in the MP BH geometry~\cite{Garbiso:2020puw,Cartwright:2021qpp,Amano:2022mlu}. 
These QNM results are holographically dual to the dispersion relations of hydrodynamic and nonhydrodynamic modes in the dual fluid.  
Technical details of the numerical methods developed and used are gathered in the appendices.

\subsection{QNMs dual to hydrodynamic and non-hydrodynamic eigenmodes}\label{sec:QNMresults}
QNMs are eigenmodes of non-hermitean operators that are in general complex-valued. In the context of linear perturbations in general relativity, the hermiticity of the linear operator given in eq.~\eqref{eq:pertgenericeom} is lost due to the ingoing boundary conditions, as modes may fall into the BH but cannot come out. We provide here an overview of the QNM spectrum of perturbations whose basis consists of eigenfunctions whose eigenvalues under the $W_3$ operator are $\mathcal{K}=\mathcal{J}$, $\mathcal{K}=\mathcal{J}-1$, or $\mathcal{K}=\mathcal{J}-2$ as originally analyzed in~\cite{Garbiso:2020puw}. 

Unique to this set of perturbations is a limiting case whose spectrum is that of a boosted Schwarzschild black brane. 
Reminiscent of the classification of metric fluctuations under rotations in the Schwarzschild black brane, these ``sectors'' are referred to as tensor, vector, and scalar sectors respectively. Given that there is a large parameter space one can explore, we limit our discussion to cases where the background spacetime configuration is away from the Hawking-Page phase transition which is dual to the deconfinement transition~\cite{Hawking:1982dh}, and we choose the outer horizon radius such that $T\ell > 1/2$. Below we work in units where the AdS radius is set to unity, $\ell=1$, and the QNM frequency is normalized by the outer horizon radius so that $\nu = \omega/(2r_+)$. 

\subsubsection{Previous QNM results for large BHs}
\label{sec:previousResults}
In the present article, QNMs of rotating MP BHs with small as well as large horizon radius are examined. 
A previous analysis of these QNMs was focused on the large temperature or large BH limit~\cite{Garbiso:2020puw}. It is so called because the dual field theory temperature, $T$, is proportional to the horizon radius, $r_+$, which was taken large.
This limit was taken directly on the level of the equations of motion, eq.~\eqref{eq:pertgenericeom}, and is defined by
\begin{equation}\label{eq:LBH1}
r\to \alpha r \,, \qquad 
r_+\to \alpha r_+ \,, \qquad
\alpha\to\infty \,,
\end{equation}
with
\begin{equation}\label{eq:LBH2}
\omega\to\alpha 2\nu r_+/\ell \,, \qquad
\mathcal{J}\to\alpha j r_+/\ell \,, \qquad
\alpha\to\infty \,,
\end{equation}
where only terms leading order in $\alpha$ are kept in the fluctuation equations. QNM frequencies are reported in terms of the dimensionless frequency, $\nu$, which is a function of the dimensionless angular momentum parameter, $j$.

As $\omega$ and $\mathcal J$ scale with $r_+$, as $r_+$ is taken large, fluctuation frequency $\nu$ and angular momentum $j$ are necessarily small at any finite values of $\omega$ and $\mathcal J$.
In this limit,  
instead of decoupling according to $(\mathcal J, \mathcal K)$, modes decouple according to the $\mathcal K$ charge of the $\sigma$ basis, e.g. the $\mathcal{K}$ charge of $h_{++} \rightarrow 2$ and $h_{3-} \rightarrow -1$.
Up to parity, this results in three sectors: tensor, scalar, and vector.\footnote{In~\cite{Garbiso:2020puw} there was a mistake in obtaining the fluctuation equations for the boosted black brane, which is referred to as case (iii) in~\cite{Garbiso:2020puw}. Thus, in~\cite{Garbiso:2020puw} it was not recognized that the metric fluctuation equations in the large BH limit~\eqref{eq:LBH1} and~\eqref{eq:LBH2} are identical to those around a boosted black brane. Already in~\cite{Cartwright:2021qpp} it was realized that the large BH limit leads to this identification. We stress, however, that the metric in this limit is {\it not} that of a boosted black brane, but that of a nontrivially rotating BH. 
}
Below we reproduce figures from~\cite{Garbiso:2020puw} that show the dispersion relations of hydrodynamic and low-lying non-hydrodynamic QNMs in the vector and scalar sector in the large BH limit.
\begin{figure}[ht]
    \centering
    \includegraphics[width=1\textwidth]{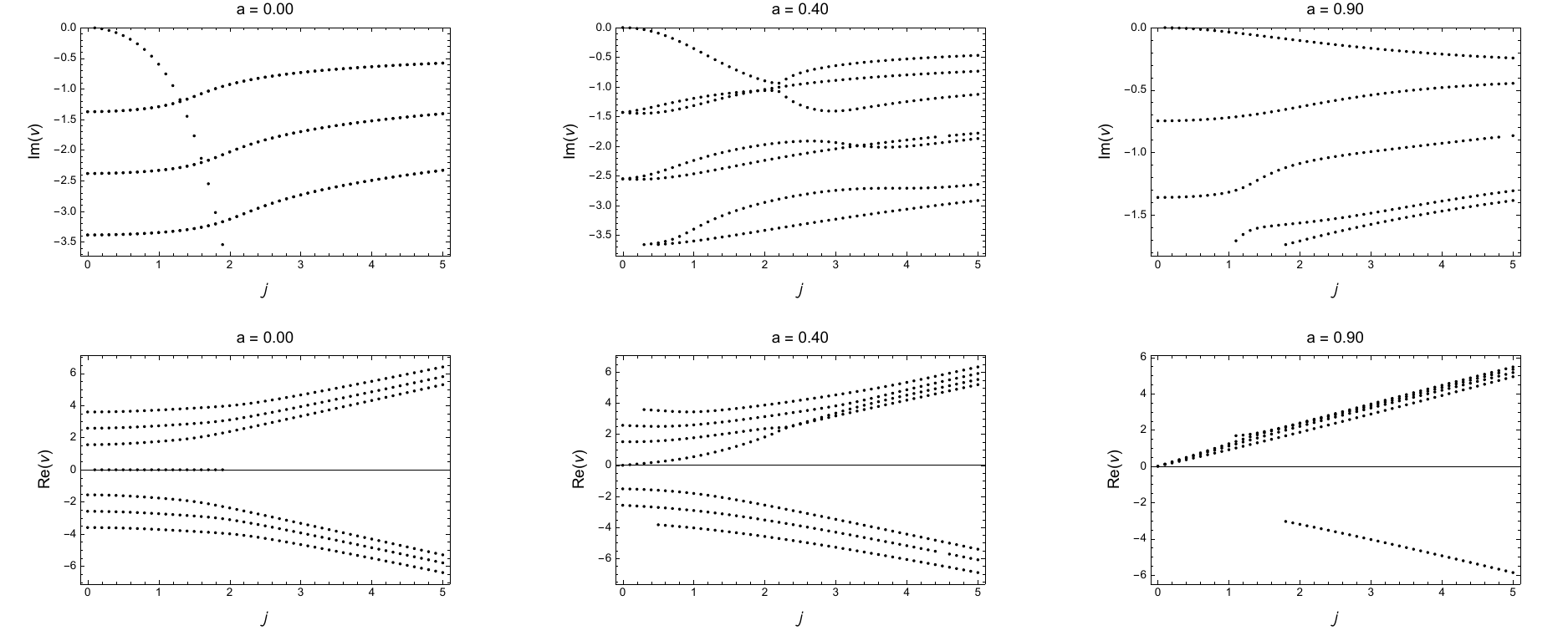}
    \caption{\label{fig:qnfsh3+t+} 
    Vector sector dispersion relations. 
    Displayed here are the real and imaginary parts of quasinormal frequencies of $h_{3+}$ and $h_{t+}$  for large BHs in case (ii) as a function of $j$.
    Modes displayed here were found with precision of at least $d\nu = \nu 10^{-5}$. 
    Missing modes did not converge to sufficient precision and have been filtered out by the numerical routine.
    }
\end{figure}
\begin{figure}[ht]
    \centering
    \includegraphics[width=1\textwidth]{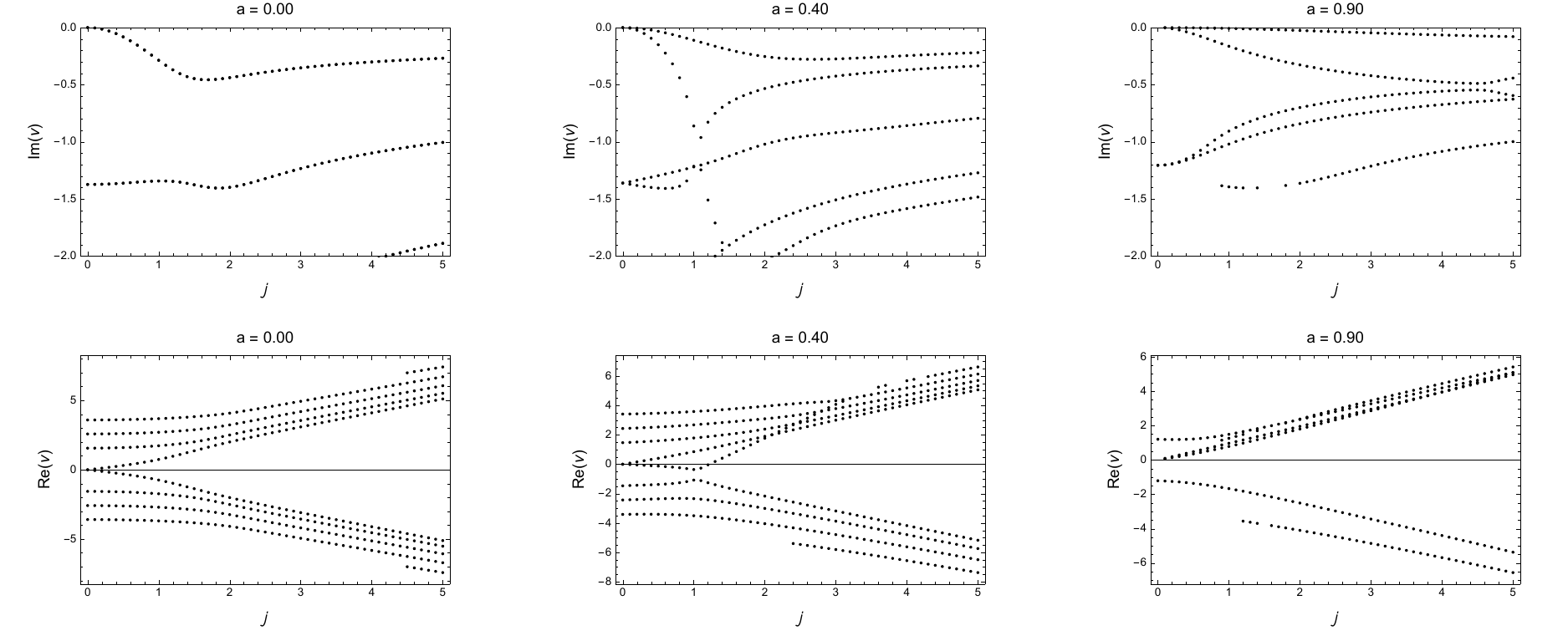}
    \caption{
    Scalar QNMs of the simply spinning BH in the large BH limit.
    There are two hydrodynamic modes, which we identify as sounds modes, which mirror each other regarding their real part at $a/\ell=0$, and both have the same imaginary part. 
    At $a/\ell=0$ these two modes behave differently. 
    Modes displayed here were found with precision of at least $d\nu = \nu 10^{-5}$. 
    Missing modes did not converge to sufficient precision and have been filtered out by the numerical routine.
    } 
    \label{fig:qnfsscalar}
\end{figure}
Note the notation used in~\cite{Garbiso:2020puw} is slightly different: in Fig.~\ref{fig:qnfsh3+t+} and~\ref{fig:qnfsscalar}, $a \equiv a/\ell$ is implicit as $\ell = 1$. Note also there is a sign flip in the values of $a$ used in~\cite{Garbiso:2020puw}, so the real part of the QNMs found there has the opposite sign when compared to the results in this work below. 

\begin{figure}
    \centering
    \includegraphics[width=1\textwidth]{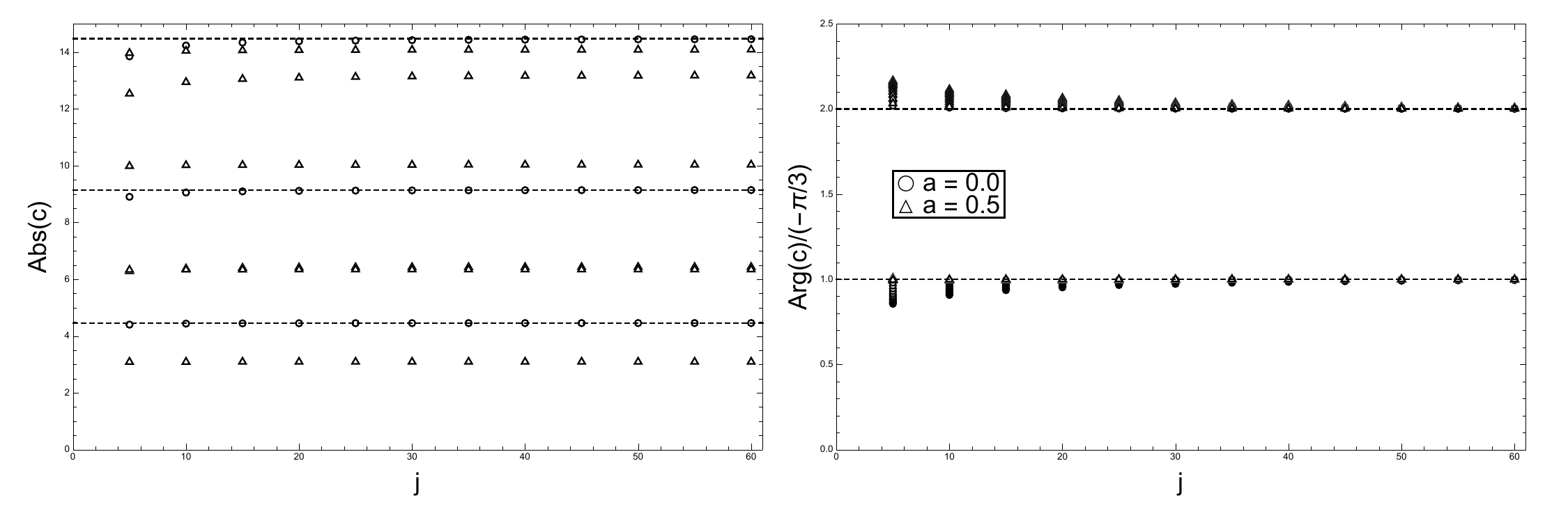}
    \caption{  \label{fig:hpplargej}
    Large $j$ (eikonal limit) analysis. Shown here are the quantities $\text{Abs}(c)$ and $\text{Arg}(c)/(-\pi/3)$, The dashed lines are the expected values according to~\cite{Festuccia:2008zx,Fuini:2016qsc}. For large $j$, the values at $a/\ell=0$ (circles) and $a/\ell=0.5$ (triangles) in the right plot converge to the predicted ones. In the left plot, the same is true for the case $a/\ell=0$ (circles). But for $a/\ell \neq 0$ (triangles), the large $j$ asymptotes change depending on the value of $a/\ell$. These plots are generated from data computed in the large BH limit. 
    }
\end{figure}
To further study the role of the angular momentum parameter $j$, 
QNMs for large $j$ 
were calculated in~\cite{Garbiso:2020puw}. In the {\it eikonal limit}, which is defined by $j\to\infty$, a general analytic expectation for the behavior of QNM frequencies was known~\cite{Festuccia:2008zx,Fuini:2016qsc}. Reproduced in Fig.~\ref{fig:hpplargej} is the figure from~\cite{Garbiso:2020puw} that shows the large BH limit tensor sector result confirming the known scaling in~\cite{Festuccia:2008zx,Fuini:2016qsc}.\footnote{Again, $a \equiv a/\ell$ as $\ell=1$ is implicit here.}
As $j$ is equivalent to the linear momentum $k$~\cite{Cartwright:2021qpp}, the quantity $c(k)\to c(j)$  defined in~\cite{Festuccia:2008zx,Fuini:2016qsc} satisfies the predicted behavior at large values of $j$. The numerical results in~\cite{Garbiso:2020puw} imply the following large-$j$ expansion for the $n$th mode: $2 \nu = 2j+c_n e^{-i (\pi /3)} (2 j)^{-1/3}+O((2 j)^{-1})$, where the conjugate frequency $-2 \nu^* = -2 j+c_n e^{-i (2\pi /3)} (2 j)^{-1/3}$ is also included in our analysis, and 
$c_n$ are real numbers in agreement with~\cite{Fuini:2016qsc}.\footnote{In order to relate to~\cite{Fuini:2016qsc}, the following identifications are needed: $\omega_F\equiv2\nu$ and $q_F\equiv2 j$ with subscript ``F'' indicating quantities in~\cite{Fuini:2016qsc}.} 
This observation further confirmed that the angular momentum $j$ is associated with the linear momentum $q$ from~\cite{Policastro:2002se}. As we now know, in the large BH limit 
the two are related by $j=q/(2\pi T)$. 
 
Summarizing the remaining results from~\cite{Garbiso:2020puw}, mostly the hydrodynamic behavior in large simply spinning five-dimensional AdS BHs was analyzed in the large BH limit~\eqref{eq:LBH1} and~\eqref{eq:LBH2}. These are holographically dual to spinning quantum fluids. Due to the spatial anisotropy introduced by
the angular momentum, hydrodynamic transport coefficients are split into groups longitudinal
or transverse to the angular momentum, and aligned or anti-aligned with it. Analytic
expressions are provided for the two shear viscosities, $\eta_{||}$ and $\eta_{\perp}$, the longitudinal momentum diffusion coefficient, $D_{||}$, two speeds of sound, $v_\pm$, and two sound attenuation coefficients, $\Gamma_\pm$ in the momentum diffusion dispersion relation $\nu = {v}_{||} j - \mathcal{D}_{||} j^2$ and the sound dispersion relation $\nu=v_{s,\pm} \, j - i \Gamma_{s,\pm} \,j^2$~\cite{Garbiso:2020puw}: 
\begin{eqnarray}
 \eta_{\perp}(a) &=& \eta_0 \frac{1}{\sqrt{1-a^2}}\, , \label{eq:etaPerpSum} \\
  \eta_{||}(a) &=& \eta_0 {\sqrt{1-a^2}}\, , \label{eq:etaLSum}\\
  v_{||}(a) & = & a \, , \label{eq:vLSum}\\
   \mathcal{D}_{||}(a) &=&  (2\pi T)\, \mathcal{D} \, (1-a^2)^{3/2} = \frac{1}{2} (1-a^2)^{3/2} \, , {\label{eq:DLSum4}}\\ 
   v_{\pm} (a) & = & v_{0} \frac{\sqrt{3} a\pm 1}{1\pm \frac{a}{\sqrt{3}}}\, , \label{eq:vsSum}\\
 \Gamma_{\pm}(a) & = & \Gamma_0 \frac{\left(1-a^2\right)^{3/2}}{\left(1\pm \frac{a}{\sqrt{3}}\right)^3} \, ,{\label{eq:GsSum}}
\end{eqnarray}
where for the holographic ($\mathcal{N}=4$ SYM) fluid the quantities at vanishing angular momentum, $a=0$, are given by: $\eta_0=N^2 \pi T_0^3/8$, conformal speed of sound $v_{0}=1/\sqrt{3}$, and conformal sound attenuation $\Gamma_0=1/3$ . 
Known relations between these coefficients are generalized to include dependence on angular momentum: 
\begin{eqnarray}\label{eq:DLSum}
 \mathcal{D}_{||}(a)
    &=&2 \pi T_0 \frac{\eta_{||}(a)}{\epsilon(a) + P_\perp(a)}\, , \quad (\text{Einstein relation})\\
    \label{eq:GammaSum}
    \Gamma_\pm(a)&=&\frac{2\eta_{||}(a)}{3(\epsilon(a) + P_\perp(a))} \frac{1}{(1\pm a/\sqrt{3})^3} \, .
\end{eqnarray}
The shear viscosity to entropy density ratio varies between zero and $1/(4\pi)$, depending on the direction of the shear, varying from transverse to the anisotropy of the fluid $\frac{\eta_\perp}{s}= \frac{1}{4\pi}$ to perpendicular to that anisotropy $\frac{\eta_{||}}{s}= \frac{1}{4\pi} (1-a^2)$ within $-1< a < 1$. 

As pointed out above, these relations~\eqref{eq:etaPerpSum} to~\eqref{eq:GammaSum} are those for shear transport, momentum diffusion and sound propagation in a boosted fluid, as discussed in~\cite{Kovtun:2019hdm,Hoult:2020eho}. The holographic fluid considered here is truly rotating (not just a boosted fluid), however, when probed by low energy fluctuations (hydrodynamic fluctuations), then their dispersion relations look like those of a fluid boosted with boost parameter $a$. This statement gets corrected at any finite horizon radius when the large BH limit is not taken, as discussed in Sec.~\ref{sec:HorizonDependence} and Sec.~\ref{sec:critical_Points}.  

These results furthermore demonstrate that large simply spinning five-dimensional MP BHs in the large BH limit are perturbatively stable for all angular momenta which we tested below extremality (which occurs at the same angular momentum $a=\ell$ in this limit as the superadiance instabilities). Away from the large BH limit, this statement is limited by superradiance instabilities which then occur at lower values of $a$ than extremality, as discussed in Sec.~\ref{sec:instabilities}. 
Beyond the large BH limit, also the QNMs for metric fluctuations around the full rotating BH metric were considered, merely evaluated at a large value for $r_+=100$, this is referred to as ``case (i)'' in~\cite{Garbiso:2020puw}. Qualitatively and quantitatively, the results from the large BH limit were confirmed by this computation. 

One major task in the present article is to go beyond large BHs and examine intermediate to small BHs, which is mainly discussed in section~\ref{sec:HorizonDependence}.

\subsubsection{$\mathcal{K}=\mathcal{J}$ --- Tensor fluctuations}\label{sec:tensor_sector}
\begin{figure}
\centering
    \includegraphics[width=0.75\textwidth]{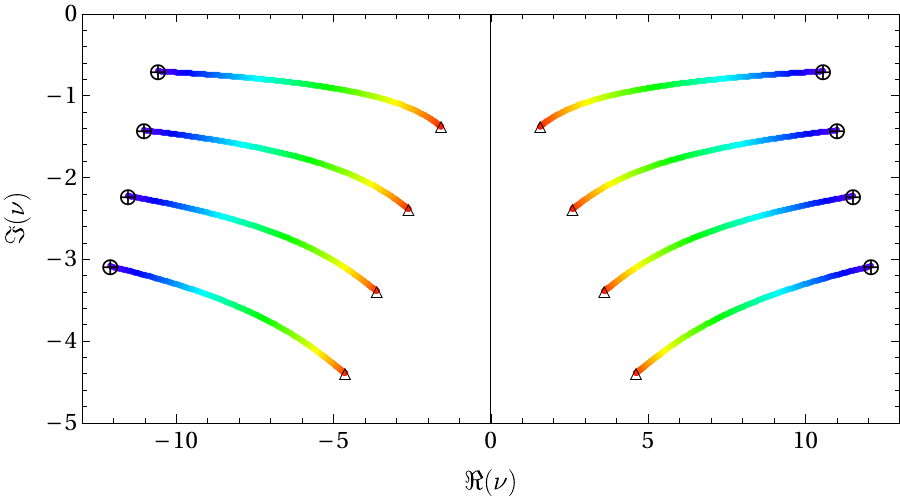} \\
    \includegraphics[width=0.75\textwidth]{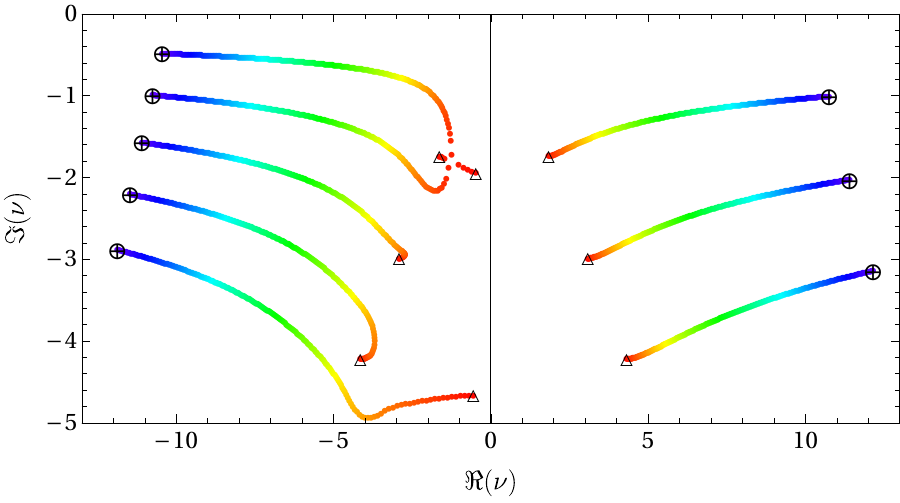} \\
    \includegraphics[width=0.75\textwidth]{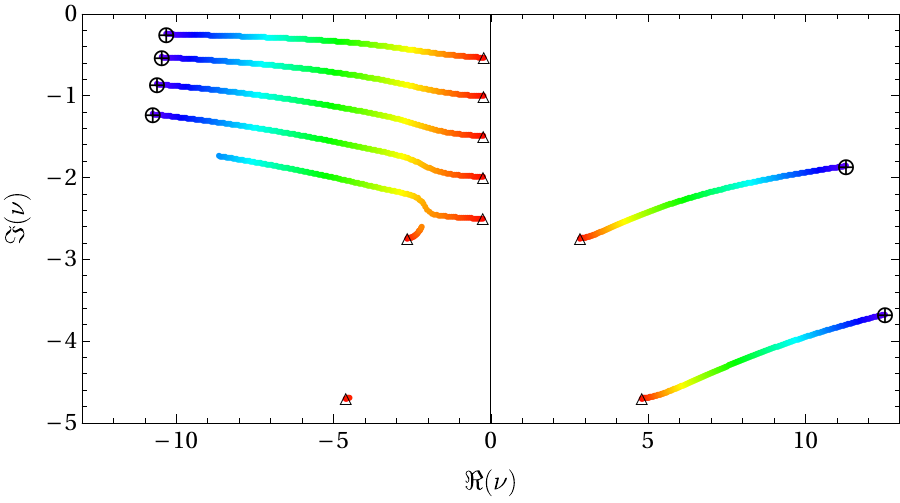}
    \caption{Scaled tensor modes, $\nu$, for three values of $a/\ell \in \{0, 1/2, 9/10\}$ at $r_+/\ell = 10$. \textit{Top:} $a/\ell=0$. \textit{Middle:} $a/\ell=1/2$. \textit{Bottom:} $a/\ell=9/10$. The total angular momentum of the fluctuation $\mathcal{J} = 0,1/2,1,\ldots,199/2,100$ increases in steps of $1/2$. The curves begin at $\mathcal{J}=0$ represented by an empty triangle ($\Delta$), and end at $\mathcal{J}=100$, represented by a circle with a plus mark ($\bigoplus$). 
    }
    \label{fig:tensor_modes_over_a}
\end{figure}
For $\mathcal{K}=\mathcal{J}$, the fluctuation $h_{++}$ decouples from all other fields and hence it is a naturally gauge invariant master variable for this class of fluctuations. In Fig.~\ref{fig:tensor_modes_over_a} we display the spectrum for three different values of angular momentum per mass $a/\ell=0,1/2,9/10$, for total angular momentum of the fluctuation $\mathcal{J}=0,1/2,1,\ldots,100$ (increasing in steps of $1/2$), and horizon radius $r_+/\ell=10$.

From Fig.~\ref{fig:tensor_modes_over_a} it is immediately clear that there is no mode with frequency $\nu$ for which $\lim_{\mathcal{J}\rightarrow 0}\nu=0$, and hence there are no ``hydrodynamic'' modes in the tensor sector. This is quite remarkable as the $h_{++}$ fluctuation couples to other fluctuations at nonzero $a$. 

At $a/\ell=0$ the modes are mirror symmetric about the imaginary axis. This is due to the time reversal invariance in the background geometry present when $a/\ell=0$, but broken by the angular momentum when $a/\ell \neq 0$, which then separates modes into those rotating with and against the BH/fluid. 
Indeed, as $a/\ell$ becomes positive definite and increases towards one, the trajectories of modes as $\mathcal{J}$ varies move down and further apart for $\Re(\nu)>0$, but they move up and closer together for $\Re(\nu)<0$. Furthermore, QNMs from deep down in the $\Im(\nu)<0$ complex half plane are pushed up vertically towards $\Im(\nu)=0$, moving up like a "bowl of a spoon" that then joins up with the "handle of the spoon" provided by the $\mathcal{J}$-trajectory of each successively higher mode. This behavior as $a/\ell$ increases resembles that of the purely imaginary QNMs in the Reissner-Nordstr\"om BH as the charge is pushed towards extremality~\cite{Janiszewski:2015ura}.

\begin{figure}
    \centering
    \includegraphics[width=0.75\textwidth]{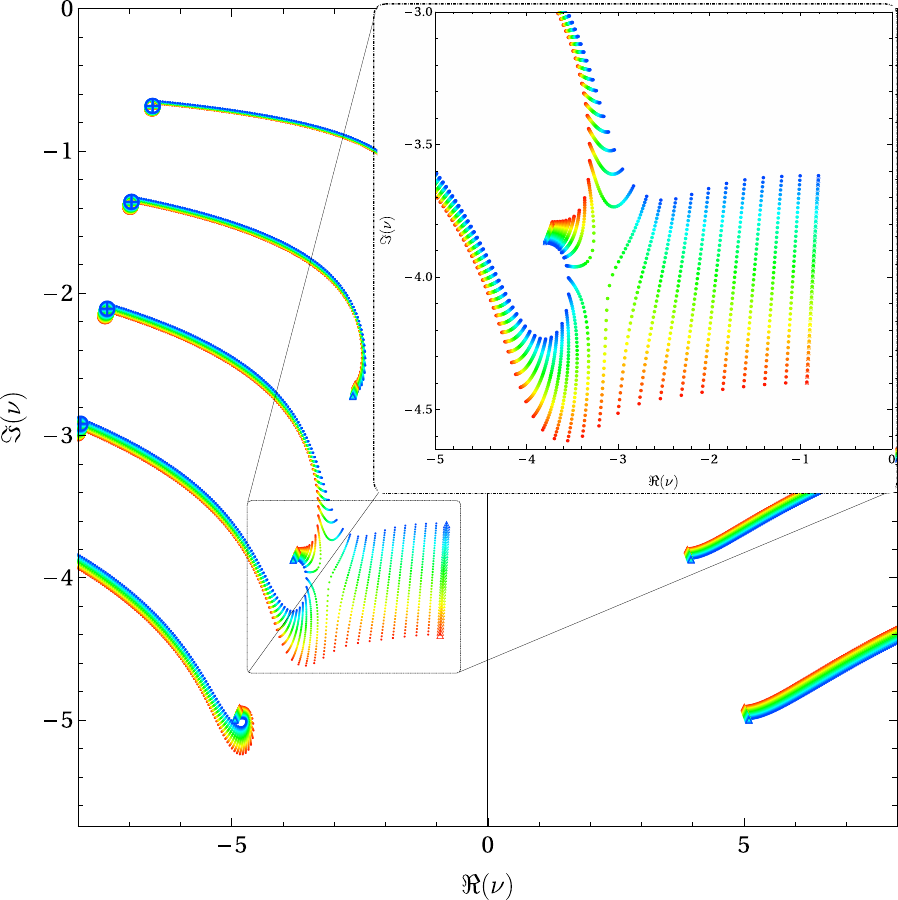} \\
      \includegraphics[width=0.75\textwidth]{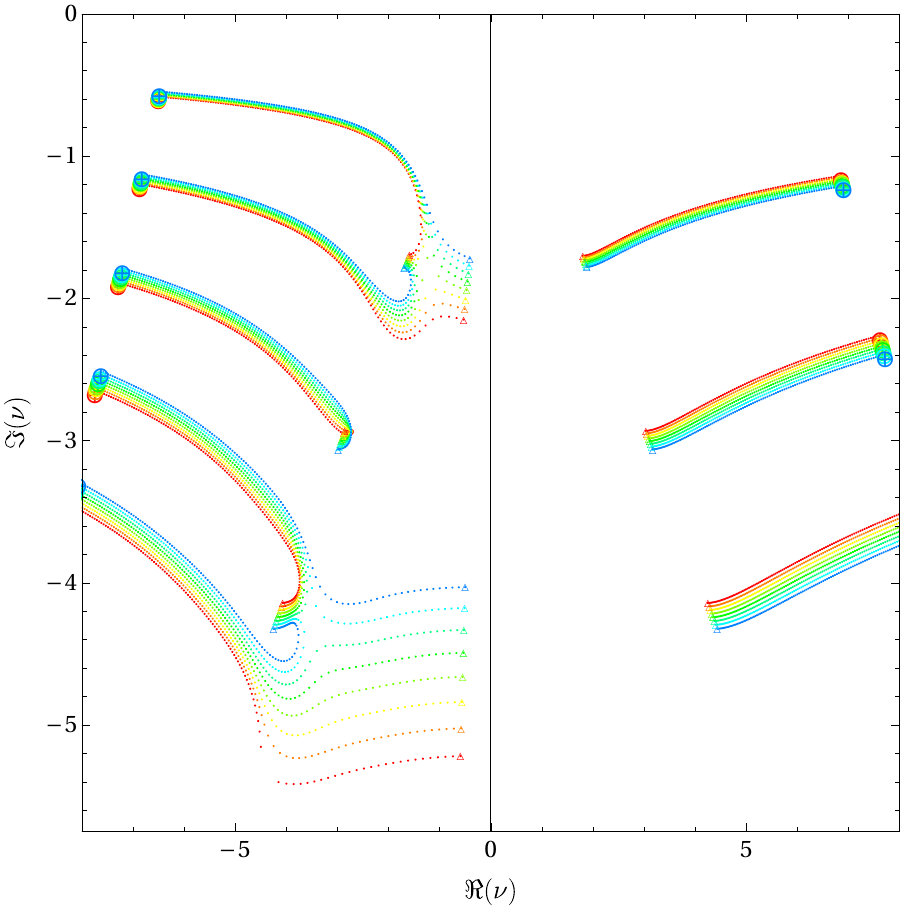}
    \caption{Scaled tensor modes, $\nu$, for two different ranges of $a/\ell$. \textit{Top:} $a/\ell\in[0.3,0.34]$ in steps of $\Delta a/\ell=0.001$. \textit{Bottom:} $a/\ell\in[0.47,0.54]$ in steps of $\Delta a/\ell=0.01$. Both images were computed with $r_+/\ell = 10$ and the total angular momentum $\mathcal{J}=0,1/2,\ldots,60$ increasing in steps of $1/2$. Colors denote the different values of $a/\ell$, \textcolor{red}{red} corresponds to the minimum value of the $a/\ell$ in the range and \textcolor{blue}{blue} corresponds to the maximum value in the range. For each curve triangles ($\Delta$) correspond to $\mathcal{J}=0$ while a circle with a plus mark ($\bigoplus$) corresponds to $\mathcal{J}=60$. The inset of the top image zooms in on the region containing the dovetailing behavior. An animated GIF showing this in real time is included in the supplementary material. 
    \label{fig:vary_A_small_scale}}
\end{figure}
To show in more detail how the QNM spectrum changes in the tensor sector, and in particular this movement of QNMs coming up from the deep of the $\Im(\nu)<0$ complex half plane as $a/\ell$ changes, we display in the top of Fig.~\ref{fig:vary_A_small_scale} the spectrum for $a/\ell\in[0.30,0.34]$ in steps of $\Delta a/\ell=0.001$. We see from the inset a behavior that looks like "level touching", which is quite remarkable as $\mathcal{J}$ is discrete here: first a spoon is formed with the $\mathcal{J}$-trajectory of a lower mode, but as the bowl of the spoon is pushed up with increasing $a/\ell$, at some point the "neck of the spoon" curls up so much it breaks off and a new spoon is formed with the upper mode. This "spoon formation" continues as $a/\ell$ is increased to ever higher values until there is a spoon for each of the modes, starting with the highest. A snapshot of the spoon forming process in the early and late stages are shown in the bottom of Fig.~\ref{fig:vary_A_small_scale} and Fig.~\ref{fig:tensor_modes_over_a} respectively. An animated GIF showing the this process in real time is also provided in the supplementary material.

Another feature to note from Figs.~\ref{fig:tensor_modes_over_a} and~\ref{fig:vary_A_small_scale} is that as $a/\ell$ increases and the spoons form, the $\mathcal{J}=0$ mode approaches the imaginary axis. This behavior is perhaps not surprising as a branch cut is expected to form when linear differential equations contain essential singularities~\footnote{For a general discussion see~\cite{OrszagBender}, while specific examples of this in the fluctuations around Reissner-Nordstr{\"o}m BHs can be found in~\cite{Edalati:2009bi,Edalati:2010hk}.}, and this can only happen if modes can come back onto the imaginary axis and coalesce.

\subsubsection{$\mathcal{K}=\mathcal{J}-1$ --- Vector fluctuations}\label{sec:vector_sector}
\begin{figure}
\centering
    \includegraphics[width=0.85\textwidth]{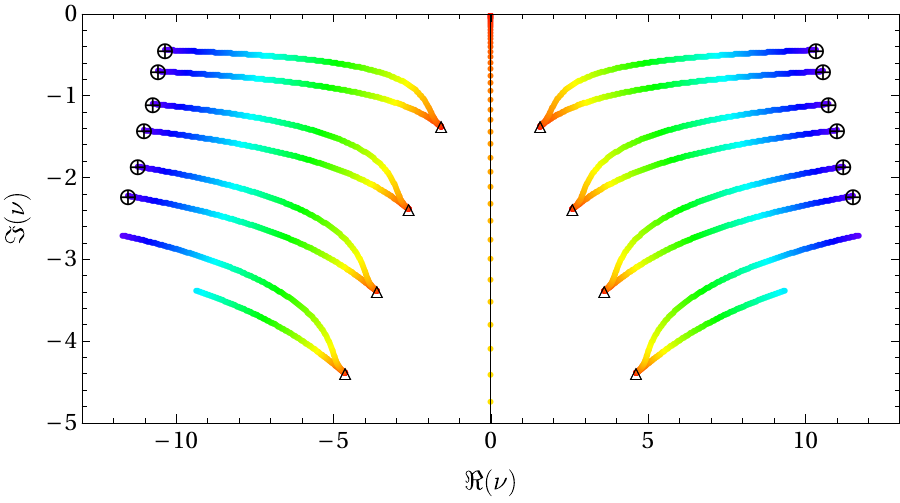} \\
    \includegraphics[width=0.85\textwidth]{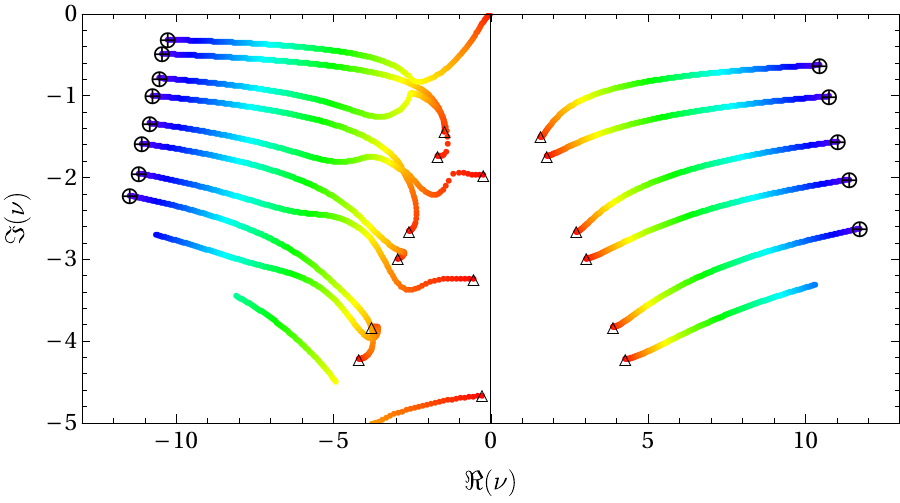} \\
    \includegraphics[width=0.85\textwidth]{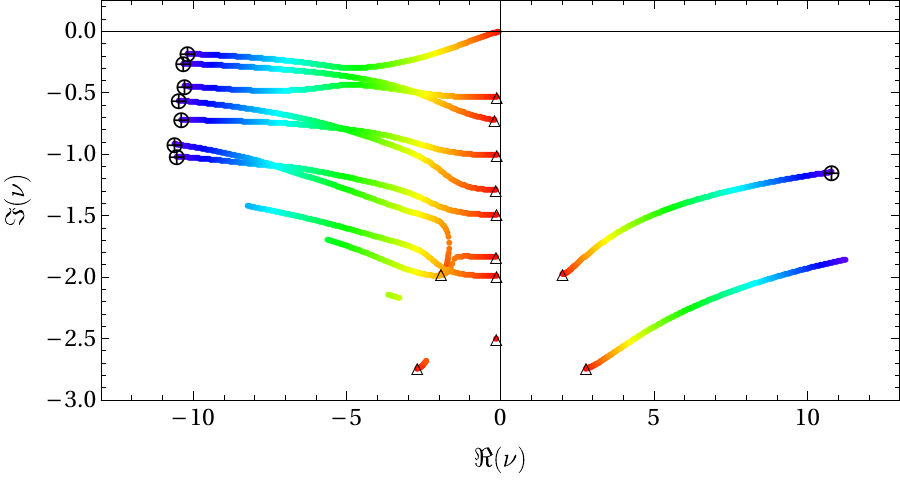}
    \caption{Scaled vector modes, $\nu$, for three values of $a/\ell \in \{0, 1/2, 9/10\}$ at $r_+/\ell = 10$. \textit{Top:} $a/\ell = 0$. \textit{Middle:} $a/\ell = 1/2$. \textit{Bottom:} $a/\ell = 9/10$. The total angular momentum of the fluctuation $\mathcal{J} = 0,1/2,1,\ldots,199/2,100$ increases in steps of $1/2$. The curves begin at $\mathcal{J}=0$ represented by an empty triangle ($\Delta$), and end at $\mathcal{J}=100$, represented by a circle with a plus mark ($\bigoplus$).}
    \label{fig:vector_modes_over_a}
\end{figure}
From the ansatz given in eq.~(\ref{eq:pertsimplygeneric}) one can see that for the choice of $\mathcal{K}=\mathcal{J}-1$ the modes $h_{+t}$, $h_{+3}$, and $h_{++}$ couple together. This occurs for all modes for which $\mathcal{K}<\mathcal{J}$, and all modes for $\mathcal{K}\in [n, \mathcal{J}]$ will couple. In Fig.~\ref{fig:vector_modes_over_a} we display the spectrum for three different values of angular momentum per mass $a/\ell=0,1/2,9/10$, for total angular momentum of the fluctuation $\mathcal{J}=0,1/2,1,\cdots 100$ (increasing in steps of $1/2$), and horizon radius $r_+/\ell=10$. 

Like in the tensor sector, the trajectories of modes are mirror symmetric about the imaginary axis at $a/\ell=0$, but once $a/\ell \neq 0$ this symmetry (due to time reversal invariance) is broken by the angular momentum, and the $\Re(\nu)>0$ trajectories are pushed down relative to the $\Re(\nu)<0$ ones.

Unlike the tensor sector though, there does exist a mode $\nu$ for which $\lim_{\mathcal{J}\rightarrow 0}\nu=0$ and hence a potential hydrodynamic mode in this sector. At $a/\ell=0$, this is the expected shear diffusion mode residing on the imaginary axis. Furthermore, we observe at $a/\ell=0$ each non-hydrodynamic QNM splits into two trajectories, one for $\mathcal{K}=\mathcal{J}$ and the other $\mathcal{K}=\mathcal{J}-1$. 

As $a/\ell$ becomes positive definite and increases, we see from Fig.~\ref{fig:vector_modes_over_a} again this behavior of QNMs deep down in the $\Im(\nu)<0$ complex half plane being pushed up and forming spoons with the $\mathcal{J}$-trajectories of the vector modes. What is interesting here is that as the QNMs move up for $a/\ell > 0$, the two branches previously joined at $\mathcal{J}=0$ when $a/\ell = 0$ split open to form spoons. Even more so is that not only is there "level touching" seen in the middle of Fig.~\ref{fig:vector_modes_over_a} when $a/\ell = 1/2$, there is also "level crossing" seen in the bottom of Fig.~\ref{fig:vector_modes_over_a} when $a/\ell = 9/10$. As $a/\ell$ approaches one, we see again the $\mathcal{J}=0$ modes approaching the imaginary axis as seen in the tensor sector.

At $a/\ell = 0$, the spectrum contains, in addition to the expected vector modes, also an almost exact copy of the tensor modes from Fig.~\ref{fig:tensor_modes_over_a}. This is perhaps surprising given that the fields $h_{++}$, $h_{+t}$, and $h_{+3}$ all couple together and the QNMs frequencies of $h_{++}$ are expected to be modified by this coupling compared to the tensor case discussed above. 
However, away from $a/\ell=0$ one observes strong effects  from the coupling of modes, as discussed above, there is a level crossing between modes formerly from the tensor sector and those from the vector sector. Furthermore, additional modes from the tensor sector continue to appear in the vector sector, although they are no longer a copy of the modes found for $\mathcal{K}=\mathcal{J}$.

\subsubsection{$\mathcal{K}=\mathcal{J}-2$ --- Scalar fluctuations}\label{sec:scalar_sector}
\begin{figure}
    \centering
    \includegraphics[width=0.85\textwidth]{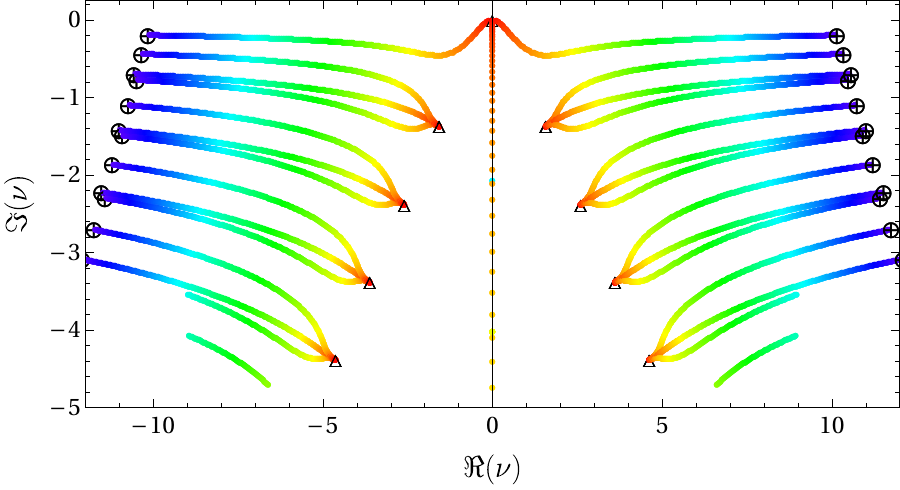}
    \includegraphics[width=0.85\textwidth]{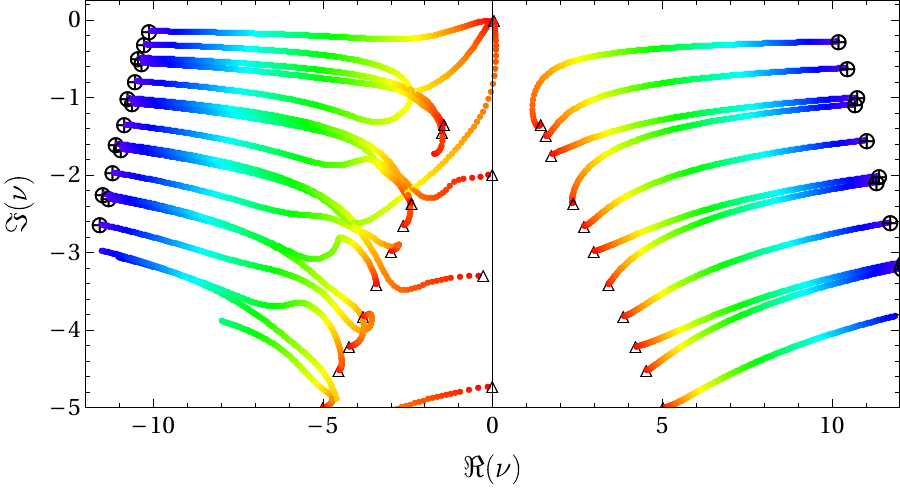}
    \includegraphics[width=0.85\textwidth]{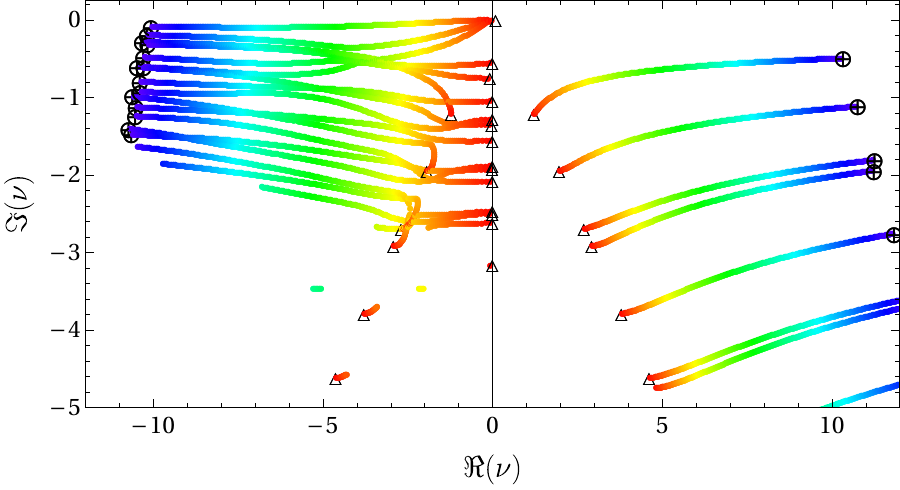}
    \caption{Scaled scalar modes, $\nu$, for three values of $a/\ell\in\{0, 1/2, 9/10\}$ at $r_+/\ell = 10$.
    \textit{Top:} $a/\ell = 0$. \textit{Middle:} $a/\ell = 1/2$. \textit{Bottom:} $a/\ell = 9/10$. The total angular momentum of the fluctuation $\mathcal{J} = 0,1/2,1,\ldots,199/2,100$ increases in steps of $1/2$. The curves begin at $\mathcal{J}=0$ represented by an empty triangle ($\Delta$), and end at $\mathcal{J}=100$, represented by a circle with a plus mark ($\bigoplus$). 
    }
    \label{fig:scalar_modes_over_a}
\end{figure}

As discussed in the beginning of section~\ref{sec:vector_sector}, lowering the eigenvalue of the operator $W_3$ couples fields to all higher $W_3$ eigenfunctions. Thus choosing $\mathcal{K}=\mathcal{J}-2$ introduces four additional fields $h_{33}$, $h_{tt}$, $h_{t3}$, and $h_{+-}$, which couple to $h_{+t}$, $h_{+3}$, and $h_{++}$. In Fig.~\ref{fig:scalar_modes_over_a} we display the spectrum for three different values of angular momentum per mass $a/\ell=0,1/2,9/10$, for total angular momentum of the fluctuation $\mathcal{J}=0,1/2,1,\ldots,100$ (increasing in steps of $1/2$), and horizon radius $r_+/\ell=10$. 

Like in the vector sector, there is also a mode $\nu$ for which $\lim_{\mathcal{J}\rightarrow 0}\nu =0$, and so a potential hydrodynamic mode in the scalar sector. 
In addition, at $a/\ell=0$ the spectrum contains an almost exact copy of both the tensor and vector spectra. Again, this is somewhat unexpected given the nontrivial coupling between the fields, from which one would expect more changes in the mode frequencies. As a result, the scalar sector contains three hydrodynamic modes: two are sound modes new in this sector, and one corresponds to the diffusion mode encountered in the vector sector discussed above. For $a/\ell > 0$, we see again the diffusion mode shifts away from the imaginary axis with an increasingly large negative real part for increasing $\mathcal{J}$. 

Also like in the vector sector, the coupling of the fields in the scalar sector at $a/\ell=0$ leads to bands of non-hydrodynamic QNM modes degenerate at $\mathcal{J}=0$, but split up for $\mathcal{J} > 0$. While there are only bands of two modes in the vector sector, this is enhanced to bands of three in the scalar sector, one for each of the possibilities: $\mathcal{K}=\mathcal{J},\mathcal{J}-1,\mathcal{J}-2$. 
While $\mathcal{J} > 0$ lifts the degeneracy at $a/\ell=0$, this is automatically lifted for $a/\ell > 0$. Indeed, this "split open" of the bands at $\mathcal{J} > 0$ when $a/\ell > 0$ happens as  QNMs deep down in the $\Im(\nu)<0$ complex half plane move up and form spoons with the $\mathcal{J}$-trajectories of the scalar modes. 

Analogous to the vector sector, we see again the "level crossing" behavior in Fig.~\ref{fig:scalar_modes_over_a} as the spoons form with increasing $a/\ell$. As the spoons are locked into place for each of the mode trajectories displayed (starting from the highest), the $\mathcal{J} = 0$ modes again approach the imaginary axis as $a/\ell$ approaches one.

\subsubsection{Horizon radius dependence \& the emergence of hydrodynamics}
\label{sec:HorizonDependence}
In all previous sections (except section~\ref{sec:previousResults}), the horizon radius was chosen to be $r_+=10$ (in units of the AdS radius, $\ell=1$). In this section, we discuss how different values of the horizon radius influence the QNM spectra. 
In particular, in the large BH limit~\eqref{eq:large_Temp_limit}, the dispersion relations of the lowest gapless QNMs display the hydrodynamic behavior expected for $\mathcal{N}=4$ SYM theory in a boosted fluid state. 

In order to study the dispersion relations of the lowest gapless modes, we choose the momentum diffusion sector as an example. Dispersion relations of the lowest gapless QNM in that channel have been computed for various values of the horizon radius $r_+=\{10, 100, 1000, 10^4, 10^5, 10^6, 10^7\}$. The momentum of the mode has been chosen in half-integer steps $\mathcal{J}=0, 1/2, 1, ..., \mathcal{J}_{max}$ up to a maximal value of $\mathcal{J}=\mathcal{J}_{max}$, which is determined by the relation $\mathcal{J}_{max}/r_+ = j_{max} = 0.1$.\footnote{This value of $j$ is well within the radius of convergence indicated by $R_{c}$ as shown in Sec.~\ref{sec:critical_Points}.} 
These values will be used as data points that will be fitted to the expected form of the hydrodynamic dispersion relation below. It is important to note that with this definition of a fit window for the momentum $0<j\le j_{max}$, there are fewer data points to fit at small $r_+$ than at larger $r_+$. This is because $j=\mathcal{J}/r_+$, where $\mathcal{J}$ takes discrete values. This introduces an increasing systematic error for this fit at decreasing $r_+$.  

\begin{figure}
\centering
    \includegraphics[width=0.7\textwidth]{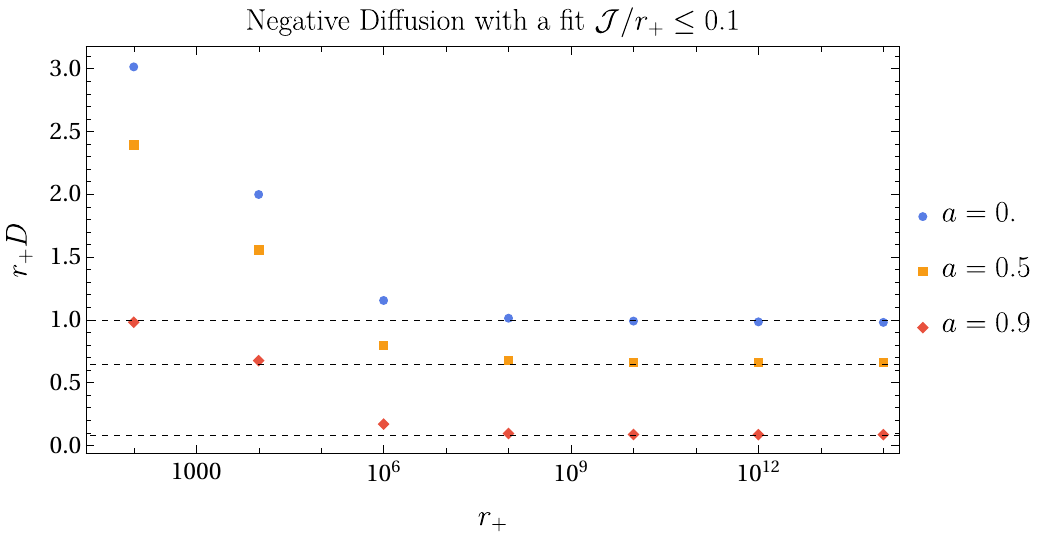} \\
    \includegraphics[width=0.7\textwidth]{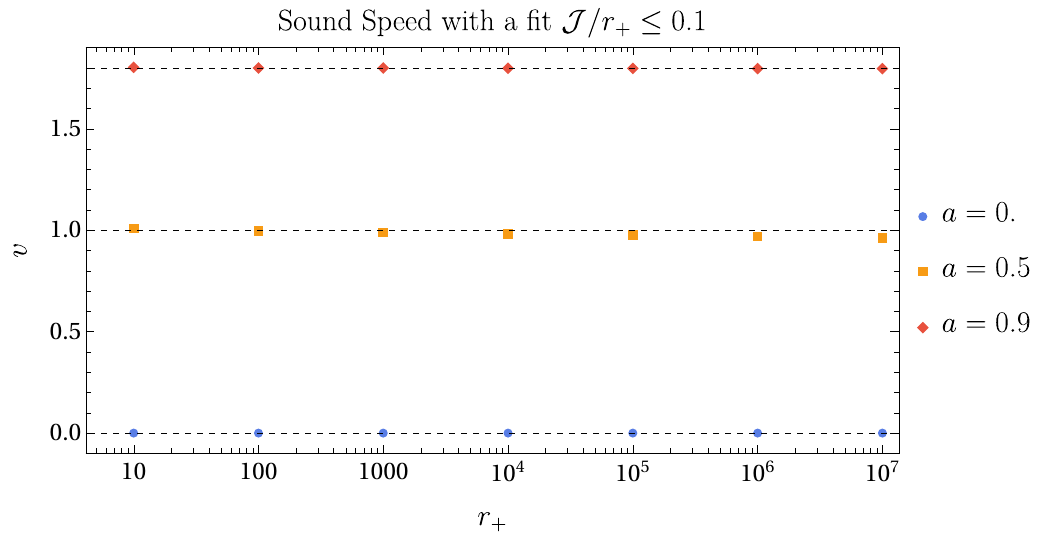} \\
    \includegraphics[width=0.7\textwidth]{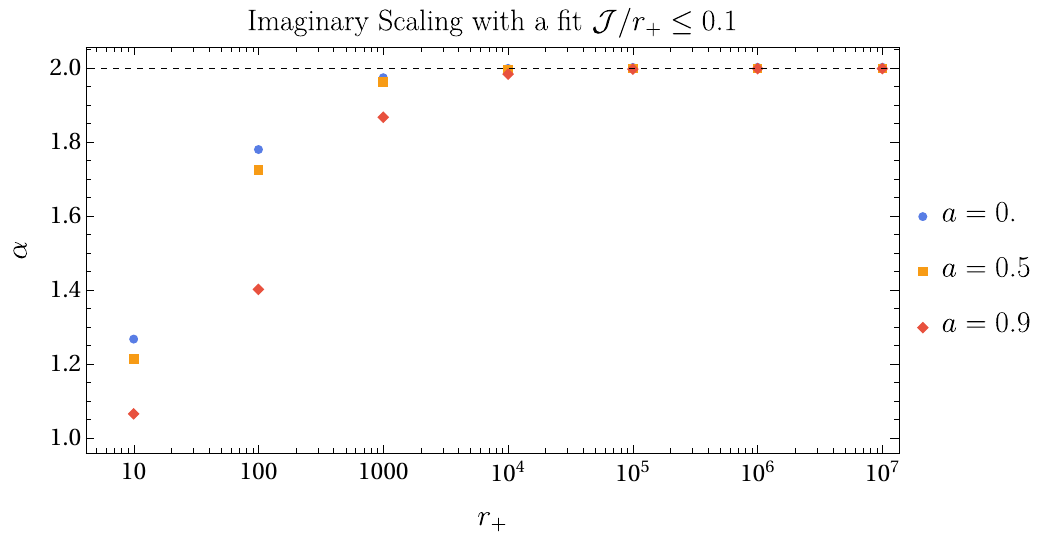}
    \includegraphics[width=0.7\textwidth]{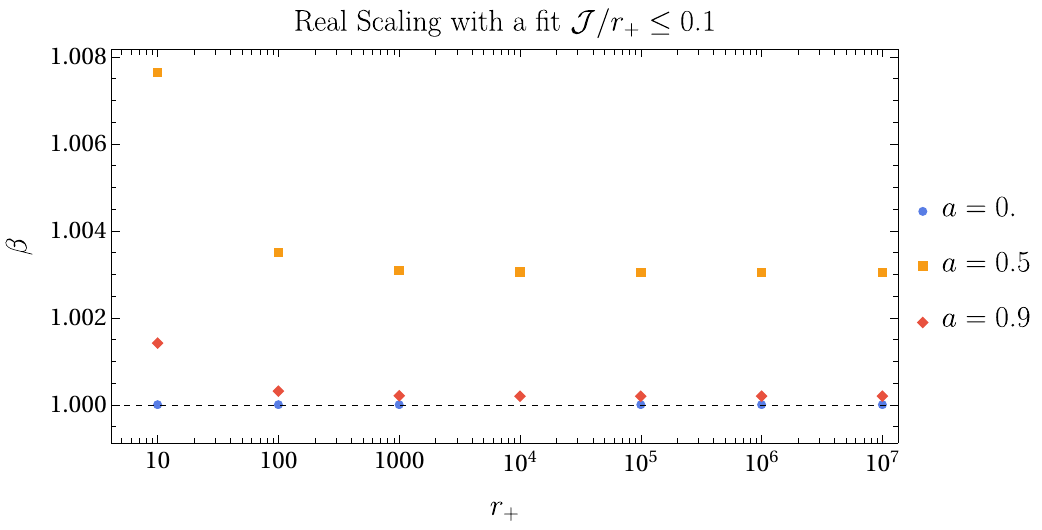}    
    \caption{Horizon radius dependence \& emergence of hydrodynamic scaling. The dashed horizontal lines indicate the values of diffusion coefficient $r_+ D=r_+ \mathcal{D}/4$ and diffusion speed $v$ expected for a fluid boosted with velocity $a=0, 0.5, 0.9$ (taken from Table~\ref{tab:expectedDv} derived in appendix~\ref{sec:hydroDispersion}). 
    Dashed horizontal lines at $\alpha=2$ and $\beta=1$ indicate the expected hydrodynamic dispersion scaling (linear propagation $\propto\mathcal{J}$, quadratic damping $\propto-i \mathcal{J}^2$).}
\label{fig:horizonRadiusDependence}
\end{figure}

Now this QNM data is fitted to the dispersion relation:
\begin{equation}
    \omega = v \mathcal{J}^\beta - i D \mathcal{J}^\alpha \, ,
\end{equation}
where we expect a diffusion mode of $\mathcal{N}=4$ SYM theory at rest ($a=0$) to take the values $\alpha =2,\, \beta=1,\, v=0,\, D=1/(\pi T) = 1/r_+$. Here, $D$ is related to the coefficient defined in~\cite{Policastro:2002se} by $\mathcal{D}=D/4$. The result of this fit is shown in Fig.~\ref{fig:horizonRadiusDependence}. This figure shows that in the fit window $\mathcal{J}/r_+\le j_{max}=0.1$ the expected diffusion scaling $\alpha=2$ occurs for $r_+\approx 1000$ while the linear velocity term scaling $\beta=1$ occurs already at much smaller radii $r_+\approx 10$. For larger values of $a$, these integer scaling values $\alpha=2,\, \beta=1$ are reached at larger $r_+$ values. 

As pointed out in Appendix~\ref{sec:hydroDispersion}, for a boosted fluid with boost velocity $a$ one expects $r_+ D = (1-a^2)^{3/2}$ and $v= 2 a$~\cite{Kovtun:2019hdm,Hoult:2020eho,Garbiso:2020puw}. As seen in Fig.~\ref{fig:horizonRadiusDependence}, these values are approached at large horizon radius $r_+\approx 10^6$ for the diffusion coefficient $r_+ D$, but already at small radii $r_+\approx 10$ for the propagation speed $v$ of this former diffusion mode.   

Finally, we point out two distinct features displayed in Fig.~\ref{fig:horizonRadiusDependence} as the horizon radius (and thus the temperature in the dual field theory) are increased:
\begin{itemize}
    \item The temperature $T$ becomes the largest scale in the system, therefore the hydrodynamic approximation becomes more applicable.
    \item For fluctuations at a fixed energy $\omega$, increasing the horizon radius leads to the large BH limit given by eqs.~\eqref{eq:LBH1} and~\eqref{eq:LBH2}, and such fluctuations become hydrodynamic ($\omega, j\ll T$). This occurs as these fluctuations see less and less of the rotating fluid and perceive it rather as a boosted fluid with boost velocity $a$. 
\end{itemize}

In the case of the strict large rotating BH limit (applying eqs.~\eqref{eq:LBH1} and~\eqref{eq:LBH2}, keeping only the leading order in scaling coefficient $\alpha$), the metric given in eq.~\eqref{eq:metric5dmpadssigpm3} simplifies slightly. But it retains the topology $S^3\times \mathbb{R}$, and remains in a rotating state with nonvanishing angular momentum. In particular, it does {\it not} simplify to a boosted black brane, and the dual fluid is {\it not} merely a boosted fluid, but a {\it nontrivially rotating fluid}. If one now chooses to probe this slightly simplified rotating background metric with metric fluctuations having {\it small frequency and momentum compared to the horizon radius} (i.e. hydrodynamic fluctuations), then these fluctuations effectively {\it see} a boosted black brane metric. 
This is explicitly seen in the equations of motion obeyed by the metric fluctuations around the rotating background metric in the large BH limit given by eqs.~\eqref{eq:LBH1} and~\eqref{eq:LBH2}, which we have shown to be identical to those of the metric fluctuations around a boosted black brane with boost parameter $a$.\footnote{This had not been recognized in the originally published version of~\cite{Garbiso:2020puw}, which is to be updated shortly.}

{\bf Summary:} While there are gapless QNMs of rotating BHs at any value of the angular momentum parameter $a$, these modes apparently only start following hydrodynamic dispersion relations with integer powers (e.g. $\omega=v \mathcal{J} - i D \mathcal{J}^2 + \mathcal{O}(3)$) at larger values of the horizon radius $r_+\approx 1000$. This may merely be due to the growing systematic error in the fit window $0\le \mathcal{J}/r_+\le 0.1$ at small $r_+$, as discussed above, or due to the discreteness of the momentum in principle. It is remarkable, that the mode at $r_+=1000$ scales hydrodynamically, $\alpha\approx 1,\, \beta\approx 2$, but the diffusion coefficient has not yet taken the value expected for a boosted fluid; that value it only takes at very large horizon values $r_+>10^6$. {\it Therefore, there is an interesting window of horizon values, approximately $1000<r_+<10^7$, in which these holographic rotating fluids display hydrodynamic behavior which is distinct from a boosted fluid.} In this window, we expect to see how hydrodynamic transport coefficients are changed by rotation (and not only by a boost).\footnote{We note that the dispersion relations analyzed in a regime around $r_+\approx 10$ may also be described by a derivative expansion. In this regime, there are less numerical data points available for the fit in the window $0\le \mathcal{J}/r_+\le 0.1$. Thus the fit values $\alpha, \beta, D, v$ displayed for $r_+=10$f in Fig.~\ref{fig:horizonRadiusDependence} have a large systematic error. Thus, hydrodynamic scaling $\alpha=2,\, \beta=1$ may persist to these small horizon values. Alternatively, in that regime there may be a scaling with non-integer exponents $\alpha$ and $\beta$; for examples of such a behavior see~\cite{Abbasi:2021fcz,Abbasi:2022aao}.} 
In Sec.~\ref{sec:critical_Points}, the convergence radius of the hydrodynamic expansion in momentum space is going to be computed for a range of horizon radii $100\le r_+ \le 10^6$, overlapping with the window in which we expect hydrodynamic scaling with nontrivially modified transport coefficients due to rotation.   

\subsubsection{Perturbative instabilities - superradiant and Gregory-Laflamme}
\label{sec:instabilities}

It was first found by~\cite{Murata:2008xr} that the simply spinning five-dimensional AdS BH suffered from a linear superradiant instability.
This instability is caused by the presence of the conformal boundary in combination with the Penrose process. It was found in~\cite{Murata:2008xr} that modes of high enough $\mathcal J$ would induce such an instability. It is still an open question what is the final state of such an unstable system.

\begin{figure}
    \begin{center}
        \includegraphics[width=0.75\textwidth]{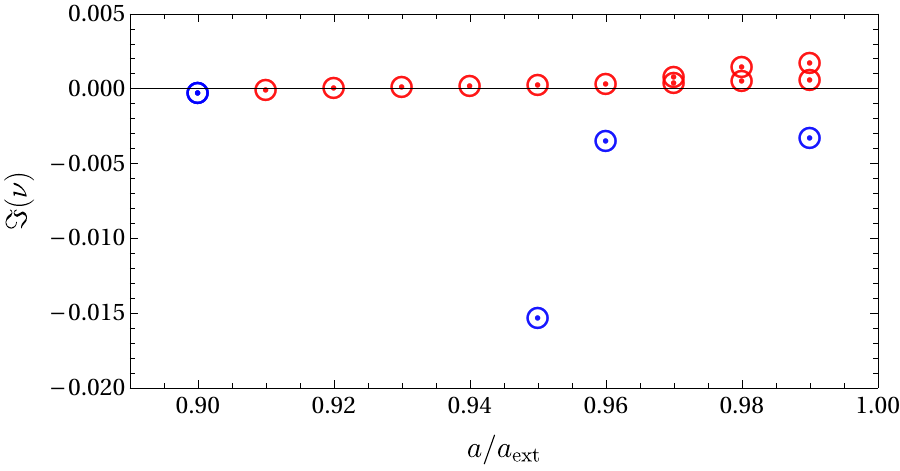}
    \end{center}
    \caption{
    $\Im(\nu)$ vs $a/a_{\text{ext}}$ in the tensor sector for $a\lesssim a_{\text{ext}}$, $r_+ = \ell$, and $\mathcal J = 5$. 
    The red points indicate unstable modes, while the blue points mark stable modes. 
    }
\label{fig:tensor_unstable_im_part}
\end{figure}
Several meta-stable states have been found~\cite{Ishii:2021ncf, Ishii:2018oms, Dias:2017tjg, Dias:2015rxy, Ishii:2021xmn}. 
Such states have an angular momentum but bear neither $U(1)$ axial nor time translational symmetry.
Instead, what is present is a helical symmetry that is a combination of the time translation and axial rotation.
In Figs.~\ref{fig:tensor_modes_over_a}, \ref{fig:vector_modes_over_a}, and~\ref{fig:scalar_modes_over_a}, the modes to the left of the imaginary axis approach the real axis as $a$ and $\mathcal J$ are increased.
For a large enough but non-extremal $a$ and $J$, these modes cross over.
For example, in the tensor sector, for $r_+/\ell = 1$, $\mathcal J = 5$, and $a/a_{ext} = 91/100$ there is an unstable mode with a frequency of $\nu \approx -15.56 + i 0.00004541$.
This behavior of modes can further be seen in Fig.~\ref{fig:tensor_unstable_im_part}.
It was shown in~\cite{Murata:2008xr} that superradiant radiant modes occur when
\begin{equation}\label{eq:superradiant_a}
    a_{\text{unstable}} = \frac{r_+^2/\ell}{1 + r_+^2/\ell^2} < a < a_{\text{ext}} = \frac{r_+^2/\ell}{1 + r_+^2/\ell^2}
    \sqrt{1 + \frac{\ell^2}{2r_+^2}} \,.
\end{equation}
In the regime of large $r_+$ and thus large temperature, $a_{\text{unstable}}$ and $a_{\text{ext}}$ would coincide as $r_+\rightarrow\infty$. 
The reason behind such an instability is similar to Cherenkov radiation.
The constant time-directed vector field 
\begin{equation}
    v_t := \partial_t - 2a\left( \frac{1+r_+^2/\ell^2}{r_+^2} \right) \partial_\psi
\end{equation}
measures the angular velocity of the horizon up to an overall constant normalization.
For values of $a$ in the superradiant regime, this timelike vector cannot be normalized such that it remains timelike at the AdS boundary. Rather, it becomes spacelike, and perturbations of the BH 
can be dragged faster than the local speed of light, thus leading to an instability.

\subsubsection{Cross Sector Spectrum Comparison}\label{sec:a_not_a_splitting}

\begin{figure}
    \begin{center}
        \includegraphics[width=0.75\textwidth]{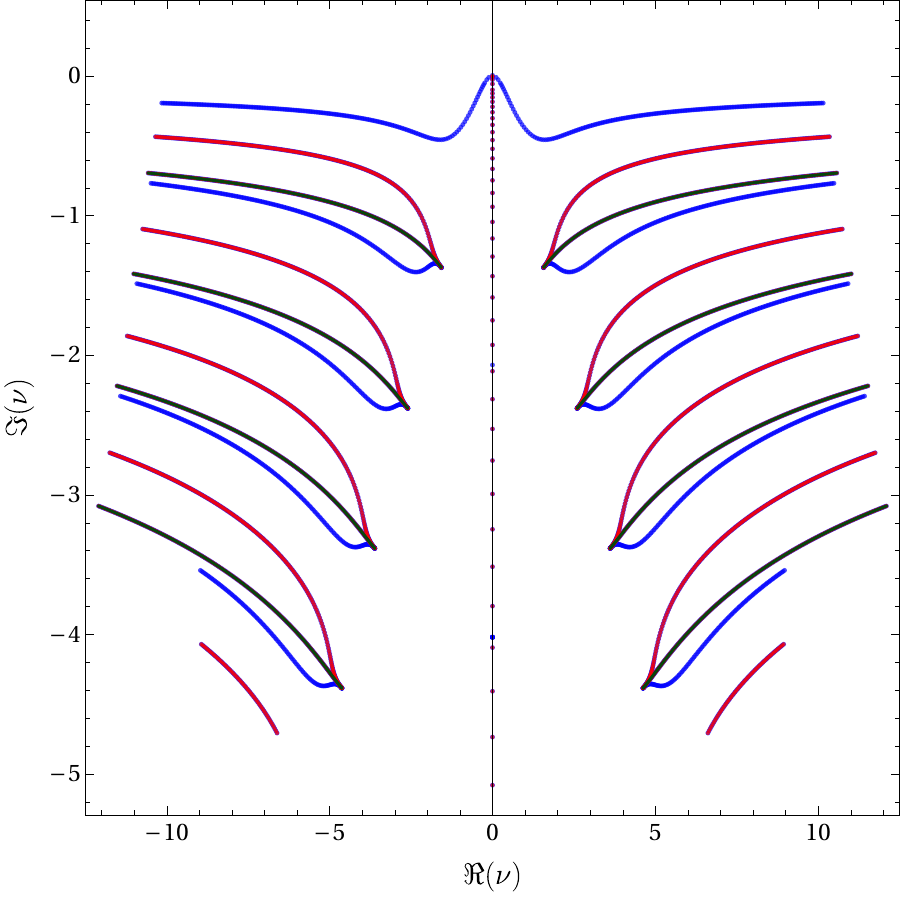}
        \includegraphics[width=0.75\textwidth]{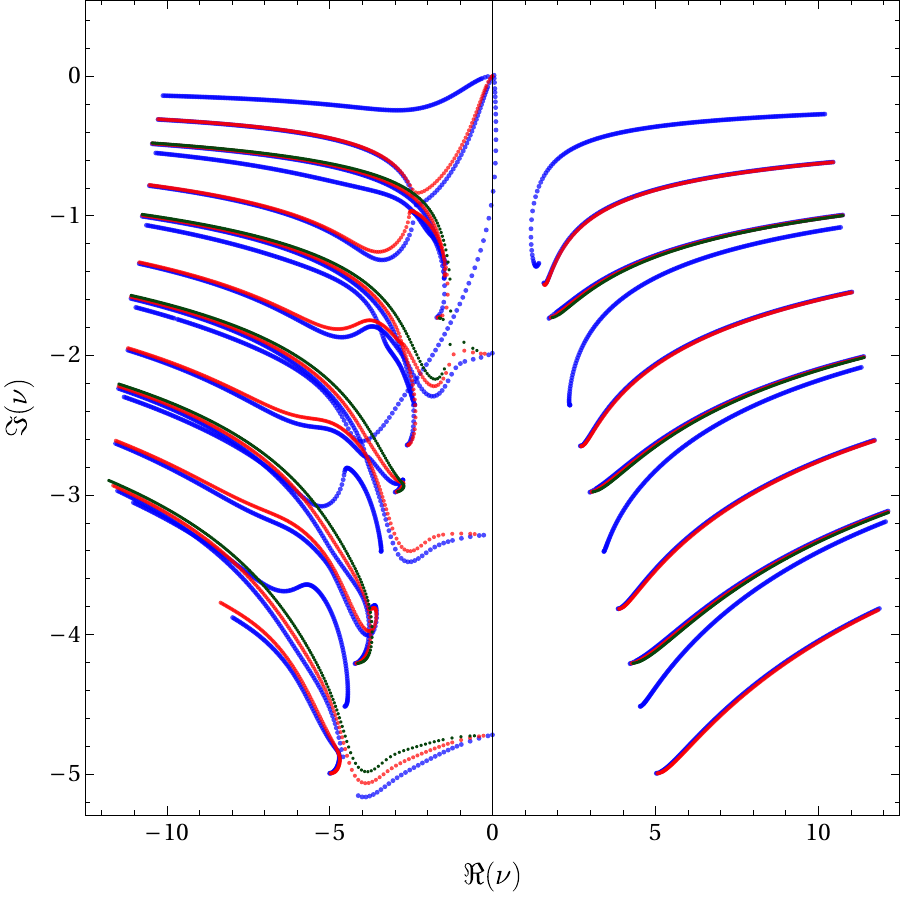}
        \end{center}
        \caption{
        Combined spectra of the sectors considered in sections~\ref{sec:tensor_sector}-\ref{sec:scalar_sector} at $r_+/l = 10$.
        The green depicts the tensor sector, red the vector sector, and blue the scalar sector.
        The modes are plotted for total angular momentum of the fluctuation $\mathcal{J} = 0,1/2,1,\ldots,199/2,100$ increasing in steps of $1/2$.
        \textit{Top:} $a/\ell = 0$. \textit{Bottom:} $a/\ell = 1/2$.
        \label{fig:all_sectors_compared_a_0_not_0}
        }
    \end{figure}

For $\mathcal J \geq 2$ there is non-trivial coupling in tensor fields in the vector sector equation of motion and tensor/vector fields in the scalar sector.
Non-trivial coupling means, that despite enhanced $U(2)$ angular symmetry, linearized field equations do not couple cleanly by component.
It is thus interesting to compare the spectra of the tensor, scalar, and tensor sector to each other.
Graphically, this has been done in Fig.~\ref{fig:all_sectors_compared_a_0_not_0}.
The first observation is, for $a=0$, the tensor spectrum appears in the spectra of the vector and scalar sectors. 
Similarly, the vector spectrum appears in the scalar spectrum.
This points to the fact that for $a=0$, the full spherical symmetry of the spacetime is restored, and the tensor, vector, and scalar fields decouple from each other for all $\mathcal J$ and $\mathcal K$. 
A similar decoupling was seen in the large temperature \cite{Garbiso:2020puw} limit as defined in Sec.~\ref{sec:previousResults}.
The second observation is that the once degenerate modes of the $a=0$ split in the $a \neq 0 $ case, as seen in the second plot of Fig.~\ref{fig:all_sectors_compared_a_0_not_0}.
Because the splitting occurs for $a\neq0$ and for finite $r_+$, it is reasonable to conclude that curvature scale associated with the spherical geometry of the BH causes this.
    
\subsection{Critical points}\label{sec:critical_Points}
In this section we give an overview of the recent work~\cite{Cartwright:2021qpp}, whose goal was to study the behavior of bounds on the convergence of linearized hydrodynamic expansions when the fluid is in a rotating state. Taking the MP geometry in eq.~(\ref{eq:Simple_Spin_MP_Metric}) as a concrete example, such bound for $\mathcal{N}=4$ SYM theory in a rotating state was found in~\cite{Cartwright:2021qpp} by calculating critical points (as introduced in~\cite{Grozdanov:2019uhi}). 

\subsubsection{Implicit functions and their series expansions}
As discussed in~\cite{Grozdanov:2019uhi}, a relativistic fluid supports, in general, collective excitations -- linearized fluctuations of energy and momentum -- around its equilibrium state, and these are referred to as hydrodynamic modes. The frequency of these modes $\omega$ are related to the momentum via a dispersion relation which takes the form,
\begin{equation}\label{eq:general_dispersion}
    \omega(\mathbf{q})=\sum_{c=0}^\infty d_n \mathbf{q}^{c/m} \,,
\end{equation}
where $\mathbf{q}$ is the wave vector, $d\in\mathbb{C}$, and $m\in\mathbb{N}$. The number of terms in the series is significant as it indicates the number of terms kept in the derivative expansion. Indeed, the series representation of the dispersion relation $\omega(\mathbf{q})$ truncates when the hydrodynamics is considered up to a certain finite order, and it is infinite only when hydrodynamic expansion to all orders is considered. The coefficients of the series can naturally be matched to the transport coefficients at each order. 

It is important to understand how dispersion relations such as eq.~(\ref{eq:general_dispersion}) arise. One begins by linearizing around an equilibrium state ($T_{\mu\nu}+\delta T_{\mu\nu}$) and making a choice of the hydrodynamic degrees of freedom. The derivative expansion is then contained in the constitutive relations: e.g. by expressing the spatial stress in terms of the energy and momentum density, these relations then contain up to $k$ derivatives of the energy and momentum density and so correspond to $k$-th order hydrodynamics. The resulting linear system, when expressed in terms of Fourier modes proportional to $\exp(-i\omega t+i\mathbf{q}\cdot x)$, contains non-trivial solutions provided its determinant vanishes. This determinant defines a non-trivial polynomial in $\omega$ and $\mathbf{q}$ and thus a complex algebraic curve for general $(\omega,\mathbf{q})\in\mathbb{C}^2$, and may hence be regarded as an implicit function with the form $F(\mathbf{q},\omega)=0$. 

Given that we can construct this implicit function, we are now in a position to borrow some basic results from the theory of algebraic curves. Consider an implicit function $f(x,y)=0$ defining a curve. Points on the curve can be classified by how it intersects with a line $L$. Following~\cite{walker:1950alg}, by choosing a point $p_0=(x_0,y_0)$ on the curve the line $L$ can be parametrized by $(x_0+\lambda t,y_0+\mu t)$. Intersections of the line with the curve are then roots of 
\begin{equation}
    f(x_0+\lambda t, y_0+ \mu t)=0\, .
\end{equation}
Taylor expanding this expression around $t=0$ leads to the following equation for the intersection
\begin{equation}
    (\partial_x f\lambda +\partial_y f\mu)t+ \frac{1}{2}(\partial^2_xf\lambda^2+2\partial^2_{xy}f \lambda \mu +\partial^2_y f\mu^2)t^2+\cdots =0 \, ,
\end{equation}
from which we can classify the number of intersections of the line and the curve. \\

\noindent \textit{Single intersections:} Consider first the term linear in $t$, 
$\left. df/dt\right|_{t=0}$. If either or both of the partial derivatives are non-zero, then every line through $p_0$ has a single intersection with the curve. There is however an exception: it is possible that $\mu/\lambda$ takes a value such that $\partial_x f\lambda +\partial_y f\mu=0$, and this line is the tangent to the curve at the point $p_0$. \\

\noindent \textit{Double intersections:} Consider now the term quadratic in $t$, 
$\left. d^2f/dt^2\right|_{t=0}$, supposing that $\partial_x f=\partial_y f=0$. Suppose further that not all of the second partial derivative are zero. Then every line through $p_0$ has at least $2$ intersections at $p_0$. As before, there is again an exception: it is possible that $\mu/\lambda$ takes a value such that $\partial^2_xf\lambda^2+2\partial^2_{xy}f \lambda \mu +\partial^2_y f\mu^2=0$, and these lines are the tangents to the curve at the point $p_0$. \\

\noindent \textit{$r^{th}$ intersections:} Consider now the $r$-th term in $t$, 
$\left. d^rf/dt^r\right|_{t=0}$, supposing that all partial derivatives up to and including the $(r-1)$ order vanish. Again, suppose further that not all of the $r$-th partial derivative are zero. Then every line through $p_0$ has at least $r$ intersections at $p_0$. The exceptional case still remains that it is possible that $\mu/\lambda$ takes a value such that 
\begin{equation}
    \partial^r_xf\lambda^r+\binom{r}{1} \partial^{r-1}_x\partial_yf \lambda^{r-1} \mu +\cdots +\binom{r}{r}\partial^r_y f\mu^r=0.
\end{equation}
Again these lines are the tangents to the curve at the point $p_0$. \\

In each of these examples $p_0$ is said to be an $r$-fold point, or a point of multiplicity $r$; a point with multiplicity two or more is referred to as singular. It is clear that for a point to be singular it must be the case that $f(x_0,y_0)=\partial_x f(x_0,y_0)=\partial_y f(x_0,y_0)=0$. We are most interested in $1$-fold points, which are referred to as critical points in~\cite{Grozdanov:2019uhi}. One can prove that the point $p_0$ is an $r$-fold point of the curve if and only if all of the $(r-1)$-th derivatives $f$ vanish at $p_0$ but not all the $r$-th derivatives~\cite{walker:1950alg}. 

With this in hand we are now in a position to understand the derivation of dispersion relations we are interested in. If the curve $f(x,y)=0$ has a non-singular point at $(x_0,y_0)$, then either $\partial_x f$ or $\partial_y f$ is non-zero. Without loss of generality we can assume it to be $\partial_y f$, and this allows the use of the implicit function theorem. \\

\noindent\textit{Analytic implicit function theorem:} 
Suppose $f(x_0,y_0)=0$ and $\partial_y f(x_0,y_0)\neq 0$. Then there exist $\epsilon>0$ and $\delta>0$ such that $\mathbb{D}_\epsilon(x_0)\times\mathbb{D}_\delta(y_0)$ is in the neighborhood of where $f$ is defined. In addition there exists an injective function $g:\mathbb{D}_\epsilon(x_0)\rightarrow \mathbb{D}_\delta(y_0)$ such that $f(x_0,g(x_0))=0$. Furthermore, for each $x\in \mathbb{D}_\epsilon(x_0)$, $g(x)$ is the unique solution of $f(x,g(x))=0$. The mapping $g(x)$ is analytic in $\mathbb{D}_\epsilon(x_0)$ and that
\begin{equation}
    g'(x)=- \frac{\partial_x f(x,g(x))}{\partial_y f(x,g(x))}\, .
\end{equation}
Higher order derivatives can be obtained in a similar way, and hence the implicit function theorem  guarantees that an expression for $y(x)$ exists locally when expressed as a power series of $(x-x_0)$,  and it converges in a neighborhood of $x_0$. \\

Unfortunately this theorem applies only to a non-singular point, and for the hydrodynamic dispersion relations we are interested in, there can be in principle 1-fold points at the origin of the complex frequency and momentum plane.
However, as discussed in~\cite{walker:1950alg,wall_2004}, the analytic implicit function theorem can be extended to points $(x_0,y_0)\in\mathbb{C}^2$, singular or not. In particular, there exist branches of functions $(x(t),y(t))$ parameterized by t that are analytic in a neighborhood of $t=0$ such that $(x(0),y(0))=(x_0,y_0)$ is a root, viz. $f(x_0,y_0)=0$. Furthermore, for every other point $(x,y)$ there is a unique branch in a suitable neighborhood of $(x_0,y_0)$ for which $x=x(t)$ and $y=y(t)$. The proof of this is typically attributed to Puiseux and arises from the proof that the field of Puiseux series is algebraically closed. The construction of such a Puiseux series solution to $f(x,y)=0$ as well as the pertinent proof can be found in any text on algebraic curves including~\cite{walker:1950alg,wall_2004}~\footnote{A particularly easy to follow description of the construction of such a Peuiseux series solution is given in~\cite{Grozdanov:2019uhi}.}, and we refer interested readers to them for further details.

The most important result of all this, as was pointed out in~\cite{Grozdanov:2019uhi}, is that for a branch given in the neighborhood of $(x_0,y_0)$, the series is convergent with a radius of convergence determined by the location of next closest critical point. We have thus a method of determining the radius of convergence of a hydrodynamic series without having to calculate a large number of coefficients of the series expansion, which may be computationally expensive or difficult.

\subsubsection{Determination of the critical point }
With the tools discussed above in hand we can proceed to compute the locations of the critical points of the linearized hydrodynamic expansion of a rotating fluid. For this purpose, we focus on the fluid behavior of the $\mathcal{N}=4$ SYM theory in a rotating state dual to the MP BH given by eq.~(\ref{eq:Simple_Spin_MP_Metric}), as was done in~\cite{Cartwright:2021qpp}. From the previous discussion it is clear we must search for $1$-fold points, which are solutions to the equations
\begin{equation}\label{eq:crit_Points}
    P(j,\nu)=0 \,,  \qquad \partial_\nu P(j,\nu)=0 \, ,
\end{equation}
where $P$ is the implicit function encoding the dispersion relations $\nu(j)$, and $j=\mathcal{J}/r_+$ here. To obtain $P$ we recall the arrangement of the field equations in a generalized eigenvalue problem as discussed in section~\ref{sec:QNMresults}. The differential operator $M$ is a function of the frequency and momentum, encoding the QNM spectrum. While in section~\ref{sec:QNMresults} we directly solve the generalized eigenvalue problem, an alternative method of obtaining the QNMs is to evaluate the determinant of this operator, $P(\nu,j)=\det(M(\nu,j))$, leading to an implicit function in terms of $(j,\nu)$, whose roots are precisely the QNM spectrum. Crucially, this is precisely the implicit function which encodes the dispersion relations of both hydrodynamic and non-hydrodynamic modes of collective excitations about the equilibrium of the CFT dual to this BH geometry. 

\begin{figure}
    \begin{center}
    \includegraphics[width=0.85\textwidth]{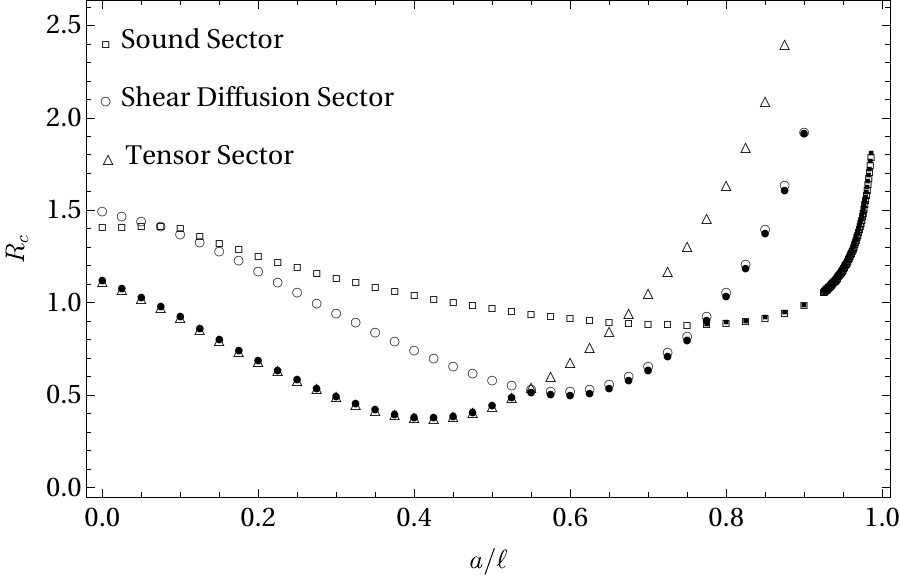}
    \end{center}
    \caption{The convergence radius, $R_c$, of the hydrodynamic expansion as a function of the angular momentum parameter $a/\ell$. Hollow shapes indicate the value of the distance from the origin ($j=0$) to the nearest critical point $j=j_{c}$ in the strict large BH limit, while solid shapes indicate this value in the full MP BH with horizon radius $r_+/\ell=100$. The sector with the smallest absolute value determines $R_c$. Adapted from~\cite{Cartwright:2021qpp}. 
    }
    \label{fig:convergenceRadius}
\end{figure}

In Fig.~\ref{fig:convergenceRadius} we present the distance from the origin (recall hydrodynamic dispersion relations, constructed as Puiseux series, are roots of the implicit function around the origin of complex frequency-momentum plane) to the nearest critical point of the implicit function $P$ encoding the dispersion relations for the MP BH in eq.~(\ref{eq:Simple_Spin_MP_Metric}). In the figure we display the nearest critical point for each of the three sectors discussed in section~\ref{sec:QNMresults} as hollow shapes. Interestingly, as already discussed in section~\ref{sec:QNMresults} sectors with successively lower values of $\mathcal{K}$, i.e. for instance the scalar sector defined by $\mathcal{K}=\mathcal{J}-2$ contain the fields of all higher sectors. As a result, when constructing the implicit function $P$ one can find numerically that the critical points obtained in, say the tensor sector, also appear in the scalar sector. An organizational strategy is then needed to understand the appearance of the modes. A simple one is to organize by the importance of contributions in a large $r_+$ expansion. In the strict large BH limit where we consider only the leading term in the limit $r_+\rightarrow \infty$ as shown in~\cite{Garbiso:2020puw,Cartwright:2021qpp}, the behavior of the fluctuation equations is that of a boosted planar black brane~\footnote{An example of the relation between the Wigner d functions and Fourier transformations in the large BH limit is discussed in appendix~\ref{sec:WignerD}.}. In this limit the sectors decouple, hence the vector sector does not interact with tensor sector, nor does the scalar sector interact with the vector and tensor sector. Furthermore, the discrete momentum $j=\mathcal{J}/r_+$ becomes a continuous variable. Hence the hollow shapes in Fig.~\ref{fig:convergenceRadius} are those obtained in this limit. A consequence of this limit is that one can interpret these critical points like in~\cite{Grozdanov:2019uhi} as radii of convergence of the dispersion relations obtained from the implicit function representing the hydrodynamic expansion. Its for this reason we label this plots vertical axis as $R_c$, the convergence radius of the corresponding sector~\footnote{It should be noted that in this large BH limit there are no hydrodynamic modes in the tensor sector. However these play a role as one moves away from the large BH limit.}. 

\begin{figure}
    \begin{center}
    \includegraphics[width=0.85\textwidth]{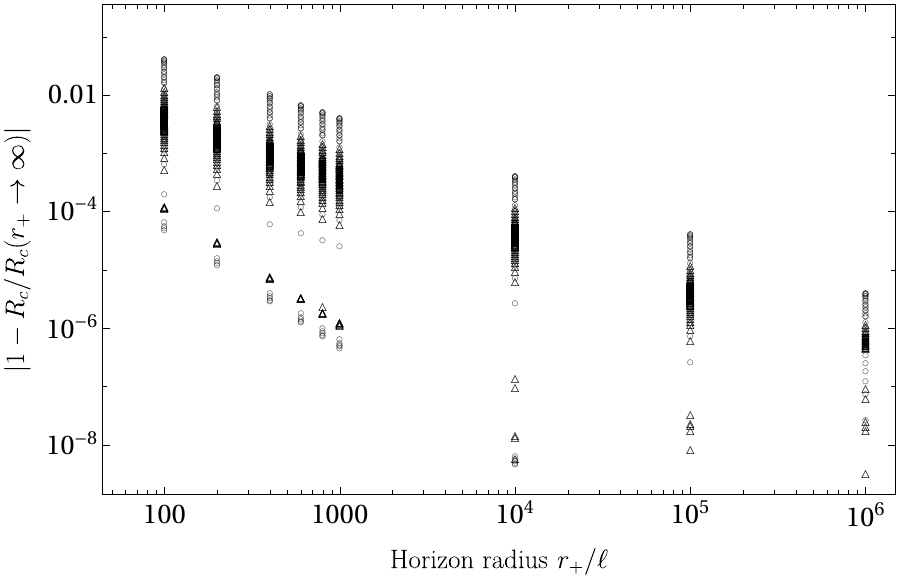}
    \end{center}
    \caption{The deviation of the convergence radius, $R_c$, of the hydrodynamic expansion as a function of the horizon radius normalized to its value in the large BH limit. The various points represent different values of the angular momentum parameter $a/\ell$. The vertical width of the bands displays the ``strength" of rotational effects on the radius of convergence. The plot markers are the same as those used in Fig.~\ref{fig:convergenceRadius}. Adapted from~\cite{Cartwright:2021qpp}. 
    }
    \label{fig:Horizon_dep}
\end{figure}

It was already shown in~\cite{Garbiso:2020puw} that the QNM spectrum behaves, roughly speaking, hydrodynamically for $r_+/\ell\approx O(10^2)$. Therefore one can think of the large BH limit, in a sense, as a hydrodynamic limit of the rotating MP BH solution. In~\cite{Cartwright:2021qpp} it was shown that the QNMs, which are also critical points, already behave hydrodynamically for $r_+/\ell\approx O(10^2)$. This is displayed in Fig.~\ref{fig:Horizon_dep}, which shows the ratio of the nearest critical point to the origin for a finite horizon radius to that of the nearest mode in the strict $r_+\rightarrow\infty$ limit. It is interesting to note that differences (on the percent level) in the radius of convergence only begin to appear for $r_+/\ell\approx O(10^2)$. 

Interestingly, the effect which more severely limits the radius of convergence of the hydrodynamic series away from the large BH limit is the interaction of modes of different values of $\mathcal{K}$. While the tensor sector ($\mathcal{K}=\mathcal{J}-2$) does not contain any hydrodynamic modes, the implicit function encoding the dispersion relations does contain critical points, and these points are closer to the origin in the complex momentum plane. This is reflected in the radius of convergence displayed as the solid shapes in Fig.~\ref{fig:convergenceRadius}. The resulting radius of convergence is still non-zero and finite, but its behavior has changed dramatically as a result of the appearance of the inter-sector couplings implied by the Einstein equations.  


\subsection{Pole-Skipping}\label{sec:pole_skipping}
In this section, we give an overview of the pole-skipping phenomenon in the MP background described in section~\ref{sec:model} and discussed in recent work~\cite{Amano:2022mlu}. While the critical points discussed in the previous section can be shown to arise as the result of mode collisions in the complex frequency-momentum plane, pole-skipping modes arise when the residue of a Greens function ``collides'' with the pole, and as a result the would-be pole is ``skipped''. The study of pole-skipping has garnered much attention in recent years due to the discovery in~\cite{Schalm:2018pjh}. The authors of~\cite{Schalm:2018pjh} found that in the gravitational shockwave calculations~\cite{Shenker:2013pqa,Shenker:2015keq}, the scrambling rate of the dual strongly-coupled quantum field theory is directly related to the hydrodynamic sound modes of the theory. This realization quickly led to the development of effective descriptions of quantum chaos, and the general prediction of pole-skipping in the energy-energy correlation functions as a ``smoking gun'' for the hydrodynamic origin for the chaotic mode~\cite{Blake:2017ris,Blake:2018leo}. When this occurs, it provides a potentially simpler method of obtaining information to characterize the chaotic dynamics (maximal Lyanpunov exponents and butterfly velocities) than the computation of OTOCs, a standard candle to diagnose operator growth in chaotic quantum field theories. Although much is known about these quantities in simple settings (insert examples), less is known about their behavior in non-trivial states of chaotic quantum field theories, such as the rotating state dual to the MP BH we consider. Earlier works have considered the rotating BTZ BH dual to a (1+1)-dimensional CFT with a chemical potential for rotation~\cite{Liu:2020poq,Mezei:2019dfv, Jahnke:2019heq}, and the field theory dual to a (3+1)-dimensional rotating Kerr-AdS BH~\cite{Blake:2021hjj}.

\subsubsection{Gravitational shockwaves}\label{sec:shockwave_setup}
In~\cite{Amano:2022mlu} we focused on computations of OTOCs in the field theory dual to eq.~(\ref{eq:Simple_Spin_MP_Metric}) of the form
\begin{equation}
    \langle \hat{W}(t_2, \Psi_2, \Phi_2, \Theta_2) \hat{V}(t_1, \Psi_1, \Phi_1, \Theta_1) \hat{W}(t_2, \Psi_2, \Phi_2, \Theta_2) \hat{V}(t_1, \Psi_1, \Phi_1, \Theta_1)  \rangle\, ,
    \label{eq:OTOC}
\end{equation}
where capital letters denote boundary coordinates, the difference in operator insertions is large compared to the inverse temperature $t_2 - t_1 \gg \beta$ and the angled brackets denote the trace with respect to the thermal density matrix $e^{-\beta H}/Z$ in the dual field theory. The gravitational process dual to this OTOC is the scattering amplitude between a particle of momentum $p_1^{U}$ travelling along the $V=0$ horizon of the BH (corresponding to quanta created by the $\hat{V}$ operator) and a particle of momentum $p_2^{V}$ travelling along the $U=0$ horizon (corresponding to quanta created by the $\hat{W}$ operator). The OTOC is then given by an integral, over momenta and angular coordinates, of $e^{i\delta}$ (a two-two scattering amplitude) weighted by bulk-boundary wave functions, which describe the distribution of quanta along at the horizon~\cite{Shenker:2015keq}.

The computation is made feasible by working in the eikonal approximation~\cite{Kabat:1992tb} which consists of linearizing the gravitational Lagrangian and treating the two particles (created by $\hat{V}$ and $\hat{W}$) as fixed energy momentum sources following their classical trajectories. The scattering amplitude for this process, $e^{i\delta}$, is given by evaluating the Einstein-Hilbert action, expanded to quadratic order, on the classical solutions sourced by the $\hat{V}$ and $\hat{W}$ operators \cite{Kabat:1992tb,Shenker:2013pqa,Shenker:2015keq}
\begin{equation}
    \delta = \frac{1}{2} \int d^5 x \sqrt{-g} (\delta g_{UU}\mathscr{D}^2\delta g_{VV}\delta g_{UU} T^{UU} + \delta g_{VV}T^{VV})\, ,
    \label{eq:eikonal_delta}
\end{equation}
where $\mathscr{D}^2$ is related to the inverse of the graviton propagator and as noted in~\cite{Shenker:2015keq}, the first term is equal in magnitude but opposite in sign as the other terms in eq. (\ref{eq:eikonal_delta} ).  

Before discussing the extraction of the OTOC from the gravitational amplitude we begin by computing the linearized backreaction of the two particles following (near) geodesic trajectories on the geometry~\footnote{This calculation is most easily done in Kruskal coordinates. These are detailed in the original paper~\cite{Amano:2022mlu} but are included in appendix~\ref{sec:Null_coordinates} for convenience.}. These solutions are examples of the well known shockwave geometries~\cite{Aichelburg:1970dh,Dray:1984ha,Sfetsos:1994xa}. The trajectory of the particle of momentum $p_1^{U}$ approximates the null geodesic given by $(U = U(\tau), V=0, \tilde{\psi}_1, \phi_1, \theta_1)$. The only non-zero component of the stress tensor of this particle is given by
\begin{equation}
    T^{UU} = \frac{1}{\sqrt{-g}}p_1^U \delta(V)\delta(\tilde{\psi} - \tilde{\psi}_1)\delta(\phi - \phi_1)\delta(\theta - \theta_1)\, ,
    \label{V-quant}
\end{equation}
while the particle of momentum $p_1^{V}$ follows the null-geodesic $(U=0, V = V(\tau), \tilde{\psi}_2, \phi_2, \theta_2)$, with the stress tensor
\begin{equation}
    T^{VV} = \frac{1}{\sqrt{-g}}p_2^V \delta(U)\delta(\tilde{\psi}-\tilde{\psi}_2)\delta(\phi - \phi_2)\delta(\theta - \theta_2).  
    \label{W-quant}
\end{equation}
The only non-zero component of the stress tensor sourced by the $\hat{V}$ quanta is $T_{VV} = A(0)^2 T^{UU}/4$.  The ansatz for the gravitational backreaction of this source takes the form of a shockwave across the $U = 0$ horizon, $U \to U + f_1(\tilde{\psi}, \phi, \theta)$ which can be written to leading order as
\begin{equation}
    \delta g_{VV} = -A(0) f_1(\tilde{\psi}, \phi, \theta)\delta (V). 
    \label{eq:Shock-ansatz}
\end{equation}
Inserting the ansatz (\ref{eq:Shock-ansatz}) into Einstein's equations, one finds that the $VV$-component is the only non-trivial equation, which reduces to the following partial differential equation for the angular profile $f_1(\tilde{\psi}, \phi, \theta)$
\begin{equation}
   \mathcal{D} f_1(\tilde{\psi}, \phi, \theta ) =(\square  + \lambda_1 + \lambda_2 \partial_{\tilde{\psi}} + \lambda_3 \partial^2_{\tilde{\psi}})f= \frac{16 \pi G_N}{\mathrm{sin}(\theta_1)}\frac{K}{L r_+^2} p_1^U \delta(\tilde{\psi} - \tilde{\psi}_1)\delta(\phi - \phi_1)\delta(\theta - \theta_1), 
   \label{eq:shockwave_differential_equation}
\end{equation}
with $\square$ given by~\footnote{It turns out that this operator is one quarter of the Laplacian on the unit 3-sphere written in terms of the Hopf fibration.}
\begin{equation}
    \square =  \mathrm{cot}(\theta)\partial_{\theta} +  \partial_{\theta}^2 + \mathrm{csc}^2(\theta) \partial_{\tilde{\psi}}^2 +\mathrm{csc}^2(\theta)\partial_{\phi}^2 -2\mathrm{cot}(\theta)\mathrm{csc}(\theta)\partial_{\tilde{\psi}}\partial_{\phi},
\end{equation}
where $\lambda_1, \lambda_2, \lambda_3$ are constants given by the expressions
\begin{equation}
    \lambda_1 =  \frac{(2K^2 + L^2 r_+^2)(2a^2(r_+^2 + L^2)^2 -\Delta)}{4L^4r_+^2 K^2}, \hspace{0.5cm} \lambda_2 = -\frac{2 a (r_+^2 + L^2) K}{L^3 r_+^2}, \hspace{0.5cm} \lambda_3 = -\frac{a^2(r_+^2 + L^2)}{L^2 r_+^2}.
    \label{eq:lambda_coefficients}
\end{equation}
In these expressions, eq. (\ref{eq:shockwave_differential_equation}) and eq. (\ref{eq:lambda_coefficients}) we have made use of two quantities $\Delta$ and $K$ to shorten the expressions, these constants are defined as
\begin{equation}
     \Delta = L^2r_+^2(L^2 + 2r_+^2), \hspace{1cm} K = \sqrt{L^2r_+^2 - a^2(L^2 + r_+^2)}
\end{equation}

Repeating these steps we can now consider the $W$ quanta travelling on the $U = 0$ horizon. The calculation goes through as above, only now the geometry is shifted across the $U=0$ horizon $V \to V + f_2(\tilde{\psi}, \phi, \theta)$. To linear order in $f_2$ this leads to the requirement that $f_2(\tilde{\psi}, \phi, \theta)$ satisfy 
\begin{equation}
    \tilde{\mathcal{D}} f_2(\tilde{\psi}, \phi, \theta) =\frac{16 \pi G_N}{\mathrm{sin}(\theta_2)} \frac{K}{L r_+^2} p_2^V \delta(\tilde{\psi} - \tilde{\psi}_2)\delta(\phi - \phi_2)\delta(\theta - \theta_2),
   \label{shock-diff2} 
\end{equation}
where $\tilde{\mathcal{D}}$ is the same differential operator given in (\ref{eq:shockwave_differential_equation}) but with the replacements $\partial_{\tilde{\psi}} \to -\partial_{\tilde{\psi}}$ and $\partial_\phi \to -\partial_\phi$.

As was first demonstrated in~\cite{Amano:2022mlu}, we can find a general analytic solution to the angular profile of the shockwave. Introducing $f(\tilde{\psi}, \phi, \theta, \theta')$ as the Green's function of the Laplacian on $S^3$ with normalised delta-function source
\begin{equation}
\mathcal{D} f(\tilde{\psi}, \phi, \theta, \theta') = -\frac{1}{2 \sin(\theta')} \delta(\tilde{\psi}) \delta(\phi) \delta(\theta - \theta')
\label{eq:normalisedshock}
\end{equation}
we can expand the delta functions in terms of Wigner D-functions using the completeness relation given in eq.(\ref{eq:completeness}). Using the fact that the Wigner D-functions satisfy
\begin{equation}
\square D^{\cal J}_{{\cal K}{\cal M}} + {\cal J}({\cal J }+ 1)  D^{\cal J}_{{\cal K}{\cal M}} = 0 \, ,
\label{eq:wignerdeqn}
\end{equation}
we can solve eq. (\ref{eq:normalisedshock}) for the normalised shockwave profile 
\begin{equation}
f(\tilde{\psi}, \phi, \theta, \theta^{\prime}) =  \sum_{{\cal J}=0,1/2,1,\dots}^{\infty} \sum_{{\cal K}=-{\cal J}}^{{\cal J}}  \sum_{{\cal M}=-{\cal J}}^{{\cal J}}     \frac{2{\cal J}+1}{16  \pi^2}  \frac{d^{{\cal J}}_{{\cal K} {\cal M}}(\theta^{\prime})  d^{{\cal J}}_{{\cal K} {\cal M}}(\theta)}{{\cal J}({\cal J}+1)- \lambda_1 - i \lambda_2 {\cal K} + \lambda_3 {\cal K}^2}  e^{i {\cal K} \tilde{\psi} + i {\cal M} \phi}\, ,
\label{eq:exactshock}
\end{equation}
where we have made use of the decomposition of the Wigner D-functions in terms of the Wigner d-functions $D^{\cal J}_{{\cal K}{\cal M}}(\tilde{\psi}, \phi,  \theta) = d^{\cal J}_{{\cal K} {\cal M}}(\theta) e^{i \mathcal{K} \tilde{\psi} + i \mathcal{M} \tilde{\phi}}$. Finally, by using the general Greens function, we compute the eikonal phase\footnote{Note here we have used rotational symmetry in the $\phi, \tilde{\psi}$ directions and that $\tilde{f}(\tilde{\psi}_1 - \tilde{\psi}_2, \phi_1 - \phi_2, \theta_1, \theta_2) = f(\tilde{\psi}_2 - \tilde{\psi}_1, \phi_2 - \phi_1, \theta_2, \theta_1)$ with $\tilde{f}$ the solution to the analogous equation to eq. (\ref{eq:normalisedshock}) with $\mathcal{D}  \to \tilde{\mathcal{D}}$.} from the action in eq. (\ref{eq:eikonal_delta})
 \begin{equation}
     \delta = \frac{16 \pi G_N  K}{L r_+^2} A(0) p_1^{U} p_2^{V} f(\psi_2 - \psi_1, \phi_2 - \phi_1, \theta_2, \theta_1) \, .
 \end{equation}
Unfortunately, for general configurations $(\psi, \phi, \theta, \theta^{\prime})$ we have not encountered a simple way to evaluate the sum in eq. (\ref{eq:exactshock}). However, there do exist configurations for which one can obtain an explicit closed form of the OTOC in the large BH limit. Making use of the existence of radially in-falling geodesics which satisfy $\dot{\theta} = \dot{\phi} = 0$ \cite{Rocha:2014gza} we can consider configurations for which the operators $\hat{V}$ and $\hat{W}$ lie on a Hopf circle (and hence are only separated in the direction of the fibre coordinate $\psi$, parallel to the rotation). Such operator configurations with $\Theta_1=\Theta_2=\Theta$ and $\Phi_1=\Phi_2=\Phi$ can be mapped by an isometry to the North pole, i.e. without loss of generality we consider configurations for which $\Theta = \Phi = 0$. The OTOC then takes the form
 \begin{equation}
     H(t, \Psi) = \langle W(t_2, \Psi_2,0,0)V(t_1, \Psi_1, 0, 0)W(t_2, \Psi_2, 0,0)V(t_1, \Psi_1, 0,0\rangle
 \end{equation}
where $t = t_2 - t_1$ and $\Psi = \Psi_2 - \Psi_1$. 

To obtain the closed form expression of the OTOC we work in the large BH limit $r_+ \gg \ell$ where the bulk to boundary wave-functions describing the distribution of quanta on the horizon will be sharply peaked in the angular coordinates on the horizon around $\tilde{\psi}_2 - \tilde{\psi}_1 = \Psi - \overline{\Omega} t$, $\theta_1 \approx \theta_2 \approx 0$, $\phi_1 \approx \phi_2 \approx 0$. Expanding the standard expression from \cite{Shenker:2015keq} for the OTOC to leading order in $G_N$, and using that the bulk-to-boundary wave-functions are sharply peaked around~\footnote{we note that in the large BH limit the expressions for the temperature and angular velocity reduce to
\begin{equation}
2 \pi T = \frac{2 r_+}{\ell^2} \sqrt{1 - a^2/\ell^2}, \hspace{1.0cm} \overline{\Omega} = -2 a/\ell^2.
\end{equation}} $p_1^{U} p_2^{V} \sim e^{2 \pi T t}$, the OTOC then takes the form 
\begin{equation}
\label{eq:otocsimple}
    H(t, \Psi) \sim 1 - c G_N e^{2\pi T t}f(\Psi - \overline{\Omega} t)\, ,
\end{equation}
where $c$ is a constant and $f(\tilde{\psi})$ is the solution to equation (\ref{eq:normalisedshock}) with $\theta  = \theta' = \phi = 0$ .
As was shown in the appendix of~\cite{Amano:2022mlu} in the large BH limit the expression \eqref{eq:exactshock} can be approximated by an integral, leading to the expression 
\begin{equation}
f(\tilde{\psi}) =  \int_{0}^{\infty} d {\cal J} \int_{-{\cal J}}^{{\cal J}} d {\cal K}   \frac{{\cal J}}{4 \pi^2}  \frac{e^{i {\cal K} \tilde{\psi}}}{{\cal J}^2- \lambda_1 - i  \lambda_2 {\cal K} + \lambda_3 {\cal K}^2}
\end{equation}
for the shock-wave profile on Hopf circles. This integral can be computed exactly by contour integration (see~\cite{Amano:2022mlu} for details)
\begin{equation} \label{eq:hopfotocF}
 f(\tilde{\psi})=-\frac{1}{4\pi |\tilde{\psi}|}
\begin{cases}
      e^{-k_+ \tilde{\psi}},&\tilde{\psi} > 0\\
     e^{k_- \tilde{\psi}}, & \tilde{\psi} < 0
\end{cases}
\end{equation}
where 
\begin{equation}
    k_{\pm} =  \frac{r_+}{\ell\sqrt{1 - a^2/\ell^2}}\left(\sqrt{\frac{3}{2}} \mp \frac{a}{\ell}\right). 
\end{equation}

In co-rotating coordinates, the OTOC is therefore exponentially growing in time with an exponent $2 \pi T t$ and exponentially decaying in the $\tilde{\psi}$ direction governed by $k_{\pm}$. Hence we can now arrive the one of the main results of~\cite{Amano:2022mlu} and this sections over, given in terms of the fixed boundary coordinates $(t, \Psi)$ the OTOC takes the functional form 
\begin{equation}
    H(t, \Psi)\sim 1- \frac{c G_N}{4\pi |\Psi - \Omega t|}
    \begin{cases}
     \exp{\bigg(2 \pi T_+\bigg(t - \frac{L \Psi}{2 v_B^{+}}\bigg)}\bigg)\, , &  \Psi - \overline{\Omega} t > 0, \\
     \exp{\bigg(2 \pi T_-\bigg(t - \frac{L \Psi}{2 v_B^{-}}\bigg)}\bigg)\, , & \Psi - \overline{\Omega} t < 0
 \end{cases}
    \label{eq:hopfotoc}
\end{equation}
where $\ell \Psi/2$ corresponds to the spatial distance between the operators $\hat{V}, \hat{W}$ and the exponents $T_{\pm}, v_B^{\pm}$ are given by 
\begin{align}
2 \pi T_{\pm} &= \frac{2 r_+}{\ell^2 \sqrt{1 - a^2/\ell^2}} \left(1 \mp \sqrt{\frac{3}{2}} \frac{a}{\ell}\right),
\label{rotatinglyapunov} \\
\frac{2 \pi T_{\pm}}{v_B^{\pm}} &=  \frac{2 r_+}{\ell^2\sqrt{1 - a^2/\ell^2}}\left(\sqrt{\frac{3}{2}} \mp \frac{a}{\ell}\right). 
\label{rotatingvb}
\end{align}
A quick inspection of eq. (\ref{rotatinglyapunov}) and eq. (\ref{rotatingvb}) reveals that these expressions can be obtained by applying a boost of velocity $v = -\overline{\Omega} \ell/2 = a/\ell$ to the expression $H_4 \sim \exp(2 \pi T_0(t - \ell \Psi/2 v_B^{(0)}))$ for the OTOC of the AdS$_5$-Schwarzschild black brane. Performing a Lorentz transformation of this form gives
\begin{equation}
2 \pi T_{\pm} = 2 \pi T_0 \gamma_v\bigg(1 \mp \frac{v}{v_B^{(0)}}\bigg), \hspace{1.0cm} \frac{2 \pi T_{\pm}}{v_B^{\pm}} = 2 \pi T_0 \gamma_v\bigg(\frac{1}{v_B^{(0)}} \mp v \bigg), 
\end{equation}
which are equivalent to \eqref{rotatinglyapunov} and \eqref{rotatingvb} upon using the expressions $2 \pi T_0 = 2 r_+/\ell^2$ and $v_B^{(0)} = \sqrt{2/3}$, with $v_B^{(0)}$ the butterfly velocity of the static Schwarzschild-AdS$_5$ black brane. 

Before demonstrating that this butterfly also has a hydrodynamic origin, as discussed in the introduction to this section, we note that the Lyapunov exponent $\lambda_L$ associated to the OTOC on Hopf circles can be extracted from $H(t, 0)$ in eq. (\ref{eq:hopfotoc}) (see for instance \cite{Mezei:2019dfv}). Doing so gives $\lambda_L = 2 \pi T_+ = \mathrm{Im}(\omega_+)$ (for $a>0$) and $\lambda_L = 2 \pi T_- = \mathrm{Im}(\omega_-)$ (for $a<0$). We therefore have that
\begin{equation}
 \lambda_L = 2\pi T \left(1-\sqrt{\frac{3}{2}}\frac{|a|}{\ell}\right) = 2 \pi T \left(1-|v|/v_B^{(0)} \right)
\end{equation}
which, as discussed in~\cite{Amano:2022mlu}, saturates a velocity-dependent generalization of the Maldacena/Shenker/Stanford chaos bound \cite{Maldacena:2015waa} orginally proposed in~\cite{Mezei:2019dfv}. Furthermore, this naturally saturates the generalized bound for the positive operators $\theta_a Q_a$  proposed in~\cite{Mezei:2019dfv}, given by
\begin{equation}\label{eq:gen_bound}
    \frac{\left| \theta_a[Q_a]^\mu \partial_\mu H \right|}{1-H}\leq 2\pi  \, .
\end{equation}
where $H$ is the OTOC, $\theta_a$ are chemical potentials and $Q_a$ are the corresponding generators. In our case the operators are given by $\theta_a Q_a=\beta(\tilde{H}+\overline{\Omega} J)$ (with $\beta$ the inverse temperature, $\tilde{H}$ the Hamiltonian, $J$ the angular momentum and $\overline{\Omega}$ the angular velocity) and in this case the generalized bound given above in equation (\ref{eq:gen_bound}) reduces to
\begin{equation}
    \frac{\left| \partial_t H(t,\psi)+\overline{\Omega} \partial_\psi H(t,\psi)\right|}{1-H(t,\psi)}\leq 2\pi T \, .
\end{equation}
Using the definition of the OTOC given in equation (\ref{eq:otocsimple}) one finds the bound is saturated~\footnote{ 
It was pointed out already in~\cite{Blake:2021hjj}, that rotating systems in 1+1 d (e.g.~\cite{Jahnke:2019heq}) have been noticed to satisfy the generalized MSS bound of~\cite{Mezei:2019dfv}. }

\subsubsection{Near horizon metric fluctuations}
As demonstrated in~\cite{Amano:2022mlu} and discussed in the introduction of this section, we now turn to the hydrodynamic origin of chaos in our holographic system, and we give an overview of the calculation done to extract information about the retarded Green's function, $G^{R}_{T^{00}T^{00}}$, from the study of infalling metric perturbations. We will see that this exhibits the characteristic features of pole-skipping as identified in~\cite{Schalm:2018pjh,Blake:2017ris, Blake:2018leo}. To be concrete, at a special value of the frequency $\omega = i2\pi T$ the linearized Einstein equations admit an additional infalling mode when the angular profile of the metric component $\delta g_{vv}$ at the horizon is solution to the (sourceless) shockwave equation (\ref{eq:shockwave_differential_equation}) that governs the form of the OTOC. 

Unlike the discussion of the previous section we now take the infalling Eddington-Finkelstein coordinates~\footnote{Note we still choose the co-rotating coordinate defined in eq. (\ref{eq:co-rot}).} (see appendix~\ref{sec:Null_coordinates}) $(v, r, \tilde{\psi}, \phi, \theta)$. We then study linearized perturbations of the metric of the form
\begin{equation}
\label{metpert}
\delta g_{\mu \nu}(v, r, \tilde{\psi}, \phi, \theta ) = e^{i (k \tilde{\psi} -\omega v)}\delta g_{\mu \nu}(r, \phi, \theta)
\end{equation}
Note that we have not yet employed the full machinery of the Wigner D-functions as developed in section~\ref{sec:model}. We initially refrain from this in order to demonstrate the equivalence between the linearized Einstein equations near the horizon and the shockwave computation. Inserting this ansatz into the linearized Einstein equations will give rise to a coupled set of coupled PDEs for $\delta g_{\mu \nu}(r, \phi, \theta)$. In particular, as in previous examples of pole-skipping, we find that at $\omega = i2\pi T$ the $vv$ component of the Einstein equations at $r = r_+$ reduces to a decoupled equation for the angular profile of $\delta g_{vv}$ evaluated at the horizon. Furthermore, all other metric perturbation decouple from this component of the Einstein equations which takes the form
\begin{align}\label{eq:horEqn}
       \bigg\{
       \Big[
       \mathrm{cot}(\theta)\partial_{\theta} + \partial_{\theta}^2 &- \mathrm{csc}^2(\theta) k^2 + \mathrm{csc}^2(\theta)\partial_{\phi}^2 - 2i\mathrm{cot}(\theta)\mathrm{csc}(\theta)k\partial_{\phi}
       \Big] \notag \\
       &+\lambda_1 + i\lambda_2 k - \lambda_3 k^2
       \bigg\}
       \delta g_{vv}(r_+, \phi, \theta) = 0\, ,
\end{align}
where $\lambda_1$, $\lambda_2$ and $\lambda_3$ the same constants defined in equation (\ref{eq:lambda_coefficients}). As advertised, eq. (\ref{eq:horEqn}) is precisely the source-less version of the shockwave equation governing the form of OTOC (\ref{eq:shockwave_differential_equation}), after a Fourier transform with respect to the $\psi$ coordinate. The key point is that the metric perturbation $\delta g_{vv}(r_0, \phi, \theta)$ must either have a specific angular profile on the horizon (i.e. be a non-trivial solution to eq.(\ref{eq:horEqn})) or it must be zero at the horizon. As such, for metric perturbations with a particular angular profile related to the shockwave equation we expect an extra infalling mode, and the pole-skipping phenomenon. 

We can go further by recalling that the Wigner D-functions are related to the Wigner d-functions as $\mathcal{D}^{{\cal J}}_{{\cal K} {\cal M}}(\tilde{\psi}, \theta, \phi) = d^{{\cal J}}_{{\cal K} {\cal M}}(\theta) e^{i {\cal K} \tilde{\psi} + i \mathcal{M} \phi}$. With this in mind we can can parameterize $\delta g_{vv}(r_+,\phi,\theta)=e^{i\mathcal{K}\phi}d^{\mathcal{J}}_{\mathcal{KM}}(\theta)$, inserting this into eq. (\ref{eq:horEqn}) leads to the statement that at pole-skipping points the quantum numbers ${\cal K}$ and ${\cal J}$ are related~\footnote{It is important to understand that for integer and half-integer ${\cal J}$ with $|{\cal M}|, |{\cal K}| \leq {\cal J}$ there is a unique solution to eq. (\ref{eq:wignerdeqn}) that is regular on $S^3$. In order to consider pole-skipping, it is necessary to relax the condition that ${\cal M}, {\cal K}, {\cal J}$ be integers and consider them to be arbitrary complex parameters. This corresponds to considering angular profiles of the metric components that are not necessarily regular on the three-sphere. } by
\begin{equation}
\label{eq:poleskippingW}
\lambda_1 + i \lambda_2 {\cal K} - \lambda_3 {\cal K}^2 = {\cal J}({\cal J} + 1). 
\end{equation}
It is worth pointing out that as a result of the equivalence between the horizon equation and shockwave equation, this relationship corresponds exactly to the locations of the poles in the shockwave profile given in eq. (\ref{eq:exactshock}).

To compare to eq. (\ref{eq:hopfotoc}) we take the large BH limit $r_+/\ell \gg 1$ while keeping $a/\ell \sim \mathcal{O}(1)$. In the this limit the leading order behavior of the constants $\lambda_1$, $\lambda_2$, $\lambda_3$ is
\begin{equation}
\lambda_1 = \frac{r_+^2}{2\ell^4}(2a^2 -3\ell^2), \hspace{1cm} \lambda_2 = -\frac{2ar_+}{\ell^2} \sqrt{1-\frac{a^2}{\ell^2}} \hspace{1cm} \lambda_3 = -\frac{a^2}{\ell^2}. \nonumber
\end{equation}
As we discussed in section~\ref{sec:QNMresults}, sectors in the MP BH are determined by a choice of the quantum number $\mathcal{K}$, here we will consider the class of pole-skipping points where ${\cal K}$ is parallel to the rotation direction i.\ e.\ taking its maximal value (${\cal K} = {\cal J}$). Further, we recall that in the large BH limit we also choose the momentum to scale with the horizon radius as ${\cal J} \sim r_+/\ell \gg 1$. Equipped with a behavior of the momentum in the large BH limit we find that eq. (\ref{eq:poleskippingW}) reduces to the following quadratic equation for ${\cal J}$
\begin{equation}
\left(1 - \frac{a^2}{\ell^2} \right){\cal J}^2 + \frac{2 i a r_+}{\ell^2} \sqrt{1 -\frac{a^2}{\ell^2}} {\cal J} + \frac{r_+^2}{2\ell^4}\left(3 \ell^2 - 2a^2\right) = 0\, ,
\end{equation}
whose solutions are given by
\begin{equation}
\label{eq:wavevectors}
{\cal J} = \pm \frac{i r_+}{\ell\sqrt{1 - a^2/\ell^2}}\bigg(\sqrt{\frac{3}{2}} \mp \frac{a}{\ell} \bigg) = \pm i k_{\pm}. 
\end{equation}
Finally, we see these modes exhibit pole-skipping at a frequency $\omega = i 2 \pi T$ and at precisely the same wave-vectors $k_{\pm}$ that appear in the exponential form of the OTOC when the operators lie on a Hopf circles (see eq. (\ref{eq:hopfotoc})). 

So far we have been working in terms of co-rotating coordinates, hence the boundary response is characterized by coordinates $(t, \tilde{\psi}, \theta, \phi)$. If, however, we transform these results to the standard coordinates $(t, \psi, \theta, \phi)$ for the three sphere, then we find that the wave-vectors $k_{\pm}$ at which pole-skipping occurs are unchanged, while the frequencies are modified to 
\begin{equation} \label{eq:poleSkippingPointsRestFrame}
\omega_{\pm} = i(2\pi T \pm k_{\pm}\overline{\Omega}) = \frac{2 ir_+}{\ell^2\sqrt{1 - a^2/\ell^2}}\bigg(1 \mp \sqrt{\frac{3}{2}} \frac{a}{\ell}\bigg). 
\end{equation}

Before closing this section it is worth discussing one of the main conclusions of this section in more detail. Namely, we have see that the pole-skipping locations uncovered in the near horizon analysis of the Einstein equations match precisely the poles of the summand in the expression for the angular profile of the shockwave geometry (compare eq. (\ref{eq:poleskippingW}) and eq. (\ref{eq:exactshock})). And, furthermore, in the large BH limit, where we can obtain a closed form expression for the OTOC, the pole-skipping points match precisely the wave vector and exponential growth rate of the OTOC (compare eq. (\ref{eq:poleSkippingPointsRestFrame}) and eq. (\ref{eq:wavevectors}) to eq. (\ref{rotatinglyapunov}) and eq. (\ref{rotatingvb})). As already remarked, the form of the OTOC in this limit can be obtained directly from the equivalent calculation in a Schwarzschild black brane by performing a Lorentz transformation. To be concrete, for a non-rotating Schwarzschild AdS black brane geometry it is the pole-skipping points in the scalar channel which precisely match the wave vector and exponential growth rate of the OTOC, these are given by~\cite{Shenker:2015keq}
\begin{equation}\label{eq:restingPoleSkipping}
    \mathfrak{w} = i,\quad \mathfrak{q}= \pm  \sqrt{\frac{3}{2}} i\, .
\end{equation}
Therefore, we should expect to see explicitly that the equations of motion for the fluctuations in the scalar channel of the MP BH reduce to those of a boosted Schwarzschild black brane in the large BH limit and hence we should be able to numerically verify, directly, that the mode given by eq. (\ref{eq:restingPoleSkipping}) when boosted appropriately, is contained in the spectrum of fluctuations. 

To do see these observations we turn to the scalar channel as discussed in section~\ref{sec:QNMresults} and take the large BH limit defined in eq. (\ref{eq:LBH1}) and eq. (\ref{eq:LBH2}) keeping only the leading term in $(r_+/\ell)$. Completing this procedure one finds that the fluctuations $(h_{33}, h_{tt}, h_{t3}, h_{+-}, h_{+t}, h_{+3}, h_{++})$ decouple into three sectors a scalar channel $(h_{33}, h_{tt}, h_{t3}, h_{+-})$, vector channel $(h_{+t}, h_{+3})$ and tensor channel $(h_{++})$. Again, focusing on the scalar channel, one finds 7 equations, which can be arranged as four dynamical equations and three constraints. Transforming to the coordinate $u=(r_+/r)^2$ (and setting $\ell=1$ for simplicity), an analysis of residual gauge symmetry leads to a master variable $Z_0(u)$ which can be constructed from the fields $(h_{33}, h_{tt}, h_{t3}, h_{+-})$ as
\begin{align}
     Z_0(u) &= (2  \left(j^2 \left(a^2-z^4-1\right)-a^2 \nu ^2 \left(z^4+1\right)-2 a j \nu  z^4+\nu ^2\right) h_{+-}(u)\nonumber\\
        &+\left(a^2-1\right)  (j^2 h_{tt}(u) - 4j \nu  h_{t3}(u) +4\nu^2 h_{33}(u))) \, .
\end{align}
In terms of this master variable the scalar channel can be reduced to a single equation 
\begin{equation}\label{eq:scalar_sector}
    \begin{aligned}
       0 &= \frac{\left(a^2 \left(4 u^2-u \nu^2-4\right)-2 a j u \nu-u \left(j^2+4 u\right)+4\right)}{a^2-1}Z_0(u) + \\
         &\quad\quad\frac{f(u)}{u}\left(3 u^2-5\right) Z_0'(u) +\frac{f(u)^2}{u^2} Z_0''(u) ,   
    \end{aligned}
\end{equation}
At this point, as was shown originally in~\cite{Amano:2022mlu}, we can demonstrate that the explicit dependence on the rotation parameter $a$ can be removed by the following Lorentz boost
\begin{equation}\label{eq:criticalPointTrajectory}
    \begin{aligned}
         \mathfrak{q} &= \frac{a \nu + j }{\sqrt{1-a^2}}\,, \qquad \mathfrak{q} = \frac k {2 \pi T},\,\\
         \mathfrak{w} &= \frac{\nu + a j }{\sqrt{1-a^2}}\,.  \qquad
         \mathfrak{w} = \frac \omega {2 \pi T}\ \, ,  
    \end{aligned}
\end{equation}
after which the resulting $a$-independent equation is equivalent to the equation for the master field governing (scalar sector) metric perturbations around a Schwarzschild AdS black brane~\cite{Kovtun:2005ev}. Hence, \eqref{eq:scalar_sector} is simply the master equation~\footnote{In fact, it was demonstrated in~\cite{Amano:2022mlu} that in all three sectors (scalar, vector, tensor) of the MP BH, expanded in the large BH limit, the metric perturbations be reduced in the same way to fluctuations around a boosted Schwarzschild AdS black brane. } for scalar fluctuations around a boosted black brane, where the momentum of the fluctuations is aligned with the boost direction. 
 
\begin{figure}
    \centering
    \includegraphics[width=12cm]{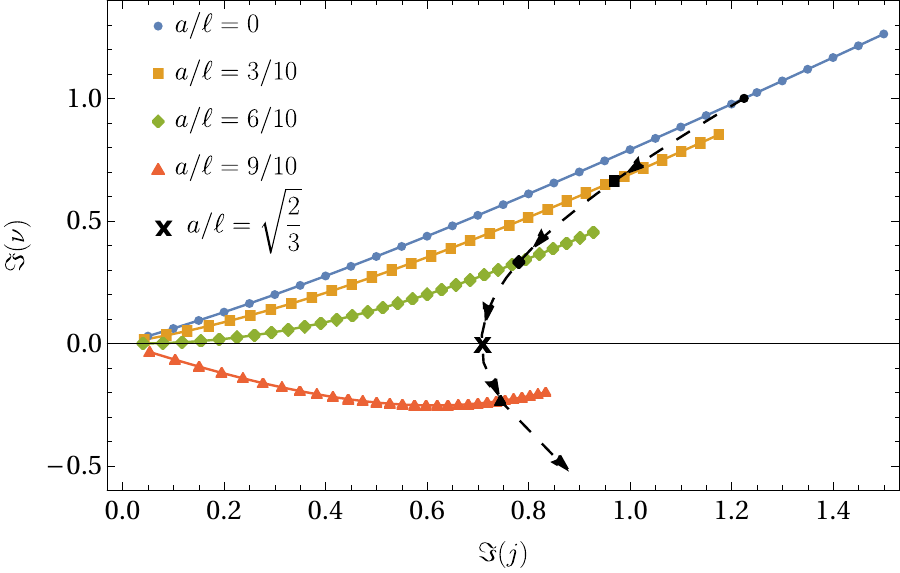}
    \caption{
        \textit{Chaos in the sound dispersion.}  
            The complex-valued dispersion relation $\nu(j)$ for the rotating BH in the large BH limit. Displayed is the imaginary part of the frequency parameter $\Im(\nu)$ of the perturbation as a function of the imaginary part of its angular momentum $\Im(j)$. 
            Each color (and different symbol) represents a different value for the angular momentum per mass $a/\ell$ in incremental steps of $3/10$. 
            The black line indicates the trajectory of the pole-skipping location as the angular momentum per mass, $a/\ell$, varies.
            The cross, $\mathbf x$, highlights the point where one of the pole-skipping points associated to this dispersion relation ceases to be located in the upper half plane, corresponding to $a/\ell = \sqrt{\frac{2}{3}}$. Adapted from~\cite{Amano:2022mlu}. 
    \label{fig:dispersion} 
    }
\end{figure}

Having confirmed explicitly that the fluctuation equations in the large BH limit reduce to those of a boosted Schwarzschild black brane it is then a simple matter to apply eq. (\ref{eq:criticalPointTrajectory}) to eq. (\ref{eq:restingPoleSkipping}) to confirm that it gives that the pole-skipping points in eq. (\ref{eq:poleSkippingPointsRestFrame}) and eq. (\ref{eq:wavevectors}). Furthermore, in~\cite{Amano:2022mlu} pole-skipping at these locations was verified explicitly by numerically~\footnote{This was done using two independent methods from two independently constructed routines, first, from a static Schwarzschild black brane whose modes were then boosted and second, directly from the large BH limit of the rotating BH (for further details see the Appendix of~\cite{Amano:2022mlu}).} solving for the dispersion relation of the (sound) QNM in the scalar sector of perturbations for each value of $a/\ell$. These numerical results are illustrated in ~\ref{fig:dispersion}. The pole-skipping locations are indicated with black symbols in Fig.~\ref{fig:dispersion} for increasing values of rotation $a/\ell$. As noted in~\cite{Amano:2022mlu}, the rotation parameter $a/\ell$ affects both the frequency and wave-vector of the pole-skipping points~\footnote{This is in contrast to static planar BHs for which only the wave-vector changes as one varies microscopic parameters at a fixed temperature (e.g. as in the general results of ~\cite{Blake:2018leo, Blake:2017ris} and the Reissner-Nordstrom-AdS geometry of~\cite{Abbasi:2020ykq})}. At the special value $a/\ell=\sqrt{2/3}=v_B^{(0)}$ i.e. the boost speed is equivalent to the conformal butterfly velocity, the pole-skipping point associated with $(\omega, {\cal K}) = (\omega_+, i  k_+)$ crosses over into the lower half plane of the complex frequency plane (which has been highlighted in the figure with a cross).

\begin{figure}
\begin{center}
    \includegraphics[width=12cm]{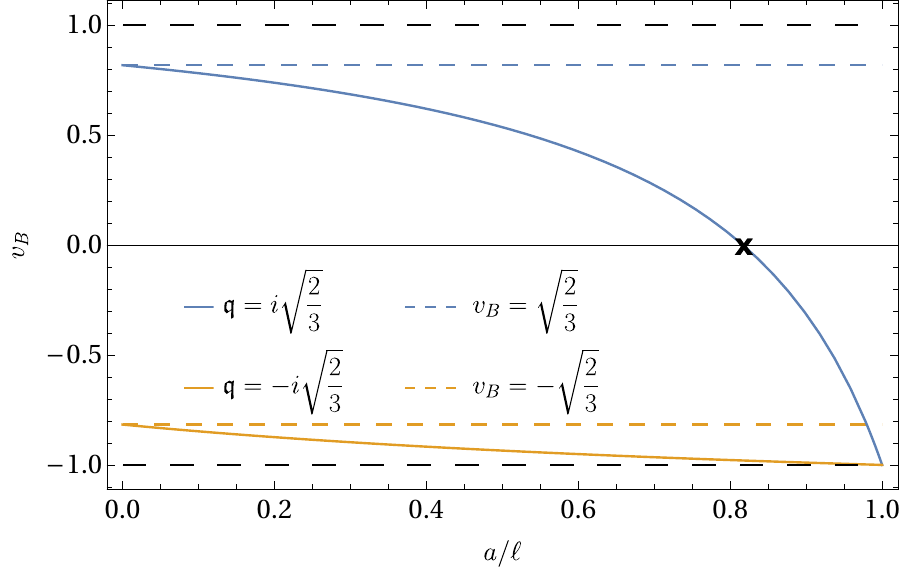}
\end{center}
    \caption{
    {\it Butterfly velocity in a rotating plasma.} The butterfly velocities $v_B^{\pm}$ given in \eqref{eq:vtildeB} are displayed as a function of the rotation parameter $a/\ell$.
    The colored dashed lines indicate the conformal value $v_B^{(0)}=\pm \sqrt{2/3}$ whilst the black dashed lines indicate the speed of light. 
    Notice that the ``upstream'' butterfly velocity $v_B^+$ crosses zero at exactly the same value of rotation $a/\ell = \sqrt{\frac{2}{3}}$ for which the pole-skipping point plotted in Fig.~\ref{fig:dispersion} ceases to be located in the upper half complex frequency plane. Adapted from~\cite{Amano:2022mlu}. 
    \label{fig:butterfly}}
\end{figure}

The fact that the pole-skipping location $(\omega, {\cal K}) = (\omega_+,i k_+)$ enters the lower half complex frequency plane when the boost associated to rotation exceeds $v_B^{(0)}$ has an interesting implication for the associated butterfly velocity. Extracting a butterfly velocity from the pole-skipping locations (or equivalently through the OTOC profile in eq. (\ref{rotatingvb}) by $\omega_{\pm}/k_{\pm} = v_B^{\pm}$ we obtain
\begin{equation}\label{eq:vtildeB}
    v_B^{\pm}=\frac{v_B^{(0)}\mp \frac a \ell }{1\mp v_B^{(0)}\frac a \ell} \, ,
\end{equation}
which, as expected, is nothing but relativistic addition of velocities obtained by applying a boost with speed $a/\ell$ parallel to $\pm v_B^{(0)}$ \footnote{This was noted earlier independently in private communication between the authors, Casey Cartwright and Matthias Kaminski, with Navid Abbasi.}. From this formula one can note that the butterfly velocity $v_B^+$ vanishes at $a/\ell=v_B^{(0)}$  precisely at same time that the pole-skipping location $(\omega, {\cal K}) = (\omega_+, i k_+)$ passes into the lower half complex frequency plane. This is displayed in Fig.~\ref{fig:butterfly} with a cross denoting the crossing location. This phenomena is quite natural since, a perturbation traveling upstream against a uniform fluid flow in a conformal theory will be experienced by an observer at rest as sitting still exactly when the perturbation has the speed $|a/\ell|$. Thus, a perturbation moving at the conformal butterfly speed $v_B^{(0)}$ appears to sit still when the fluid streams with velocity $|a/\ell|=v_B^{(0)}$. When the fluid flows faster, $|a/\ell|>v_B^{(0)}$, this  perturbation is dragged downstream. Although natural, this phenomenon could not be observed in previous studies of OTOCs and pole-skipping in rotating BHs. For instance, in the BTZ BH case~\cite{Liu:2020poq,Jahnke:2019heq} the butterfly velocity is equal to the speed of light, and while in the Kerr-AdS case~\cite{Blake:2021hjj} the authors were restricted to the slowly rotating limit, a limit in the which the rotation speed was parametrically smaller than the corresponding conformal butterfly velocity.

\section{Outlook}
\label{sec:discussion}
\noindent A summary was already given in the introduction sections for the fluid results~\ref{sec:introSummary} and for the gravity results~\ref{sec:gravityResults}. Here, we focus on the outlook:
\begin{enumerate}
    \item The general case with unequal angular momenta $J_\phi \neq J_\psi$ ($a\neq b$) or the singly spinning configurations $J_\phi \neq 0,  J_\psi = 0$ and $J_\phi = 0,  J_\psi \neq 0$ could be explored. 
    \begin{enumerate}
        \item The singly spinning case contains an axis of rotation, so large temperature expansions around such a state would be interesting.
        \item Both cases involve solving 2D partial differential equations in order to find the QNMs.
    \end{enumerate}
    \item Explore the intermediate horizon radius regime around $10^3<r_+<10^7$ in which nontrivial hydrodynamic transport is expected (which is not simply that of a boosted fluid, as argued in Sec.~\ref{sec:HorizonDependence}). 
    \item What is the physical relevance of the level crossings between non-hydrodynamic modes (see Fig.~\ref{fig:vary_A_small_scale})? Are these crossing points related to the level crossings and mode collisions determining critical points at complexified momentum~\cite{Grozdanov:2019uhi,Grozdanov:2019kge}?
    \item Only a subset of fluctuations was explored in this article, namely the ones that reduce in the large BH limit to the fluctuations with linear momentum aligned with the anisotropy created in the fluid due to its rotation. More general fluctuations should be considered, which would for example allow one to compute the perpendicular diffusion coefficient $\mathcal{D}_\perp$. 
    \item The presence of a linear instability in global AdS Kerr begs the question of what is the dual field theoretic interpretation to this instability?
    \item An implementation of rotation, polarization and spin in hydrodynamic codes is desirable for various applications ranging from modelling heavy-ion-collisions~\cite{Romatschke:2017ejr} over modelling astrophysical settings with relevance to BH mergers or neutron star mergers~\cite{LIGOScientific:2016aoc,Bailes:2021tot}, and fluids in the context of condensed matter physics, {e.g.}~\cite{Molenkamp:1994,Moll:2016,Bloch:2008,Sinova:2015}. The holographic fluid presented in this article can serve as a testing ground for such hydrodynamic codes. 
    \item It is possible to systematically construct a version of hydrodynamics around a rotating equilibrium state. This state breaks rotation invariance and thus involves a new vector, the unit vector pointing in the symmetry direction, similar to anisotropic hydrodynamics~\cite{Florkowski:2010cf, Martinez:2010sc, Strickland:2014pga}). The fluid/gravity correspondence~\cite{Bhattacharyya:2007vjd} can be used to systematically derive and test such a description. It can be used to extend the rotating black hole solutions~\eqref{eq:Simple_Spin_MP_Metric} into metric solutions which have temperature and angular momentum depend on the transverse coordinates and time. This leads to a gradient expansion in the dual field theory, locally correcting the rotating equilibrium state by gradients of temperature and angular momentum, likely modifying the hydrodynamic dispersion relations significantly. 
\end{enumerate}

It is worthwhile pointing out the logical possibility that the quark-gluon-plasma generated in heavy-ion-collisions is {\it not} carrying any significant vorticity or angular momentum. Note that a recent numerical study of a holographic fluid created in the collision of localized, lumpy shocks (modeling a granular nuclear structure) finds that most of the angular momentum after the collision resides in the periphery while only a small fraction of the total angular momentum is present in the region representing the holographic QGP~\cite{Waeber:2022vgf}. If the same should be true for the QCD QGP generated at RHIC and LHC, then the hyperon polarization can not be explained by any arguments involving any significant amount of vorticity or angular momentum in the QGP. In that case, the hyperon polarization data may be considered an open question lacking theoretical understanding. 

%

\vspace{6pt} 



\section*{Acknowledgements}
{C.C., M.A. and M.K. acknowledge the excellent discussions and collaborations with Jorge Noronha, Enrico Speranza, Mike Blake and Anthony Thompson, on which part of this work is based. 
This research was supported in part by the U.S.~Department of Energy under Grant No. DE-SC0012447 and by the National Science Foundation under Grant No. NSF PHY-1748958. C.C. was supported in part, by the Netherlands Organisation for Scientific Research (NWO) under the VICI grant VI.C.202.104.  M.K. thanks the Kavli Institute for Theoretical Physics, Santa Barbara for hospitality while part of this work was completed. }



\appendix
\section{QNM Data}
\label{sec:QNMdata}
For convenience we include the six modes closest to the line $\Im(v)=0$ in the three sectors discussed in section~\ref{sec:QNMresults} for $a/\ell=0,1/2/9/10$. In each table we fix a value of $\mathcal{J}=0,1/2,1$\ and $r_+=10$, the remaining data can be found in the supplementary material. \\
\subsection{Tensor sector}
\begin{table}[h]
    \centering
    \begin{tabular}{c|c|c|c|c|c | c }
    n & \multicolumn{2}{|c|}{$a/\ell=0$}   & \multicolumn{2}{|c|}{$a/\ell=1/2$} & \multicolumn{2}{|c}{$a/\ell=9/10$} \\
      \hline 
         & $\Re(v)$ & $\Im(v)$ & $\Re(v)$ & $\Im(v)$ & $\Re(v)$ & $\Im(v)$  \\
         \hline
    1   &  1.57792 & -1.36943 & 1.83839 & -1.73479 & -0.214354 & -0.529707 \\
2& -1.57792 & -1.36943 & -1.63407 & -1.73893 & -0.223526 & -1.00114 \\
3& 2.60488 & -2.37904 & -0.471547 & -1.94296 & -0.231857 & -1.4879 \\
4& -2.60488 & -2.37904 & -2.90868 & -2.98463 & -0.239585 & -1.98766 \\
5& -3.61756 & -3.38252 & 3.08986 & -2.98592 & -0.246234 & -2.49843 \\
6& 3.61756 & -3.38252 & -4.14395 & -4.21689 & 2.84375 & -2.73946 \\
    \end{tabular}
    \caption{ $\mathcal{J}=0$ tensor modes. $n$ indicates the $n$th furthest mode from the $\Im(\nu)=0$ axis.
    \label{tab:tensor_table_1} }
\end{table}

\begin{table}[h]
    \centering
    \begin{tabular}{c|c|c|c|c|c | c }
    n & \multicolumn{2}{|c|}{$a/\ell=0$}   & \multicolumn{2}{|c|}{$a/\ell=1/2$} & \multicolumn{2}{|c}{$a/\ell=9/10$} \\
      \hline 
         & $\Re(v)$ & $\Im(v)$ & $\Re(v)$ & $\Im(v)$ & $\Re(v)$ & $\Im(v)$  \\
         \hline
    1 & -1.58609 & -1.36696 & 1.86324 & -1.73367 & -0.267892 & -0.52861 \\
 2 & 1.58609 & -1.36696 & -1.6031 & -1.74128 & -0.279378 & -1.00015 \\
 3 & 2.6107 & -2.3774 & -0.594678 & -1.92682 & -0.289814 & -1.48695 \\
 4 & -2.6107 & -2.3774 & -2.88713 & -2.98347 & -0.299508 & -1.98678 \\
 5 & 3.62224 & -3.38124 & 3.1134 & -2.98514 & -0.307855 & -2.49776 \\
 6 & -3.62224 & -3.38124 & -4.12207 & -4.21626 & -2.62293 & -2.73856 \\
    \end{tabular}
    \caption{ $\mathcal{J}=1/2$ tensor modes. $n$ indicates the $n$th furthest mode from the $\Im(\nu)=0$ axis.
    \label{tab:tensor_table_2} }
\end{table}

\begin{table}[h]
    \centering
    \begin{tabular}{c|c|c|c|c|c | c }
    n & \multicolumn{2}{|c|}{$a/\ell=0$}   & \multicolumn{2}{|c|}{$a/\ell=1/2$} & \multicolumn{2}{|c}{$a/\ell=9/10$} \\
      \hline 
         & $\Re(v)$ & $\Im(v)$ & $\Re(v)$ & $\Im(v)$ & $\Re(v)$ & $\Im(v)$  \\
         \hline
   1 & -1.5965 & -1.36383 & 1.88859 & -1.73202 & -0.321408 & -0.527454 \\
 2 & 1.5965 & -1.36383 & -1.56789 & -1.74515 & -0.335217 & -0.999071 \\
 3 & 2.61814 & -2.37531 & -0.722979 & -1.90682 & -0.347771 & -1.48588 \\
 4 & -2.61814 & -2.37531 & -2.86612 & -2.98187 & -0.359452 & -1.9858 \\
 5 & -3.62822 & -3.37959 & 3.13731 & -2.984 & -0.369518 & -2.49702 \\
 6 & 3.62822 & -3.37959 & -4.10043 & -4.21535 & -2.59952 & -2.73727 \\
    \end{tabular}
    \caption{ $\mathcal{J}=1$ tensor modes. $n$ indicates the $n$th furthest mode from the $\Im(\nu)=0$ axis.
    \label{tab:tensor_table_3} }
\end{table}
\newpage

\subsection{Vector sector}
\begin{table}[h]
    \centering
    \begin{tabular}{c|c|c|c|c|c | c }
    n & \multicolumn{2}{|c|}{$a/\ell=0$}   & \multicolumn{2}{|c|}{$a/\ell=1/2$} & \multicolumn{2}{|c}{$a/\ell=9/10$} \\
      \hline 
         & $\Re(v)$ & $\Im(v)$ & $\Re(v)$ & $\Im(v)$ & $\Re(v)$ & $\Im(v)$  \\
         \hline
    1   &  1.57792 & -1.36943 & 1.83839 & -1.73479 & -0.214354 & -0.529707 \\
2& -1.57792 & -1.36943 & -1.63407 & -1.73893 & -0.223526 & -1.00114 \\
3& 2.60488 & -2.37904 & -0.471547 & -1.94296 & -0.231857 & -1.4879 \\
4& -2.60488 & -2.37904 & -2.90868 & -2.98463 & -0.239585 & -1.98766 \\
5& -3.61756 & -3.38252 & 3.08986 & -2.98592 & -0.246234 & -2.49843 \\
6& 3.61756 & -3.38252 & -4.14395 & -4.21689 & 2.84375 & -2.73946 \\
    \end{tabular}
    \caption{ $\mathcal{J}=0$ vector modes. $n$ indicates the $n$th furthest mode from the $\Im(\nu)=0$ axis.
    \label{tab:vector_table_1} }
\end{table}
\begin{table}[h]
    \centering
    \begin{tabular}{c|c|c|c|c|c | c }
    n & \multicolumn{2}{|c|}{$a/\ell=0$}   & \multicolumn{2}{|c|}{$a/\ell=1/2$} & \multicolumn{2}{|c}{$a/\ell=9/10$} \\
      \hline 
         & $\Re(v)$ & $\Im(v)$ & $\Re(v)$ & $\Im(v)$ & $\Re(v)$ & $\Im(v)$  \\
         \hline
   1 & -9.77489$\times 10^{-85}$ & -0.00626198 & -0.0252699 & -0.00514204 & -0.045346 & -0.00226457 \\
 2 & -1.58609 & -1.36696 & -1.46214 & -1.42053 & -0.160298 & -0.532138 \\
 3 & 1.58609 & -1.36696 & 1.59728 & -1.4975 & -0.241578 & -0.716389 \\
 4 & -1.57145 & -1.37138 & 1.81681 & -1.7323 & -0.167453 & -1.00312 \\
 5 & 1.57145 & -1.37138 & -1.66563 & -1.73445 & -0.202738 & -1.2898 \\
 6 & -2.6107 & -2.3774 & -0.350175 & -1.96503 & -0.173784 & -1.48965 \\
    \end{tabular}
    \caption{ $\mathcal{J}=1/2$ vector modes. $n$ indicates the $n$th furthest mode from the $\Im(\nu)=0$ axis.
    \label{tab:vector_table_2} }
\end{table}
\begin{table}[ht]
    \centering
    \begin{tabular}{c|c|c|c|c|c | c }
    n & \multicolumn{2}{|c|}{$a/\ell=0$}   & \multicolumn{2}{|c|}{$a/\ell=1/2$} & \multicolumn{2}{|c}{$a/\ell=9/10$} \\
      \hline 
         & $\Re(v)$ & $\Im(v)$ & $\Re(v)$ & $\Im(v)$ & $\Re(v)$ & $\Im(v)$  \\
         \hline
  1 & -6.3260$\times 10^{-94}$ & -0.0150693 & -0.0508062 & -0.0119096 & -0.090775 & -0.00472503 \\
 2 & -1.5965 & -1.36383 & -1.45984 & -1.4059 & -0.213379 & -0.532508 \\
 3 & 1.5965 & -1.36383 & 1.6058 & -1.49884 & -0.297615 & -0.713923 \\
 4 & -1.57556 & -1.37012 & 1.84404 & -1.729 & -0.223126 & -1.00332 \\
 5 & 1.57556 & -1.37012 & -1.64015 & -1.73249 & -0.262902 & -1.28889 \\
 6 & 2.61814 & -2.37531 & -0.468888 & -1.96172 & -0.23164 & -1.48972 \\
    \end{tabular}
    \caption{ $\mathcal{J}=1$ vector modes. $n$ indicates the $n$th furthest mode from the $\Im(\nu)=0$ axis.
    \label{tab:vector_table_3} }
\end{table}
\newpage

\subsection{Scalar sector}
\begin{table}[h]
    \centering
    \begin{tabular}{c|c|c|c|c|c | c }
    n & \multicolumn{2}{|c|}{$a/\ell=0$}   & \multicolumn{2}{|c|}{$a/\ell=1/2$} & \multicolumn{2}{|c}{$a/\ell=9/10$} \\
      \hline 
         & $\Re(v)$ & $\Im(v)$ & $\Re(v)$ & $\Im(v)$ & $\Re(v)$ & $\Im(v)$  \\
         \hline
         1 & -1.5779 & -1.3694 & -1.4252 & -1.3411 & -0.0010668 & -0.55268 \\
        2 & 1.5779 & -1.3694 & 1.4252 & -1.3411 & -0.074684 & -0.75314 \\
        3 & -1.5685 & -1.3723 & -1.4809 & -1.4504 & 0.0 & -0.86704 \\
        4 & 1.5685 & -1.3723 & 1.5765 & -1.4824 & 0.0 & -0.86704 \\
        5 & -1.5661 & -1.3730 & -1.7330 & -1.7340 & -0.00047209 & -1.0449 \\
        6 & 1.5661 & -1.3730 & 1.7332 & -1.7366 & -1.2191 & -1.2125 \\
    \end{tabular}
    \caption{ $\mathcal{J}=0$ scalar modes. $n$ indicates the $n$th furthest mode from the $\Im(\nu)=0$ axis.
    \label{tab:scalar_table_1} }
\end{table}
\begin{table}[h]
    \centering
    \begin{tabular}{c|c|c|c|c|c | c }
    n & \multicolumn{2}{|c|}{$a/\ell=0$}   & \multicolumn{2}{|c|}{$a/\ell=1/2$} & \multicolumn{2}{|c}{$a/\ell=9/10$} \\
      \hline 
         & $\Re(v)$ & $\Im(v)$ & $\Re(v)$ & $\Im(v)$ & $\Re(v)$ & $\Im(v)$  \\
         \hline
         1 & 0.0 & -0.0062620 & -0.011386 & -0.0060167 & -0.036860 & -0.0035422 \\
        2 & -1.5861 & -1.3670 & -1.4396 & -1.3341 & -0.053926 & -0.55468 \\
        3 & 1.5861 & -1.3670 & 1.4124 & -1.3472 & -0.13133 & -0.75273 \\
        4 & -1.5714 & -1.3714 & -1.4762 & -1.4403 & -0.045112 & -0.86704 \\
        5 & 1.5714 & -1.3714 & 1.5848 & -1.4869 & -0.045112 & -0.86704 \\
        6 & -1.5673 & -1.3726 & -1.7121 & -1.7305 & -0.045112 & -0.86704 \\
    \end{tabular}
    \caption{ $\mathcal{J}=1/2$ scalar modes. $n$ indicates the $n$th furthest mode from the $\Im(\nu)=0$ axis.
    \label{tab:scalar_table_2} }
\end{table}
\begin{table}[h]
    \centering
    \begin{tabular}{c|c|c|c|c|c | c }
    n & \multicolumn{2}{|c|}{$a/\ell=0$}   & \multicolumn{2}{|c|}{$a/\ell=1/2$} & \multicolumn{2}{|c}{$a/\ell=9/10$} \\
      \hline 
         & $\Re(v)$ & $\Im(v)$ & $\Re(v)$ & $\Im(v)$ & $\Re(v)$ & $\Im(v)$  \\
         \hline
         1 & -0.081988 & -0.0041527 & -0.095984 & -0.0014583 & -0.099872 & -0.00014121 \\
        2 & 0.081988 & -0.0041527 & 0.051356 & -0.0070246 & 0.010573 & -0.0044887 \\
        3 & 0.0 & -0.015069 & -0.028210 & -0.014144 & -0.074679 & -0.0080336 \\
        4 & -1.5965 & -1.3638 & -1.4545 & -1.3266 & -0.10681 & -0.55665 \\
        5 & 1.5965 & -1.3638 & 1.3998 & -1.3526 & -0.18799 & -0.75219 \\
        6 & -1.5756 & -1.3701 & -1.4733 & -1.4284 & -0.090225 & -0.86704 \\
    \end{tabular}
    \caption{ $\mathcal{J}=1$ scalar modes. $n$ indicates the $n$th furthest mode from the $\Im(\nu)=0$ axis.
    \label{tab:scalar_table_3} }
\end{table}


\section{Numerical methods}
\label{sec:numerics}
In general, the linearized equations of motion of the fluctuations give rise to what is known as a non-linear eigenvalue problem that takes the form
\begin{equation}\label{eq:non_linear_eignvalue_qnmmethod}
    \left(\sum_{n=0}^R M_n \omega^n\right) \Psi_0 = 0 \,,
\end{equation}
where $M_n$ are $\omega$-independent differential operators, $\Psi_0$ are the fields of interest in a given sector, and $R$ is the order of $\omega$ appearing in the equations of motion. Note $R$ is at most two because eq.~\eqref{eq:pertgenericeom} has second order time derivatives. If more than one field makes up a given sector, say $m$ of them, then $M_n$ are $m\times m$ matrices acting on a column vector of $m$ fields represented by $\Psi_0$. It is well-known how to linearize such a non-linear eigenvalue problem by introducing auxiliary variables, $\Psi_i = \omega\Psi_{i-1}$, for $i = 1,\ldots R-1$. The original equation, eq.~\eqref{eq:non_linear_eignvalue_qnmmethod}, then becomes a set of linear equations
\begin{align}\label{eq:linear_eignvalue_qnmmethod}
    \omega \left(\sum_{n=1}^{R} M_n \Psi_{n-1}\right) = - M_0 \Psi_0 \,, \qquad\qquad
    \Psi_i = \omega\Psi_{i-1} \,, \quad i = 1,\ldots R-1 \,,
\end{align}
which is now a linear generalized eigenvalue problem. We work in null coordinates (see Appendix~\ref{sec:Null_coordinates}) to avoid non-regular behavior of the fields at the horizon. 

To proceed, we discretize the resulting equations of motion by decomposing the fields in a truncated Chebyshev basis on a pseudo-spectral grid. Fields in the equations of motion are replaced with vectors of these fields evaluated at the grid points, the coefficients of these fields in the equations of motion become diagonal matrices with diagonal elements the coefficients evaluated at the grid points, and the partial derivatives turn into pseudo-spectral differentiation matrices. Appropriate boundary conditions also have to be imposed, whereby the dynamical equations are required to be regular at the horizon (null coordinates ensure this for us) and are not sourced at the conformal boundary. Finally we can solve the resulting (generically) non-linear eigenvalue problem defined on the pseudo-spectral grid using the \textbf{Eigenvalues} function in Mathematica, and these give the frequencies of QNMs. 

To ensure convergence, comparison over several orders of grid points should be performed to filter out spurious modes introduced in the spectrum due to the discretization scheme. Physical modes will move only slightly when changing the number of grid points used in the scheme unlike spurious modes. Moreover, the equations of motion contain additional unused equations, and these function as constraints that have to be satisfied. By using the \textbf{Eigensystem} function in Mathematica, eigenvalues and the corresponding eigenvectors can be obtained in one go. The eigenvectors represent the values of the fields at each of the grid points for a given QNM frequency. Inserting the eigenvector and eigenvalue back into the constraint equations gives a measure of the success of the method. We use the following conservative measure: 

Let $S^j_i$, $i = 1,\ldots,c$, represent the $c$ constraint equations at each grid point $j$. Using a Gauss-Lobatto grid of $N$ points, we compute (recall there are $m$ fields)
\begin{equation}
   \varsigma= \frac{1}{m N}\sum_{i=1}^c\sum_{j=1}^N | S^j_i| \, .
\end{equation}
What $\varsigma$ measures is how much all the constraint equations in sum differ from zero as an average over all grid points. For the tensor sector, there is no constraint, and we can verify that the QNM problem is solved at least to the level of $10^{-15}$ by directly putting the solution back into the equation. For the vector sector we keep only modes for which $\varsigma<10^{-12}$. The scalar sector is more challenging. There, we are able to find solutions for which $\varsigma \lesssim 10^{-4}$ for  $r_+=10$, $\mathcal{J}=25$, $a/\ell=1/2$ and $N=20$. At smaller $a$ or $J$, or near the large BH limit the convergence improves significantly.\footnote{Two independent codes were constructed for the purpose of computing QNMs in this work. One was initially created for the study~\cite{Garbiso:2020puw} and did not feature a verification of the constraint violation. The second was created independently by another author of the present work during the preparation of~\cite{Amano:2022mlu,Cartwright:2021qpp}, and it was later improved to measure constraint violation. In this work, primary results were obtained using the first code and then independently verified using the second code.} Work to improve this is underway. 

\textit{Critical points}--- Unfortunately, obtaining roots to eq.~(\ref{eq:crit_Points}) is not possible analytically, and numerical method is required. As discussed in~\cite{Cartwright:2021qpp}, first the relevant fields are discretized by introducing a truncated Chebyshev pseudo-spectral representation of $N$th order on a Gauss-Lobatto grid of size $N$. This means that the differential operator $M$ is now a dense, $nN\times nN$ matrix. To have the numerical computation feasible, the matrix $M$ is next $LU$-decomposed, and the spectral curve is then given by
\begin{equation}
P(\nu,j)=\det (M(\nu,j)) = \det (LU) =\det (L)\det (U)=\det (U) \,,
\end{equation}
as $\det (L)=1$ by construction. 
This is now computationally much easier to do as $U$ is an upper triangular matrix whose determinant is simply the product of the diagonal entries. 
To extract the critical points, a two-dimensional Newton-Raphson root finder can be used to search for the value of $(j,\nu)$ satisfying the two conditions in eq.~(\ref{eq:crit_Points}). 
Doing so requires a definition of the derivative of the numerical determinant $P$, and this can be done e.g. with finite differences:
\begin{equation}
\frac{\partial P }{\partial \nu}= \frac{\det (U(\nu+\delta,j))-\det (U(\nu,j))}{\delta} \,,
\end{equation}
where $\delta$ is a small number. For the results presented in~\cite{Cartwright:2021qpp} and discussed here, $\delta=10^{-15}$.

\section{Null coordinates}
\label{sec:Null_coordinates}
\subsection{Infalling Eddington-Finkelstein coordinates}
The calculation of the QNM displayed in section~\ref{sec:QNMresults} can be done in any coordinate system. 
However, a particularly useful coordinate system to choose is one based on a null slicing of the spacetime in which the coordinate induced singularity at the outer horizon radius is removed. 
The infalling Eddington–Finkelstein like coordinates used in the calculations are defined as follows
\begin{align}\label{eq:frametransef}
    d\hat{t} &= dt + \frac{\sqrt{\beta(r)}}{G(r)} dr \, ,\\
    d\hat{\psi}&= d\psi + \frac{2 h(r) \sqrt{\beta(r)}}{G(r)} dr\, ,\\
    d\hat{r} &= d r \, ,
\end{align}
where 
\begin{align}
    \beta(r) &= 1 - \frac{a^2 r_+^4\left(r_+^2+L^2\right)}{r^4 \left( a^2 r_+^2+a^2L^2-r_+^2L^2 \right)}\, ,\\
    h(r) &= \frac{a r_+^4 \left(r_+^2+L^2\right)}{a^2 \left(r_+^2+L^2\right) \left(r_+^4-r^4\right)+r^4 r_+^2L^2} \, .
\end{align}
For convenience, we will suppress the hat on $r$ unless necessary. Making this transformation of coordinates induces a frame transformation
\begin{align}
    \hat{\sigma}^3&= \sigma^3 + \frac{2 h(r) \sqrt{\beta(r)}}{G(r)} dr \, .
\end{align}
Expressed in terms of the null coordinates and the new frame, the metric given in eq.~(\ref{eq:Simple_Spin_MP_Metric}) now takes the form
\begin{align}
   ds^2 &= -\frac{G(r)}{\beta(r)} {d\hat{t}}^2 + \frac{d\hat{t}dr }{\sqrt{\beta(r)}} + \frac{r^2}{4} \left( (\sigma^1)^2 + (\sigma^2)^2  \right) \\ 
   &+ \frac{r^2}{4} \beta(r) \left(\hat\sigma^3 - 2h(r)d\hat{t}\right)^2 \nonumber
\end{align}
From eq.~\ref{eq:frametransef} one can construct the dual vectors as the following
\begin{align}\label{eq:partialtpartialthat}
    \frac\partial{\partial \hat{t}} &= \frac\partial{\partial t}\\
    \frac\partial{\partial \hat{\psi}}&= \frac\partial{\partial \psi}\nonumber\\
    \frac\partial{\partial \hat{r}} &= \frac\partial{\partial r} - \frac{\sqrt{\beta(r)}}{G(r)} \left(\frac\partial{\partial t} + 2 h(r) \frac\partial{\partial \psi} \right)\nonumber
\end{align}
Hence the notation $d\hat t$ and $d\hat\psi$ is justified because their associated vectors' Lie brackets still vanish. Furthermore, the commutation relations in eq.~\eqref{eq:WLalgebra} are preserved.

\subsection{Kruskal coordinates}
A further set of null coordinates is necessary for the study of the gravitational shockwaves in section~\ref{sec:pole_skipping}, the analogue of Kruskal coordinates. We review the transformation to this coordinate system, the full details can be found in~\cite{Amano:2022mlu}. We begin moving to the co-rotating angular coordinate $\tilde{\psi}$ and the associated one form $\tilde{\sigma}_3$ by
\begin{equation}
    \tilde{\psi} = \psi - \overline{\Omega} t, \hspace{2cm}  \tilde{\sigma}^3 = d\tilde{\psi} + \mathrm{cos}(\theta)d\phi,
    \label{eq:co-rot}
\end{equation}
with $\overline{\Omega}=-2\widetilde{\Omega}$ for $\widetilde{\Omega}$ defined in eq. (\ref{eq:thermodynamic_ang_vel_murata}). In the co-rotating coordinates $(t, r, \tilde{\psi}, \phi, \theta)$ the radial and temporal coordinates $(t,r)$ take the form
\begin{equation}
ds_{(t,r)}^2 = - F(r) dt^2 + \frac{dr^2}{G(r)},
\label{metricrt}
\end{equation}
where $F(r)$ is given as 
\begin{equation}\label{eq:new_blackening}
F(r) = 1 + \frac{r^2}{L^2}  - \frac{r_+^2}{r^2}\bigg(1 + \frac{r_+^2}{L^2}\bigg) - \frac{a^2 \Sigma^2 (r^4 -r_+^4)}{L^4 r^2 r_+^4} \, ,
\end{equation}
which vanishes at $r=r_+$ as expected in co-rotating coordinates. In eq. (\ref{eq:new_blackening}) we have introduced a new parameter $\Sigma$ defined as
\begin{equation}
\Sigma = L^2 + r_+^2\, .
\end{equation}
As is familiar we can introduce the infalling ($v$) and outgoing ($u$) Eddington=Finkelstein coordinates by
\begin{subequations}
\begin{align}
    \label{ingoing}
    dv = dt + \frac{dr}{\sqrt{F(r)G(r)}}, \\
    du = dt - \frac{dr}{\sqrt{F(r)G(r)}}. 
\end{align}   
\end{subequations}
Finally, define the Kruskal-like coordinates, $U$ and $V$
\begin{equation}
\label{kruskdef}
    U = -e^{-\alpha u}, \hspace{2cm} V = e^{\alpha v}, 
\end{equation}
where $\alpha = 2 \pi T$, for which the metric is then given by
\begin{equation}
    ds^2  = A(UV)  dUdV + B(U V)(UdV - VdU)\tilde{\sigma}^3 + \frac{r^2}{4}((\sigma^1)^2 + (\sigma^2)^2 +(\tilde{\sigma}^3)^2) + \frac{\mu a^2}{2 r^2}(\tilde{\sigma}^3)^2 \, .
    \label{Kruskal}
\end{equation}
The function $r = r(UV)$ is defined implicitly through \eqref{kruskdef} and 
\begin{equation}
A(UV) =  \frac{F(r)}{\alpha^2 UV}, \;\;\;\;\;\;\;\;  B(U V) = \frac{a}{2 \alpha U V} \frac{\Sigma(r_+^4 - r^4)}{L^2 r^2 r_+^2}. 
\end{equation}
The metric \eqref{Kruskal} is manifestly smooth at both the $U=0$ and $V=0$ horizons at which $r(0)=r_+$.

\section{Hydrodynamic dispersion relations and conventions}
\label{sec:hydroDispersion}
In this appendix, the hydrodynamic dispersion relation conventions, expected values for transport coefficients and expected scaling powers are presented in the example of the momentum diffusion channel. 
At vanishing rotation, in the momentum diffusion channel the lowest of the eigenmodes is the hydrodynamic diffusion mode with the dispersion 
\begin{equation}
    \omega = -i\mathcal{D} q^2 \, ,
\end{equation}
where the frequency $\omega$ and momentum $q$ are chosen to agree with the definitions from~\cite{Kovtun:2005ev}. In this case, $\mathcal{N}=4$ SYM theory has been shown to have $\mathcal{D}=1/(4\pi T)$~\cite{Policastro:2002se,Kovtun:2004de}. This diffusion coefficient is furthermore related to the shear viscosity over entropy ratio as follows (via an Einstein relation): $\mathcal{D}=\eta/(\epsilon+P) =\frac{1}{T} \frac{\eta}{s} = \frac{1}{T} \frac{1}{4\pi}$, where we used that $\mathcal{N}=4$ SYM has $\eta/s=1/(4\pi)$ and that $\epsilon+P=s T$. 

In this article, we consider different frequency $\nu=\omega/(2\pi T\ell)$ and momentum $j=q/(2\pi T \ell)$. For the rest of this appendix, we assume $\ell=1$, as is done in the main text unless stated otherwise. We also consider rotating states, which will modify the dispersion of the diffusion mode, so for the longitudinal diffusion mode we write
\begin{equation}
    \nu = v_{||} j - i \mathcal{D}_{||} j^2 +\mathcal{O}(j^3) \, , 
\end{equation}
with the longitudinal speed $v_{||}$ of this former diffusion mode and its modified diffusion coefficient $\mathcal{D}_{||}$. The temperature at vanishing angular momentum $a=0$ takes the value $T=r_+/\pi$. Therefore, at $a=0$, we expect 
\begin{equation}
    \mathcal{D}_{||}(a=0) = 2\pi T \mathcal{D} = \frac{1}{2} \, .
\end{equation}

Let us rewrite the dispersion relation in different frequency and momentum variables: 
\begin{eqnarray}
    \omega =& -i\mathcal{D} q^2 \, , \\
   \Leftrightarrow 2\pi T \nu  =& -i\mathcal{D} (2\pi T j )^2 \, , \\
   \Leftrightarrow \nu =& -i\underbrace{\mathcal{D} (2\pi T)}_{\mathcal{D}_{||}} j^2 \, .
\end{eqnarray}
Now, let us write the dispersion relation in terms of the variables $\omega$ and $\mathcal{J}$, which appear in the ansatz \eqref{eq:pertsimplygeneric} for the metric fluctuations:
\begin{eqnarray}
    \omega =& -i D \mathcal{J}^2 \, , \\
  \Leftrightarrow  \omega =& -i D (\pi T j)^2 \, , \\
  \Leftrightarrow  \omega =& -i D (\pi T)^2 (\frac{q}{2\pi T})^2 
  = -i \underbrace{\frac{D}{4}}_{\mathcal{D}} q^2 \, , 
\end{eqnarray}
such that we expect $D=1/(\pi T) = 1/r_+$. 

Moving on to the case of $a\neq 0$, we write 
\begin{eqnarray} \label{eq:diffusionDispersionForFit}
    \omega =& v \mathcal{J}-i D \mathcal{J}^2 \, , \\
  \Leftrightarrow  \omega =& -i D (\pi T j)^2 \, , \\
  \Leftrightarrow  \omega =& -i D (\pi T)^2 (\frac{q}{2\pi T})^2 
  = -i \underbrace{\frac{D}{4}}_{\mathcal{D}} q^2 \, , 
\end{eqnarray}
where $D$ and $v$ now depend on the angular momentum parameter $a$. In the large BH limit $r_+\to\infty$ (more precisely~\eqref{eq:LBH1} and~\eqref{eq:LBH2}), we expect $D(a)=2\frac{\mathcal{D}_{||}}{\pi T}=\frac{1}{r_+}(1-a^2)^{3/2}$ and $v=2 v_{||} = 2a$~\cite{Garbiso:2020puw}, which are the values expected for a fluid boosted with boost parameter $a$~\cite{Kovtun:2019hdm, Hoult:2020eho}. Thus, for example, the values given in Table~\ref{tab:expectedDv} are expected in this limit. 
This form~\eqref{eq:diffusionDispersionForFit} in terms of $\omega$ and $\mathcal{J}$ and allowing general exponents, $\omega = v \mathcal{J}^\beta-i D \mathcal{J}^\alpha$, was used to fit $D$, $v$, $\alpha$ and $\beta$ and the result of that fit is displayed in Fig.~\ref{fig:horizonRadiusDependence}. 
\begin{table}[]
    \centering
 \begin{tabular}{c|c|c}
 $a$   & $v(a)$ & $r_+\, D(a)$  \\
\hline
  0  & 0  &  1\\
  0.5 & 1  & 0.64952\\
  0.9  & 1.8  &  0.08282
\end{tabular}
    \caption{Example values expected for the diffusion coefficient $D$ (multiplied by the horizon radius $r_+$) and the velocity $v$ in the large BH limit $r_+\to\infty$ (more precisely~\eqref{eq:LBH1} and~\eqref{eq:LBH2}) at distinct values of angular momentum $a$.}
    \label{tab:expectedDv}
\end{table}

\section{Limits of the Wigner D function}
\label{sec:WignerD}
An important consideration which has not yet been discussed is the relation of the Wigner d functions to Fourier transformations. To begin, we consider the decomposition of the Wigner-$D$ functions in terms of the Wigner-$d$ functions~\footnote{In this appendix we drop the notation $(\mathcal{J},\mathcal{K},\mathcal{M})$ in favor of $(J,m,k)$}. 
\begin{equation}\label{eq:wigner_to_little_d}
    \tensor{D}{^J_m_k}(\theta,\phi,\psi)=e^{-i(k\phi+m\psi)} \tensor{d}{^J_m_k}(\theta)
\end{equation}
The explicit form of the Wigner-$d$ functions can be written using the Wigner formula~\cite{ Schwinger:1952,Tajima:2015owa},
\begin{equation}
    \tensor{d}{^J_m_k}(\theta)=\sum_n (-1)^nw^{Jmk}_n \cos(\theta/2)^{2J+k-m-2n}(-\sin(\theta/2))^{m-k-2n}
\end{equation}
where the coefficients $w$ can be expressed in terms of a complicated fraction of factorials of its indices. Although this expression is complicated, an important observation is that one can reduce this formula for each value of the indices by repeated operation of the power reduction trigonometric identities to $\cos(v\theta)$ ($\sin(\nu \theta)$), such that,
\begin{equation}
     \tensor{d}{^J_m_k}(\theta)=\sum_{\nu=\nu_{min}}^{\nu_{max}=J} t^{Jmk}_\nu f(\nu\theta) \, ,
\end{equation}
where,
\begin{equation}
    f=\begin{cases}
    \cos, & m-k\hspace{0.2 cm} \text{even} \\
    \sin, & m-k \hspace{0.2 cm} \text{odd} \\
    \end{cases}
\end{equation}
Hence we see that expansion for the $d$ function is in terms of $\cos$ if $J-k$ is even and $\sin$ if $J-k$ is odd.
Recalling the large BH limit studied in~\cite{Garbiso:2020puw,Cartwright:2021qpp,Amano:2022mlu} in eq. (\ref{eq:LBH1}) and eq. (\ref{eq:LBH2}) (repeated here for ease of presentation)
\begin{equation}\label{eq:large_Temp_limit} 
    J=jr_{+}/L, \quad \omega=2\nu r_{+}/L, \quad r\rightarrow \alpha r\quad  r_+\rightarrow \alpha r_+
\end{equation}
around $\alpha\rightarrow\infty$ along with the fact that sectors are distinguished by choices of $m$ in terms of J, in the large BH limit $J\approx m$. In the actual calculation all $k$ dependence of the field equations drops out and hence we are left with a freedom. Here we fix this freedom by assuming $k$ even and $J\in\mathbb{Z}$ and hence,
\begin{equation}
     \tensor{d}{^J_m_k}(\theta)=\sum_{\mu=0}^{\mu_{max}=J} t^{Jmk}_\mu \cos(\mu\theta) \label{eq:wigner_cos}
\end{equation}
From orthogonality,
\begin{equation}
    t^{Jmk}_\mu=\frac{1}{2\pi (1+\delta_{\mu 0})}\int_0^{4\pi}\exd\theta d^{J}_{mk}(\theta)f(\mu\theta), \qquad \delta_{\mu 0}=\begin{cases}1 & \text{for}\hspace{.1 cm} \mu=0\\ 0 & \text{else} \end{cases}
\end{equation}
one sees that $t^{Jmk}_{\mu}=t^{Jmk}_{-\mu}$ and hence one can rewrite eq.~(\ref{eq:wigner_cos}) as,
\begin{equation}
     \tensor{d}{^J_m_k}(\theta)=\frac{1}{2}\sum_{\mu=-J}^{J} t^{Jmk}_\mu e^{I\mu\theta}
\end{equation}
Therefore, the Wigner $d$ functions can be written as Fourier sine and cosine series~\cite{Tajima:2015owa}.

Now, let us diverge from~\cite{Tajima:2015owa} and investigate briefly what happens to this Fourier representation in the large BH limit, keeping mind that in this limit $J\approx m$. We have the following expression,
\begin{equation}
    \tensor{d}{^{\frac{j \alpha r_+}{L}}_{\frac{j  \alpha r_+}{L}}_k}(\theta)=\frac{1}{2}\sum_{\mu=-\frac{j  \alpha r_+}{L}}^{\frac{j  \alpha r_+}{L}} \frac{\Delta \mu}{2\pi}\frac{2\pi}{\Delta \mu}t^{\frac{j  \alpha r_+}{L}\frac{j  \alpha r_+}{L}k}_\mu  e^{I\mu\theta}
\end{equation}
where we have cleverly inserted a factor of $1$ in the summand. If we formally extend the range of $\theta$ to $(-\infty,\infty)$ and define,
\begin{equation}
 T_k(\mu)=\lim_{\Delta\mu\rightarrow 0}    \lim_{\alpha\rightarrow \infty}  \frac{2\pi}{\Delta \mu}t^{\frac{j \alpha r_+}{L}\frac{j \alpha r_+}{L}k}_\mu 
\end{equation}
while using the definition from calculus that,
\begin{equation}
  \lim_{\Delta\mu\rightarrow 0}    \lim_{\alpha\rightarrow \infty}   \sum_{\mu=-\frac{j \alpha r_+}{L}}^{\frac{j \alpha r_+}{L}} \frac{\Delta \mu}{2\pi}=\frac{1}{2\pi}\int_{-\infty}^{\infty}\exd \mu
\end{equation}
we find,
\begin{equation}
 \lim_{\Delta\mu\rightarrow 0}    \lim_{\alpha\rightarrow \infty}   \tensor{d}{^{\frac{j \alpha r_+}{L}}_{\frac{j \alpha r_+}{L}}_k}(\theta) = \frac{1}{4\pi}\int^{\infty}_{-\infty}\exd \mu T_k(\mu) e^{I\mu \theta}
\end{equation}
and hence the Wigner $d$ function can be interpreted as a Fourier transformation. In these equations $\theta$ should now be thought of as a coordinate along the real line $\mathbb{R}$.



\bibliographystyle{apsrev4-2}
\bibliography{bib.bib}

\begin{thebibliography}{103}%
\makeatletter
\providecommand \@ifxundefined [1]{%
 \@ifx{#1\undefined}
}%
\providecommand \@ifnum [1]{%
 \ifnum #1\expandafter \@firstoftwo
 \else \expandafter \@secondoftwo
 \fi
}%
\providecommand \@ifx [1]{%
 \ifx #1\expandafter \@firstoftwo
 \else \expandafter \@secondoftwo
 \fi
}%
\providecommand \natexlab [1]{#1}%
\providecommand \enquote  [1]{``#1''}%
\providecommand \bibnamefont  [1]{#1}%
\providecommand \bibfnamefont [1]{#1}%
\providecommand \citenamefont [1]{#1}%
\providecommand \href@noop [0]{\@secondoftwo}%
\providecommand \href [0]{\begingroup \@sanitize@url \@href}%
\providecommand \@href[1]{\@@startlink{#1}\@@href}%
\providecommand \@@href[1]{\endgroup#1\@@endlink}%
\providecommand \@sanitize@url [0]{\catcode `\\12\catcode `\$12\catcode
  `\&12\catcode `\#12\catcode `\^12\catcode `\_12\catcode `\%12\relax}%
\providecommand \@@startlink[1]{}%
\providecommand \@@endlink[0]{}%
\providecommand \url  [0]{\begingroup\@sanitize@url \@url }%
\providecommand \@url [1]{\endgroup\@href {#1}{\urlprefix }}%
\providecommand \urlprefix  [0]{URL }%
\providecommand \Eprint [0]{\href }%
\providecommand \doibase [0]{https://doi.org/}%
\providecommand \selectlanguage [0]{\@gobble}%
\providecommand \bibinfo  [0]{\@secondoftwo}%
\providecommand \bibfield  [0]{\@secondoftwo}%
\providecommand \translation [1]{[#1]}%
\providecommand \BibitemOpen [0]{}%
\providecommand \bibitemStop [0]{}%
\providecommand \bibitemNoStop [0]{.\EOS\space}%
\providecommand \EOS [0]{\spacefactor3000\relax}%
\providecommand \BibitemShut  [1]{\csname bibitem#1\endcsname}%
\let\auto@bib@innerbib\@empty
\bibitem [{\citenamefont {Adamczyk}\ \emph {et~al.}(2017)\citenamefont
  {Adamczyk} \emph {et~al.}}]{STAR:2017ckg}%
  \BibitemOpen
  \bibfield  {author} {\bibinfo {author} {\bibfnamefont {L.}~\bibnamefont
  {Adamczyk}} \emph {et~al.} (\bibinfo {collaboration} {STAR}),\ }\href
  {https://doi.org/10.1038/nature23004} {\bibfield  {journal} {\bibinfo
  {journal} {Nature}\ }\textbf {\bibinfo {volume} {548}},\ \bibinfo {pages}
  {62} (\bibinfo {year} {2017})},\ \Eprint {https://arxiv.org/abs/1701.06657}
  {arXiv:1701.06657 [nucl-ex]} \BibitemShut {NoStop}%
\bibitem [{\citenamefont {Abbott}\ \emph {et~al.}(2016)\citenamefont {Abbott}
  \emph {et~al.}}]{LIGOScientific:2016aoc}%
  \BibitemOpen
  \bibfield  {author} {\bibinfo {author} {\bibfnamefont {B.~P.}\ \bibnamefont
  {Abbott}} \emph {et~al.} (\bibinfo {collaboration} {LIGO Scientific,
  Virgo}),\ }\href {https://doi.org/10.1103/PhysRevLett.116.061102} {\bibfield
  {journal} {\bibinfo  {journal} {Phys. Rev. Lett.}\ }\textbf {\bibinfo
  {volume} {116}},\ \bibinfo {pages} {061102} (\bibinfo {year} {2016})},\
  \Eprint {https://arxiv.org/abs/1602.03837} {arXiv:1602.03837 [gr-qc]}
  \BibitemShut {NoStop}%
\bibitem [{\citenamefont {Bailes}\ \emph {et~al.}(2021)\citenamefont {Bailes}
  \emph {et~al.}}]{Bailes:2021tot}%
  \BibitemOpen
  \bibfield  {author} {\bibinfo {author} {\bibfnamefont {M.}~\bibnamefont
  {Bailes}} \emph {et~al.},\ }\href
  {https://doi.org/10.1038/s42254-021-00303-8} {\bibfield  {journal} {\bibinfo
  {journal} {Nature Rev. Phys.}\ }\textbf {\bibinfo {volume} {3}},\ \bibinfo
  {pages} {344} (\bibinfo {year} {2021})}\BibitemShut {NoStop}%
\bibitem [{\citenamefont {Karpenko}\ and\ \citenamefont
  {Becattini}(2017)}]{Karpenko:2016jyx}%
  \BibitemOpen
  \bibfield  {author} {\bibinfo {author} {\bibfnamefont {I.}~\bibnamefont
  {Karpenko}}\ and\ \bibinfo {author} {\bibfnamefont {F.}~\bibnamefont
  {Becattini}},\ }\href {https://doi.org/10.1140/epjc/s10052-017-4765-1}
  {\bibfield  {journal} {\bibinfo  {journal} {Eur. Phys. J. C}\ }\textbf
  {\bibinfo {volume} {77}},\ \bibinfo {pages} {213} (\bibinfo {year} {2017})},\
  \Eprint {https://arxiv.org/abs/1610.04717} {arXiv:1610.04717 [nucl-th]}
  \BibitemShut {NoStop}%
\bibitem [{\citenamefont {Becattini}\ and\ \citenamefont
  {Karpenko}(2018)}]{Becattini:2017gcx}%
  \BibitemOpen
  \bibfield  {author} {\bibinfo {author} {\bibfnamefont {F.}~\bibnamefont
  {Becattini}}\ and\ \bibinfo {author} {\bibfnamefont {I.}~\bibnamefont
  {Karpenko}},\ }\href {https://doi.org/10.1103/PhysRevLett.120.012302}
  {\bibfield  {journal} {\bibinfo  {journal} {Phys. Rev. Lett.}\ }\textbf
  {\bibinfo {volume} {120}},\ \bibinfo {pages} {012302} (\bibinfo {year}
  {2018})},\ \Eprint {https://arxiv.org/abs/1707.07984} {arXiv:1707.07984
  [nucl-th]} \BibitemShut {NoStop}%
\bibitem [{\citenamefont {Florkowski}\ \emph {et~al.}(2018)\citenamefont
  {Florkowski}, \citenamefont {Friman}, \citenamefont {Jaiswal},\ and\
  \citenamefont {Speranza}}]{Florkowski:2017ruc}%
  \BibitemOpen
  \bibfield  {author} {\bibinfo {author} {\bibfnamefont {W.}~\bibnamefont
  {Florkowski}}, \bibinfo {author} {\bibfnamefont {B.}~\bibnamefont {Friman}},
  \bibinfo {author} {\bibfnamefont {A.}~\bibnamefont {Jaiswal}},\ and\ \bibinfo
  {author} {\bibfnamefont {E.}~\bibnamefont {Speranza}},\ }\href
  {https://doi.org/10.1103/PhysRevC.97.041901} {\bibfield  {journal} {\bibinfo
  {journal} {Phys. Rev. C}\ }\textbf {\bibinfo {volume} {97}},\ \bibinfo
  {pages} {041901} (\bibinfo {year} {2018})},\ \Eprint
  {https://arxiv.org/abs/1705.00587} {arXiv:1705.00587 [nucl-th]} \BibitemShut
  {NoStop}%
\bibitem [{\citenamefont {Montenegro}\ and\ \citenamefont
  {Torrieri}(2020)}]{Montenegro:2020paq}%
  \BibitemOpen
  \bibfield  {author} {\bibinfo {author} {\bibfnamefont {D.}~\bibnamefont
  {Montenegro}}\ and\ \bibinfo {author} {\bibfnamefont {G.}~\bibnamefont
  {Torrieri}},\ }\href {https://doi.org/10.1103/PhysRevD.102.036007} {\bibfield
   {journal} {\bibinfo  {journal} {Phys. Rev. D}\ }\textbf {\bibinfo {volume}
  {102}},\ \bibinfo {pages} {036007} (\bibinfo {year} {2020})},\ \Eprint
  {https://arxiv.org/abs/2004.10195} {arXiv:2004.10195 [hep-th]} \BibitemShut
  {NoStop}%
\bibitem [{\citenamefont {Bhadury}\ \emph {et~al.}(2021)\citenamefont
  {Bhadury}, \citenamefont {Florkowski}, \citenamefont {Jaiswal}, \citenamefont
  {Kumar},\ and\ \citenamefont {Ryblewski}}]{Bhadury:2020cop}%
  \BibitemOpen
  \bibfield  {author} {\bibinfo {author} {\bibfnamefont {S.}~\bibnamefont
  {Bhadury}}, \bibinfo {author} {\bibfnamefont {W.}~\bibnamefont {Florkowski}},
  \bibinfo {author} {\bibfnamefont {A.}~\bibnamefont {Jaiswal}}, \bibinfo
  {author} {\bibfnamefont {A.}~\bibnamefont {Kumar}},\ and\ \bibinfo {author}
  {\bibfnamefont {R.}~\bibnamefont {Ryblewski}},\ }\href
  {https://doi.org/10.1103/PhysRevD.103.014030} {\bibfield  {journal} {\bibinfo
   {journal} {Phys. Rev. D}\ }\textbf {\bibinfo {volume} {103}},\ \bibinfo
  {pages} {014030} (\bibinfo {year} {2021})},\ \Eprint
  {https://arxiv.org/abs/2008.10976} {arXiv:2008.10976 [nucl-th]} \BibitemShut
  {NoStop}%
\bibitem [{\citenamefont {Weickgenannt}\ \emph {et~al.}(2021)\citenamefont
  {Weickgenannt}, \citenamefont {Speranza}, \citenamefont {Sheng},
  \citenamefont {Wang},\ and\ \citenamefont {Rischke}}]{Weickgenannt:2020aaf}%
  \BibitemOpen
  \bibfield  {author} {\bibinfo {author} {\bibfnamefont {N.}~\bibnamefont
  {Weickgenannt}}, \bibinfo {author} {\bibfnamefont {E.}~\bibnamefont
  {Speranza}}, \bibinfo {author} {\bibfnamefont {X.-l.}\ \bibnamefont {Sheng}},
  \bibinfo {author} {\bibfnamefont {Q.}~\bibnamefont {Wang}},\ and\ \bibinfo
  {author} {\bibfnamefont {D.~H.}\ \bibnamefont {Rischke}},\ }\href
  {https://doi.org/10.1103/PhysRevLett.127.052301} {\bibfield  {journal}
  {\bibinfo  {journal} {Phys. Rev. Lett.}\ }\textbf {\bibinfo {volume} {127}},\
  \bibinfo {pages} {052301} (\bibinfo {year} {2021})},\ \Eprint
  {https://arxiv.org/abs/2005.01506} {arXiv:2005.01506 [hep-ph]} \BibitemShut
  {NoStop}%
\bibitem [{\citenamefont {Fukushima}\ and\ \citenamefont
  {Pu}(2021)}]{Fukushima:2020ucl}%
  \BibitemOpen
  \bibfield  {author} {\bibinfo {author} {\bibfnamefont {K.}~\bibnamefont
  {Fukushima}}\ and\ \bibinfo {author} {\bibfnamefont {S.}~\bibnamefont {Pu}},\
  }\href {https://doi.org/10.1016/j.physletb.2021.136346} {\bibfield  {journal}
  {\bibinfo  {journal} {Phys. Lett. B}\ }\textbf {\bibinfo {volume} {817}},\
  \bibinfo {pages} {136346} (\bibinfo {year} {2021})},\ \Eprint
  {https://arxiv.org/abs/2010.01608} {arXiv:2010.01608 [hep-th]} \BibitemShut
  {NoStop}%
\bibitem [{\citenamefont {Li}\ \emph {et~al.}(2021)\citenamefont {Li},
  \citenamefont {Stephanov},\ and\ \citenamefont {Yee}}]{Li:2020eon}%
  \BibitemOpen
  \bibfield  {author} {\bibinfo {author} {\bibfnamefont {S.}~\bibnamefont
  {Li}}, \bibinfo {author} {\bibfnamefont {M.~A.}\ \bibnamefont {Stephanov}},\
  and\ \bibinfo {author} {\bibfnamefont {H.-U.}\ \bibnamefont {Yee}},\ }\href
  {https://doi.org/10.1103/PhysRevLett.127.082302} {\bibfield  {journal}
  {\bibinfo  {journal} {Phys. Rev. Lett.}\ }\textbf {\bibinfo {volume} {127}},\
  \bibinfo {pages} {082302} (\bibinfo {year} {2021})},\ \Eprint
  {https://arxiv.org/abs/2011.12318} {arXiv:2011.12318 [hep-th]} \BibitemShut
  {NoStop}%
\bibitem [{\citenamefont {Becattini}\ \emph {et~al.}(2021)\citenamefont
  {Becattini}, \citenamefont {Buzzegoli}, \citenamefont {Inghirami},
  \citenamefont {Karpenko},\ and\ \citenamefont {Palermo}}]{Becattini:2021iol}%
  \BibitemOpen
  \bibfield  {author} {\bibinfo {author} {\bibfnamefont {F.}~\bibnamefont
  {Becattini}}, \bibinfo {author} {\bibfnamefont {M.}~\bibnamefont
  {Buzzegoli}}, \bibinfo {author} {\bibfnamefont {G.}~\bibnamefont
  {Inghirami}}, \bibinfo {author} {\bibfnamefont {I.}~\bibnamefont
  {Karpenko}},\ and\ \bibinfo {author} {\bibfnamefont {A.}~\bibnamefont
  {Palermo}},\ }\href {https://doi.org/10.1103/PhysRevLett.127.272302}
  {\bibfield  {journal} {\bibinfo  {journal} {Phys. Rev. Lett.}\ }\textbf
  {\bibinfo {volume} {127}},\ \bibinfo {pages} {272302} (\bibinfo {year}
  {2021})},\ \Eprint {https://arxiv.org/abs/2103.14621} {arXiv:2103.14621
  [nucl-th]} \BibitemShut {NoStop}%
\bibitem [{\citenamefont {Gallegos}\ \emph {et~al.}(2021)\citenamefont
  {Gallegos}, \citenamefont {G\"ursoy},\ and\ \citenamefont
  {Yarom}}]{Gallegos:2021bzp}%
  \BibitemOpen
  \bibfield  {author} {\bibinfo {author} {\bibfnamefont {A.~D.}\ \bibnamefont
  {Gallegos}}, \bibinfo {author} {\bibfnamefont {U.}~\bibnamefont {G\"ursoy}},\
  and\ \bibinfo {author} {\bibfnamefont {A.}~\bibnamefont {Yarom}},\ }\href
  {https://doi.org/10.21468/SciPostPhys.11.2.041} {\bibfield  {journal}
  {\bibinfo  {journal} {SciPost Phys.}\ }\textbf {\bibinfo {volume} {11}},\
  \bibinfo {pages} {041} (\bibinfo {year} {2021})},\ \Eprint
  {https://arxiv.org/abs/2101.04759} {arXiv:2101.04759 [hep-th]} \BibitemShut
  {NoStop}%
\bibitem [{\citenamefont {Hongo}\ \emph {et~al.}(2021)\citenamefont {Hongo},
  \citenamefont {Huang}, \citenamefont {Kaminski}, \citenamefont {Stephanov},\
  and\ \citenamefont {Yee}}]{Hongo:2021ona}%
  \BibitemOpen
  \bibfield  {author} {\bibinfo {author} {\bibfnamefont {M.}~\bibnamefont
  {Hongo}}, \bibinfo {author} {\bibfnamefont {X.-G.}\ \bibnamefont {Huang}},
  \bibinfo {author} {\bibfnamefont {M.}~\bibnamefont {Kaminski}}, \bibinfo
  {author} {\bibfnamefont {M.}~\bibnamefont {Stephanov}},\ and\ \bibinfo
  {author} {\bibfnamefont {H.-U.}\ \bibnamefont {Yee}},\ }\href
  {https://doi.org/10.1007/JHEP11(2021)150} {\bibfield  {journal} {\bibinfo
  {journal} {JHEP}\ }\textbf {\bibinfo {volume} {11}},\ \bibinfo {pages}
  {150}},\ \Eprint {https://arxiv.org/abs/2107.14231} {arXiv:2107.14231
  [hep-th]} \BibitemShut {NoStop}%
\bibitem [{\citenamefont {Gallegos}\ \emph {et~al.}(2023)\citenamefont
  {Gallegos}, \citenamefont {Gursoy},\ and\ \citenamefont
  {Yarom}}]{Gallegos:2022jow}%
  \BibitemOpen
  \bibfield  {author} {\bibinfo {author} {\bibfnamefont {A.~D.}\ \bibnamefont
  {Gallegos}}, \bibinfo {author} {\bibfnamefont {U.}~\bibnamefont {Gursoy}},\
  and\ \bibinfo {author} {\bibfnamefont {A.}~\bibnamefont {Yarom}},\ }\href
  {https://doi.org/10.1007/JHEP05(2023)139} {\bibfield  {journal} {\bibinfo
  {journal} {JHEP}\ }\textbf {\bibinfo {volume} {05}},\ \bibinfo {pages}
  {139}},\ \Eprint {https://arxiv.org/abs/2203.05044} {arXiv:2203.05044
  [hep-th]} \BibitemShut {NoStop}%
\bibitem [{\citenamefont {Wang}\ \emph {et~al.}(2022)\citenamefont {Wang},
  \citenamefont {Xie}, \citenamefont {Fang},\ and\ \citenamefont
  {Pu}}]{Wang:2021wqq}%
  \BibitemOpen
  \bibfield  {author} {\bibinfo {author} {\bibfnamefont {D.-L.}\ \bibnamefont
  {Wang}}, \bibinfo {author} {\bibfnamefont {X.-Q.}\ \bibnamefont {Xie}},
  \bibinfo {author} {\bibfnamefont {S.}~\bibnamefont {Fang}},\ and\ \bibinfo
  {author} {\bibfnamefont {S.}~\bibnamefont {Pu}},\ }\href
  {https://doi.org/10.1103/PhysRevD.105.114050} {\bibfield  {journal} {\bibinfo
   {journal} {Phys. Rev. D}\ }\textbf {\bibinfo {volume} {105}},\ \bibinfo
  {pages} {114050} (\bibinfo {year} {2022})},\ \Eprint
  {https://arxiv.org/abs/2112.15535} {arXiv:2112.15535 [hep-ph]} \BibitemShut
  {NoStop}%
\bibitem [{\citenamefont {Weickgenannt}\ \emph
  {et~al.}(2022{\natexlab{a}})\citenamefont {Weickgenannt}, \citenamefont
  {Wagner}, \citenamefont {Speranza},\ and\ \citenamefont
  {Rischke}}]{Weickgenannt:2022zxs}%
  \BibitemOpen
  \bibfield  {author} {\bibinfo {author} {\bibfnamefont {N.}~\bibnamefont
  {Weickgenannt}}, \bibinfo {author} {\bibfnamefont {D.}~\bibnamefont
  {Wagner}}, \bibinfo {author} {\bibfnamefont {E.}~\bibnamefont {Speranza}},\
  and\ \bibinfo {author} {\bibfnamefont {D.~H.}\ \bibnamefont {Rischke}},\
  }\href {https://doi.org/10.1103/PhysRevD.106.096014} {\bibfield  {journal}
  {\bibinfo  {journal} {Phys. Rev. D}\ }\textbf {\bibinfo {volume} {106}},\
  \bibinfo {pages} {096014} (\bibinfo {year} {2022}{\natexlab{a}})},\ \Eprint
  {https://arxiv.org/abs/2203.04766} {arXiv:2203.04766 [nucl-th]} \BibitemShut
  {NoStop}%
\bibitem [{\citenamefont {Weickgenannt}\ \emph
  {et~al.}(2022{\natexlab{b}})\citenamefont {Weickgenannt}, \citenamefont
  {Wagner}, \citenamefont {Speranza},\ and\ \citenamefont
  {Rischke}}]{Weickgenannt:2022qvh}%
  \BibitemOpen
  \bibfield  {author} {\bibinfo {author} {\bibfnamefont {N.}~\bibnamefont
  {Weickgenannt}}, \bibinfo {author} {\bibfnamefont {D.}~\bibnamefont
  {Wagner}}, \bibinfo {author} {\bibfnamefont {E.}~\bibnamefont {Speranza}},\
  and\ \bibinfo {author} {\bibfnamefont {D.~H.}\ \bibnamefont {Rischke}},\
  }\href {https://doi.org/10.1103/PhysRevD.106.L091901} {\bibfield  {journal}
  {\bibinfo  {journal} {Phys. Rev. D}\ }\textbf {\bibinfo {volume} {106}},\
  \bibinfo {pages} {L091901} (\bibinfo {year} {2022}{\natexlab{b}})},\ \Eprint
  {https://arxiv.org/abs/2208.01955} {arXiv:2208.01955 [nucl-th]} \BibitemShut
  {NoStop}%
\bibitem [{\citenamefont {Myers}\ and\ \citenamefont
  {Perry}(1986)}]{Myers:1986un}%
  \BibitemOpen
  \bibfield  {author} {\bibinfo {author} {\bibfnamefont {R.~C.}\ \bibnamefont
  {Myers}}\ and\ \bibinfo {author} {\bibfnamefont {M.}~\bibnamefont {Perry}},\
  }\href {https://doi.org/10.1016/0003-4916(86)90186-7} {\bibfield  {journal}
  {\bibinfo  {journal} {Annals Phys.}\ }\textbf {\bibinfo {volume} {172}},\
  \bibinfo {pages} {304} (\bibinfo {year} {1986})}\BibitemShut {NoStop}%
\bibitem [{\citenamefont {Carter}(1968)}]{Carter:1968ks}%
  \BibitemOpen
  \bibfield  {author} {\bibinfo {author} {\bibfnamefont {B.}~\bibnamefont
  {Carter}},\ }\href {https://doi.org/10.1007/BF03399503} {\bibfield  {journal}
  {\bibinfo  {journal} {Commun. Math. Phys.}\ }\textbf {\bibinfo {volume}
  {10}},\ \bibinfo {pages} {280} (\bibinfo {year} {1968})}\BibitemShut
  {NoStop}%
\bibitem [{\citenamefont {Gibbons}\ \emph {et~al.}(2005)\citenamefont
  {Gibbons}, \citenamefont {Perry},\ and\ \citenamefont
  {Pope}}]{Gibbons:2004ai}%
  \BibitemOpen
  \bibfield  {author} {\bibinfo {author} {\bibfnamefont {G.~W.}\ \bibnamefont
  {Gibbons}}, \bibinfo {author} {\bibfnamefont {M.~J.}\ \bibnamefont {Perry}},\
  and\ \bibinfo {author} {\bibfnamefont {C.~N.}\ \bibnamefont {Pope}},\ }\href
  {https://doi.org/10.1088/0264-9381/22/9/002} {\bibfield  {journal} {\bibinfo
  {journal} {Class. Quant. Grav.}\ }\textbf {\bibinfo {volume} {22}},\ \bibinfo
  {pages} {1503} (\bibinfo {year} {2005})},\ \Eprint
  {https://arxiv.org/abs/hep-th/0408217} {arXiv:hep-th/0408217 [hep-th]}
  \BibitemShut {NoStop}%
\bibitem [{\citenamefont {Gibbons}\ \emph {et~al.}(2004)\citenamefont
  {Gibbons}, \citenamefont {Lu}, \citenamefont {Page},\ and\ \citenamefont
  {Pope}}]{Gibbons:2004js}%
  \BibitemOpen
  \bibfield  {author} {\bibinfo {author} {\bibfnamefont {G.~W.}\ \bibnamefont
  {Gibbons}}, \bibinfo {author} {\bibfnamefont {H.}~\bibnamefont {Lu}},
  \bibinfo {author} {\bibfnamefont {D.~N.}\ \bibnamefont {Page}},\ and\
  \bibinfo {author} {\bibfnamefont {C.~N.}\ \bibnamefont {Pope}},\ }\href
  {https://doi.org/10.1103/PhysRevLett.93.171102} {\bibfield  {journal}
  {\bibinfo  {journal} {Phys. Rev. Lett.}\ }\textbf {\bibinfo {volume} {93}},\
  \bibinfo {pages} {171102} (\bibinfo {year} {2004})},\ \Eprint
  {https://arxiv.org/abs/hep-th/0409155} {arXiv:hep-th/0409155 [hep-th]}
  \BibitemShut {NoStop}%
\bibitem [{\citenamefont {Hawking}\ and\ \citenamefont
  {Reall}(2000)}]{Hawking:1999dp}%
  \BibitemOpen
  \bibfield  {author} {\bibinfo {author} {\bibfnamefont {S.~W.}\ \bibnamefont
  {Hawking}}\ and\ \bibinfo {author} {\bibfnamefont {H.~S.}\ \bibnamefont
  {Reall}},\ }\href {https://doi.org/10.1103/PhysRevD.61.024014} {\bibfield
  {journal} {\bibinfo  {journal} {Phys. Rev.}\ }\textbf {\bibinfo {volume}
  {D61}},\ \bibinfo {pages} {024014} (\bibinfo {year} {2000})},\ \Eprint
  {https://arxiv.org/abs/hep-th/9908109} {arXiv:hep-th/9908109 [hep-th]}
  \BibitemShut {NoStop}%
\bibitem [{\citenamefont {Murata}(2009)}]{Murata:2008xr}%
  \BibitemOpen
  \bibfield  {author} {\bibinfo {author} {\bibfnamefont {K.}~\bibnamefont
  {Murata}},\ }\href {https://doi.org/10.1143/PTP.121.1099} {\bibfield
  {journal} {\bibinfo  {journal} {Prog. Theor. Phys.}\ }\textbf {\bibinfo
  {volume} {121}},\ \bibinfo {pages} {1099} (\bibinfo {year} {2009})},\ \Eprint
  {https://arxiv.org/abs/0812.0718} {arXiv:0812.0718 [hep-th]} \BibitemShut
  {NoStop}%
\bibitem [{\citenamefont {Murata}\ and\ \citenamefont
  {Soda}(2008{\natexlab{a}})}]{Murata:2008yx}%
  \BibitemOpen
  \bibfield  {author} {\bibinfo {author} {\bibfnamefont {K.}~\bibnamefont
  {Murata}}\ and\ \bibinfo {author} {\bibfnamefont {J.}~\bibnamefont {Soda}},\
  }\href {https://doi.org/10.1143/PTP.120.561} {\bibfield  {journal} {\bibinfo
  {journal} {Prog. Theor. Phys.}\ }\textbf {\bibinfo {volume} {120}},\ \bibinfo
  {pages} {561} (\bibinfo {year} {2008}{\natexlab{a}})},\ \Eprint
  {https://arxiv.org/abs/0803.1371} {arXiv:0803.1371 [hep-th]} \BibitemShut
  {NoStop}%
\bibitem [{\citenamefont {Murata}\ and\ \citenamefont
  {Soda}(2008{\natexlab{b}})}]{Murata:2007gv}%
  \BibitemOpen
  \bibfield  {author} {\bibinfo {author} {\bibfnamefont {K.}~\bibnamefont
  {Murata}}\ and\ \bibinfo {author} {\bibfnamefont {J.}~\bibnamefont {Soda}},\
  }\href {https://doi.org/10.1088/0264-9381/25/3/035006} {\bibfield  {journal}
  {\bibinfo  {journal} {Class. Quant. Grav.}\ }\textbf {\bibinfo {volume}
  {25}},\ \bibinfo {pages} {035006} (\bibinfo {year} {2008}{\natexlab{b}})},\
  \Eprint {https://arxiv.org/abs/0710.0221} {arXiv:0710.0221 [hep-th]}
  \BibitemShut {NoStop}%
\bibitem [{\citenamefont {Ishii}\ \emph {et~al.}(2021)\citenamefont {Ishii},
  \citenamefont {Murata}, \citenamefont {Santos},\ and\ \citenamefont
  {Way}}]{Ishii:2021xmn}%
  \BibitemOpen
  \bibfield  {author} {\bibinfo {author} {\bibfnamefont {T.}~\bibnamefont
  {Ishii}}, \bibinfo {author} {\bibfnamefont {K.}~\bibnamefont {Murata}},
  \bibinfo {author} {\bibfnamefont {J.~E.}\ \bibnamefont {Santos}},\ and\
  \bibinfo {author} {\bibfnamefont {B.}~\bibnamefont {Way}},\ }\href
  {https://doi.org/10.1007/JHEP05(2021)011} {\bibfield  {journal} {\bibinfo
  {journal} {JHEP}\ }\textbf {\bibinfo {volume} {05}},\ \bibinfo {pages}
  {011}},\ \Eprint {https://arxiv.org/abs/2101.06325} {arXiv:2101.06325
  [hep-th]} \BibitemShut {NoStop}%
\bibitem [{\citenamefont {Ishii}\ and\ \citenamefont
  {Murata}(2020)}]{Ishii:2019wfs}%
  \BibitemOpen
  \bibfield  {author} {\bibinfo {author} {\bibfnamefont {T.}~\bibnamefont
  {Ishii}}\ and\ \bibinfo {author} {\bibfnamefont {K.}~\bibnamefont {Murata}},\
  }\href {https://doi.org/10.1088/1361-6382/ab7418} {\bibfield  {journal}
  {\bibinfo  {journal} {Class. Quant. Grav.}\ }\textbf {\bibinfo {volume}
  {37}},\ \bibinfo {pages} {075009} (\bibinfo {year} {2020})},\ \Eprint
  {https://arxiv.org/abs/1910.03234} {arXiv:1910.03234 [hep-th]} \BibitemShut
  {NoStop}%
\bibitem [{\citenamefont {Cardoso}\ \emph {et~al.}(2014)\citenamefont
  {Cardoso}, \citenamefont {Dias}, \citenamefont {Hartnett}, \citenamefont
  {Lehner},\ and\ \citenamefont {Santos}}]{Cardoso:2013pza}%
  \BibitemOpen
  \bibfield  {author} {\bibinfo {author} {\bibfnamefont {V.}~\bibnamefont
  {Cardoso}}, \bibinfo {author} {\bibfnamefont {O.~J.~C.}\ \bibnamefont
  {Dias}}, \bibinfo {author} {\bibfnamefont {G.~S.}\ \bibnamefont {Hartnett}},
  \bibinfo {author} {\bibfnamefont {L.}~\bibnamefont {Lehner}},\ and\ \bibinfo
  {author} {\bibfnamefont {J.~E.}\ \bibnamefont {Santos}},\ }\href
  {https://doi.org/10.1007/JHEP04(2014)183} {\bibfield  {journal} {\bibinfo
  {journal} {JHEP}\ }\textbf {\bibinfo {volume} {04}},\ \bibinfo {pages}
  {183}},\ \Eprint {https://arxiv.org/abs/1312.5323} {arXiv:1312.5323 [hep-th]}
  \BibitemShut {NoStop}%
\bibitem [{\citenamefont {Gregory}\ \emph {et~al.}(2013)\citenamefont
  {Gregory}, \citenamefont {Kubiznak},\ and\ \citenamefont
  {Wills}}]{Gregory:2013xca}%
  \BibitemOpen
  \bibfield  {author} {\bibinfo {author} {\bibfnamefont {R.}~\bibnamefont
  {Gregory}}, \bibinfo {author} {\bibfnamefont {D.}~\bibnamefont {Kubiznak}},\
  and\ \bibinfo {author} {\bibfnamefont {D.}~\bibnamefont {Wills}},\ }\href
  {https://doi.org/10.1007/JHEP06(2013)023} {\bibfield  {journal} {\bibinfo
  {journal} {JHEP}\ }\textbf {\bibinfo {volume} {06}},\ \bibinfo {pages}
  {023}},\ \Eprint {https://arxiv.org/abs/1303.0519} {arXiv:1303.0519 [gr-qc]}
  \BibitemShut {NoStop}%
\bibitem [{\citenamefont {Hawking}\ \emph {et~al.}(1999)\citenamefont
  {Hawking}, \citenamefont {Hunter},\ and\ \citenamefont
  {Taylor}}]{Hawking:1998kw}%
  \BibitemOpen
  \bibfield  {author} {\bibinfo {author} {\bibfnamefont {S.~W.}\ \bibnamefont
  {Hawking}}, \bibinfo {author} {\bibfnamefont {C.~J.}\ \bibnamefont
  {Hunter}},\ and\ \bibinfo {author} {\bibfnamefont {M.}~\bibnamefont
  {Taylor}},\ }\href {https://doi.org/10.1103/PhysRevD.59.064005} {\bibfield
  {journal} {\bibinfo  {journal} {Phys. Rev.}\ }\textbf {\bibinfo {volume}
  {D59}},\ \bibinfo {pages} {064005} (\bibinfo {year} {1999})},\ \Eprint
  {https://arxiv.org/abs/hep-th/9811056} {arXiv:hep-th/9811056 [hep-th]}
  \BibitemShut {NoStop}%
\bibitem [{\citenamefont {Edalati}\ \emph
  {et~al.}(2010{\natexlab{a}})\citenamefont {Edalati}, \citenamefont {Jottar},\
  and\ \citenamefont {Leigh}}]{Edalati:2009bi}%
  \BibitemOpen
  \bibfield  {author} {\bibinfo {author} {\bibfnamefont {M.}~\bibnamefont
  {Edalati}}, \bibinfo {author} {\bibfnamefont {J.~I.}\ \bibnamefont
  {Jottar}},\ and\ \bibinfo {author} {\bibfnamefont {R.~G.}\ \bibnamefont
  {Leigh}},\ }\href {https://doi.org/10.1007/JHEP01(2010)018} {\bibfield
  {journal} {\bibinfo  {journal} {JHEP}\ }\textbf {\bibinfo {volume} {01}},\
  \bibinfo {pages} {018}},\ \Eprint {https://arxiv.org/abs/0910.0645}
  {arXiv:0910.0645 [hep-th]} \BibitemShut {NoStop}%
\bibitem [{\citenamefont {Edalati}\ \emph
  {et~al.}(2010{\natexlab{b}})\citenamefont {Edalati}, \citenamefont {Jottar},\
  and\ \citenamefont {Leigh}}]{Edalati:2010hk}%
  \BibitemOpen
  \bibfield  {author} {\bibinfo {author} {\bibfnamefont {M.}~\bibnamefont
  {Edalati}}, \bibinfo {author} {\bibfnamefont {J.~I.}\ \bibnamefont
  {Jottar}},\ and\ \bibinfo {author} {\bibfnamefont {R.~G.}\ \bibnamefont
  {Leigh}},\ }\href {https://doi.org/10.1007/JHEP04(2010)075} {\bibfield
  {journal} {\bibinfo  {journal} {JHEP}\ }\textbf {\bibinfo {volume} {04}},\
  \bibinfo {pages} {075}},\ \Eprint {https://arxiv.org/abs/1001.0779}
  {arXiv:1001.0779 [hep-th]} \BibitemShut {NoStop}%
\bibitem [{\citenamefont {Janiszewski}\ and\ \citenamefont
  {Kaminski}(2016)}]{Janiszewski:2015ura}%
  \BibitemOpen
  \bibfield  {author} {\bibinfo {author} {\bibfnamefont {S.}~\bibnamefont
  {Janiszewski}}\ and\ \bibinfo {author} {\bibfnamefont {M.}~\bibnamefont
  {Kaminski}},\ }\href {https://doi.org/10.1103/PhysRevD.93.025006} {\bibfield
  {journal} {\bibinfo  {journal} {Phys. Rev. D}\ }\textbf {\bibinfo {volume}
  {93}},\ \bibinfo {pages} {025006} (\bibinfo {year} {2016})},\ \Eprint
  {https://arxiv.org/abs/1508.06993} {arXiv:1508.06993 [hep-th]} \BibitemShut
  {NoStop}%
\bibitem [{\citenamefont {Blake}\ \emph
  {et~al.}(2018{\natexlab{a}})\citenamefont {Blake}, \citenamefont {Lee},\ and\
  \citenamefont {Liu}}]{Blake:2017ris}%
  \BibitemOpen
  \bibfield  {author} {\bibinfo {author} {\bibfnamefont {M.}~\bibnamefont
  {Blake}}, \bibinfo {author} {\bibfnamefont {H.}~\bibnamefont {Lee}},\ and\
  \bibinfo {author} {\bibfnamefont {H.}~\bibnamefont {Liu}},\ }\href
  {https://doi.org/10.1007/JHEP10(2018)127} {\bibfield  {journal} {\bibinfo
  {journal} {JHEP}\ }\textbf {\bibinfo {volume} {10}},\ \bibinfo {pages}
  {127}},\ \Eprint {https://arxiv.org/abs/1801.00010} {arXiv:1801.00010
  [hep-th]} \BibitemShut {NoStop}%
\bibitem [{\citenamefont {Blake}\ \emph
  {et~al.}(2018{\natexlab{b}})\citenamefont {Blake}, \citenamefont {Davison},
  \citenamefont {Grozdanov},\ and\ \citenamefont {Liu}}]{Blake:2018leo}%
  \BibitemOpen
  \bibfield  {author} {\bibinfo {author} {\bibfnamefont {M.}~\bibnamefont
  {Blake}}, \bibinfo {author} {\bibfnamefont {R.~A.}\ \bibnamefont {Davison}},
  \bibinfo {author} {\bibfnamefont {S.}~\bibnamefont {Grozdanov}},\ and\
  \bibinfo {author} {\bibfnamefont {H.}~\bibnamefont {Liu}},\ }\href
  {https://doi.org/10.1007/JHEP10(2018)035} {\bibfield  {journal} {\bibinfo
  {journal} {JHEP}\ }\textbf {\bibinfo {volume} {10}},\ \bibinfo {pages}
  {035}},\ \Eprint {https://arxiv.org/abs/1809.01169} {arXiv:1809.01169
  [hep-th]} \BibitemShut {NoStop}%
\bibitem [{\citenamefont {Garbiso}\ and\ \citenamefont
  {Kaminski}(2020)}]{Garbiso:2020puw}%
  \BibitemOpen
  \bibfield  {author} {\bibinfo {author} {\bibfnamefont {M.}~\bibnamefont
  {Garbiso}}\ and\ \bibinfo {author} {\bibfnamefont {M.}~\bibnamefont
  {Kaminski}},\ }\href {https://doi.org/10.1007/JHEP12(2020)112} {\bibfield
  {journal} {\bibinfo  {journal} {JHEP}\ }\textbf {\bibinfo {volume} {12}},\
  \bibinfo {pages} {112}},\ \Eprint {https://arxiv.org/abs/2007.04345}
  {arXiv:2007.04345 [hep-th]} \BibitemShut {NoStop}%
\bibitem [{\citenamefont {Festuccia}\ and\ \citenamefont
  {Liu}(2009)}]{Festuccia:2008zx}%
  \BibitemOpen
  \bibfield  {author} {\bibinfo {author} {\bibfnamefont {G.}~\bibnamefont
  {Festuccia}}\ and\ \bibinfo {author} {\bibfnamefont {H.}~\bibnamefont
  {Liu}},\ }\href {https://doi.org/10.1166/asl.2009.1029} {\bibfield  {journal}
  {\bibinfo  {journal} {Adv. Sci. Lett.}\ }\textbf {\bibinfo {volume} {2}},\
  \bibinfo {pages} {221} (\bibinfo {year} {2009})},\ \Eprint
  {https://arxiv.org/abs/0811.1033} {arXiv:0811.1033 [gr-qc]} \BibitemShut
  {NoStop}%
\bibitem [{\citenamefont {Fuini}\ \emph {et~al.}(2016)\citenamefont {Fuini},
  \citenamefont {Uhlemann},\ and\ \citenamefont {Yaffe}}]{Fuini:2016qsc}%
  \BibitemOpen
  \bibfield  {author} {\bibinfo {author} {\bibfnamefont {J.~F.}\ \bibnamefont
  {Fuini}}, \bibinfo {author} {\bibfnamefont {C.~F.}\ \bibnamefont
  {Uhlemann}},\ and\ \bibinfo {author} {\bibfnamefont {L.~G.}\ \bibnamefont
  {Yaffe}},\ }\href {https://doi.org/10.1007/JHEP12(2016)042} {\bibfield
  {journal} {\bibinfo  {journal} {JHEP}\ }\textbf {\bibinfo {volume} {12}},\
  \bibinfo {pages} {042}},\ \Eprint {https://arxiv.org/abs/1610.03491}
  {arXiv:1610.03491 [hep-th]} \BibitemShut {NoStop}%
\bibitem [{\citenamefont {Grozdanov}\ \emph
  {et~al.}(2019{\natexlab{a}})\citenamefont {Grozdanov}, \citenamefont
  {Kovtun}, \citenamefont {Starinets},\ and\ \citenamefont
  {Tadi\'c}}]{Grozdanov:2019uhi}%
  \BibitemOpen
  \bibfield  {author} {\bibinfo {author} {\bibfnamefont {S.}~\bibnamefont
  {Grozdanov}}, \bibinfo {author} {\bibfnamefont {P.~K.}\ \bibnamefont
  {Kovtun}}, \bibinfo {author} {\bibfnamefont {A.~O.}\ \bibnamefont
  {Starinets}},\ and\ \bibinfo {author} {\bibfnamefont {P.}~\bibnamefont
  {Tadi\'c}},\ }\href {https://doi.org/10.1007/JHEP11(2019)097} {\bibfield
  {journal} {\bibinfo  {journal} {JHEP}\ }\textbf {\bibinfo {volume} {11}},\
  \bibinfo {pages} {097}},\ \Eprint {https://arxiv.org/abs/1904.12862}
  {arXiv:1904.12862 [hep-th]} \BibitemShut {NoStop}%
\bibitem [{\citenamefont {Heller}\ \emph
  {et~al.}(2021{\natexlab{a}})\citenamefont {Heller}, \citenamefont {Serantes},
  \citenamefont {Spali\'nski}, \citenamefont {Svensson},\ and\ \citenamefont
  {Withers}}]{Heller:2020hnq}%
  \BibitemOpen
  \bibfield  {author} {\bibinfo {author} {\bibfnamefont {M.~P.}\ \bibnamefont
  {Heller}}, \bibinfo {author} {\bibfnamefont {A.}~\bibnamefont {Serantes}},
  \bibinfo {author} {\bibfnamefont {M.}~\bibnamefont {Spali\'nski}}, \bibinfo
  {author} {\bibfnamefont {V.}~\bibnamefont {Svensson}},\ and\ \bibinfo
  {author} {\bibfnamefont {B.}~\bibnamefont {Withers}},\ }\href
  {https://doi.org/10.21468/SciPostPhys.10.6.123} {\bibfield  {journal}
  {\bibinfo  {journal} {SciPost Phys.}\ }\textbf {\bibinfo {volume} {10}},\
  \bibinfo {pages} {123} (\bibinfo {year} {2021}{\natexlab{a}})},\ \Eprint
  {https://arxiv.org/abs/2012.15393} {arXiv:2012.15393 [hep-th]} \BibitemShut
  {NoStop}%
\bibitem [{\citenamefont {Heller}\ \emph
  {et~al.}(2021{\natexlab{b}})\citenamefont {Heller}, \citenamefont {Serantes},
  \citenamefont {Spali\'nski}, \citenamefont {Svensson},\ and\ \citenamefont
  {Withers}}]{Heller:2020uuy}%
  \BibitemOpen
  \bibfield  {author} {\bibinfo {author} {\bibfnamefont {M.~P.}\ \bibnamefont
  {Heller}}, \bibinfo {author} {\bibfnamefont {A.}~\bibnamefont {Serantes}},
  \bibinfo {author} {\bibfnamefont {M.}~\bibnamefont {Spali\'nski}}, \bibinfo
  {author} {\bibfnamefont {V.}~\bibnamefont {Svensson}},\ and\ \bibinfo
  {author} {\bibfnamefont {B.}~\bibnamefont {Withers}},\ }\href
  {https://doi.org/10.1103/PhysRevD.104.066002} {\bibfield  {journal} {\bibinfo
   {journal} {Phys. Rev. D}\ }\textbf {\bibinfo {volume} {104}},\ \bibinfo
  {pages} {066002} (\bibinfo {year} {2021}{\natexlab{b}})},\ \Eprint
  {https://arxiv.org/abs/2007.05524} {arXiv:2007.05524 [hep-th]} \BibitemShut
  {NoStop}%
\bibitem [{\citenamefont {Cartwright}\ \emph {et~al.}(2021)\citenamefont
  {Cartwright}, \citenamefont {Amano}, \citenamefont {Kaminski}, \citenamefont
  {Noronha},\ and\ \citenamefont {Speranza}}]{Cartwright:2021qpp}%
  \BibitemOpen
  \bibfield  {author} {\bibinfo {author} {\bibfnamefont {C.}~\bibnamefont
  {Cartwright}}, \bibinfo {author} {\bibfnamefont {M.~G.}\ \bibnamefont
  {Amano}}, \bibinfo {author} {\bibfnamefont {M.}~\bibnamefont {Kaminski}},
  \bibinfo {author} {\bibfnamefont {J.}~\bibnamefont {Noronha}},\ and\ \bibinfo
  {author} {\bibfnamefont {E.}~\bibnamefont {Speranza}},\ }\href@noop {} {\
  (\bibinfo {year} {2021})},\ \Eprint {https://arxiv.org/abs/2112.10781}
  {arXiv:2112.10781 [hep-th]} \BibitemShut {NoStop}%
\bibitem [{\citenamefont {Amano}\ \emph {et~al.}(2023)\citenamefont {Amano},
  \citenamefont {Blake}, \citenamefont {Cartwright}, \citenamefont {Kaminski},\
  and\ \citenamefont {Thompson}}]{Amano:2022mlu}%
  \BibitemOpen
  \bibfield  {author} {\bibinfo {author} {\bibfnamefont {M.~A.~G.}\
  \bibnamefont {Amano}}, \bibinfo {author} {\bibfnamefont {M.}~\bibnamefont
  {Blake}}, \bibinfo {author} {\bibfnamefont {C.}~\bibnamefont {Cartwright}},
  \bibinfo {author} {\bibfnamefont {M.}~\bibnamefont {Kaminski}},\ and\
  \bibinfo {author} {\bibfnamefont {A.~P.}\ \bibnamefont {Thompson}},\ }\href
  {https://doi.org/10.1007/JHEP02(2023)253} {\bibfield  {journal} {\bibinfo
  {journal} {JHEP}\ }\textbf {\bibinfo {volume} {02}},\ \bibinfo {pages}
  {253}},\ \Eprint {https://arxiv.org/abs/2211.00016} {arXiv:2211.00016
  [hep-th]} \BibitemShut {NoStop}%
\bibitem [{\citenamefont {Maldacena}(1999)}]{Maldacena:1997re}%
  \BibitemOpen
  \bibfield  {author} {\bibinfo {author} {\bibfnamefont {J.~M.}\ \bibnamefont
  {Maldacena}},\ }\href {https://doi.org/10.1023/A:1026654312961} {\bibfield
  {journal} {\bibinfo  {journal} {Int. J. Theor. Phys.}\ }\textbf {\bibinfo
  {volume} {38}},\ \bibinfo {pages} {1113} (\bibinfo {year} {1999})},\ \Eprint
  {https://arxiv.org/abs/hep-th/9711200} {arXiv:hep-th/9711200} \BibitemShut
  {NoStop}%
\bibitem [{\citenamefont {Kovtun}\ and\ \citenamefont
  {Starinets}(2005)}]{Kovtun:2005ev}%
  \BibitemOpen
  \bibfield  {author} {\bibinfo {author} {\bibfnamefont {P.~K.}\ \bibnamefont
  {Kovtun}}\ and\ \bibinfo {author} {\bibfnamefont {A.~O.}\ \bibnamefont
  {Starinets}},\ }\href {https://doi.org/10.1103/PhysRevD.72.086009} {\bibfield
   {journal} {\bibinfo  {journal} {Phys. Rev. D}\ }\textbf {\bibinfo {volume}
  {72}},\ \bibinfo {pages} {086009} (\bibinfo {year} {2005})},\ \Eprint
  {https://arxiv.org/abs/hep-th/0506184} {arXiv:hep-th/0506184} \BibitemShut
  {NoStop}%
\bibitem [{\citenamefont {McInnes}(2014)}]{McInnes:2014haa}%
  \BibitemOpen
  \bibfield  {author} {\bibinfo {author} {\bibfnamefont {B.}~\bibnamefont
  {McInnes}},\ }\href {https://doi.org/10.1016/j.nuclphysb.2014.08.011}
  {\bibfield  {journal} {\bibinfo  {journal} {Nucl. Phys. B}\ }\textbf
  {\bibinfo {volume} {887}},\ \bibinfo {pages} {246} (\bibinfo {year}
  {2014})},\ \Eprint {https://arxiv.org/abs/1403.3258} {arXiv:1403.3258
  [hep-th]} \BibitemShut {NoStop}%
\bibitem [{\citenamefont {McInnes}(2016)}]{McInnes:2016dwk}%
  \BibitemOpen
  \bibfield  {author} {\bibinfo {author} {\bibfnamefont {B.}~\bibnamefont
  {McInnes}},\ }\href {https://doi.org/10.1016/j.nuclphysb.2016.08.001}
  {\bibfield  {journal} {\bibinfo  {journal} {Nucl. Phys. B}\ }\textbf
  {\bibinfo {volume} {911}},\ \bibinfo {pages} {173} (\bibinfo {year}
  {2016})},\ \Eprint {https://arxiv.org/abs/1604.03669} {arXiv:1604.03669
  [hep-th]} \BibitemShut {NoStop}%
\bibitem [{\citenamefont {McInnes}(2017)}]{McInnes:2017rxu}%
  \BibitemOpen
  \bibfield  {author} {\bibinfo {author} {\bibfnamefont {B.}~\bibnamefont
  {McInnes}},\ }\href@noop {} {\  (\bibinfo {year} {2017})},\ \Eprint
  {https://arxiv.org/abs/1710.07442} {arXiv:1710.07442 [hep-ph]} \BibitemShut
  {NoStop}%
\bibitem [{\citenamefont {Chen}\ \emph {et~al.}(2020)\citenamefont {Chen},
  \citenamefont {Zhang}, \citenamefont {Li}, \citenamefont {Hou},\ and\
  \citenamefont {Huang}}]{Chen:2020ath}%
  \BibitemOpen
  \bibfield  {author} {\bibinfo {author} {\bibfnamefont {X.}~\bibnamefont
  {Chen}}, \bibinfo {author} {\bibfnamefont {L.}~\bibnamefont {Zhang}},
  \bibinfo {author} {\bibfnamefont {D.}~\bibnamefont {Li}}, \bibinfo {author}
  {\bibfnamefont {D.}~\bibnamefont {Hou}},\ and\ \bibinfo {author}
  {\bibfnamefont {M.}~\bibnamefont {Huang}},\ }\href@noop {} {\  (\bibinfo
  {year} {2020})},\ \Eprint {https://arxiv.org/abs/2010.14478}
  {arXiv:2010.14478 [hep-ph]} \BibitemShut {NoStop}%
\bibitem [{\citenamefont {Aliev}\ and\ \citenamefont
  {Delice}(2009)}]{Aliev:2008yk}%
  \BibitemOpen
  \bibfield  {author} {\bibinfo {author} {\bibfnamefont {A.~N.}\ \bibnamefont
  {Aliev}}\ and\ \bibinfo {author} {\bibfnamefont {O.}~\bibnamefont {Delice}},\
  }\href {https://doi.org/10.1103/PhysRevD.79.024013} {\bibfield  {journal}
  {\bibinfo  {journal} {Phys. Rev. D}\ }\textbf {\bibinfo {volume} {79}},\
  \bibinfo {pages} {024013} (\bibinfo {year} {2009})},\ \Eprint
  {https://arxiv.org/abs/0808.0280} {arXiv:0808.0280 [hep-th]} \BibitemShut
  {NoStop}%
\bibitem [{\citenamefont {Monteiro}\ \emph {et~al.}(2009)\citenamefont
  {Monteiro}, \citenamefont {Perry},\ and\ \citenamefont
  {Santos}}]{Monteiro:2009tc}%
  \BibitemOpen
  \bibfield  {author} {\bibinfo {author} {\bibfnamefont {R.}~\bibnamefont
  {Monteiro}}, \bibinfo {author} {\bibfnamefont {M.~J.}\ \bibnamefont
  {Perry}},\ and\ \bibinfo {author} {\bibfnamefont {J.~E.}\ \bibnamefont
  {Santos}},\ }\href {https://doi.org/10.1103/PhysRevD.80.024041} {\bibfield
  {journal} {\bibinfo  {journal} {Phys. Rev. D}\ }\textbf {\bibinfo {volume}
  {80}},\ \bibinfo {pages} {024041} (\bibinfo {year} {2009})},\ \Eprint
  {https://arxiv.org/abs/0903.3256} {arXiv:0903.3256 [gr-qc]} \BibitemShut
  {NoStop}%
\bibitem [{\citenamefont {Elvang}\ and\ \citenamefont
  {Figueras}(2007)}]{Elvang:2007rd}%
  \BibitemOpen
  \bibfield  {author} {\bibinfo {author} {\bibfnamefont {H.}~\bibnamefont
  {Elvang}}\ and\ \bibinfo {author} {\bibfnamefont {P.}~\bibnamefont
  {Figueras}},\ }\href {https://doi.org/10.1088/1126-6708/2007/05/050}
  {\bibfield  {journal} {\bibinfo  {journal} {JHEP}\ }\textbf {\bibinfo
  {volume} {05}},\ \bibinfo {pages} {050}},\ \Eprint
  {https://arxiv.org/abs/hep-th/0701035} {arXiv:hep-th/0701035} \BibitemShut
  {NoStop}%
\bibitem [{\citenamefont {Emparan}\ and\ \citenamefont
  {Reall}(2002)}]{Emparan:2001wn}%
  \BibitemOpen
  \bibfield  {author} {\bibinfo {author} {\bibfnamefont {R.}~\bibnamefont
  {Emparan}}\ and\ \bibinfo {author} {\bibfnamefont {H.~S.}\ \bibnamefont
  {Reall}},\ }\href {https://doi.org/10.1103/PhysRevLett.88.101101} {\bibfield
  {journal} {\bibinfo  {journal} {Phys. Rev. Lett.}\ }\textbf {\bibinfo
  {volume} {88}},\ \bibinfo {pages} {101101} (\bibinfo {year} {2002})},\
  \Eprint {https://arxiv.org/abs/hep-th/0110260} {arXiv:hep-th/0110260}
  \BibitemShut {NoStop}%
\bibitem [{\citenamefont {Papadimitriou}\ and\ \citenamefont
  {Skenderis}(2005)}]{Papadimitriou:2005ii}%
  \BibitemOpen
  \bibfield  {author} {\bibinfo {author} {\bibfnamefont {I.}~\bibnamefont
  {Papadimitriou}}\ and\ \bibinfo {author} {\bibfnamefont {K.}~\bibnamefont
  {Skenderis}},\ }\href {https://doi.org/10.1088/1126-6708/2005/08/004}
  {\bibfield  {journal} {\bibinfo  {journal} {JHEP}\ }\textbf {\bibinfo
  {volume} {08}},\ \bibinfo {pages} {004}},\ \Eprint
  {https://arxiv.org/abs/hep-th/0505190} {arXiv:hep-th/0505190 [hep-th]}
  \BibitemShut {NoStop}%
\bibitem [{\citenamefont {Wald}(1984)}]{Wald:1984rg}%
  \BibitemOpen
  \bibfield  {author} {\bibinfo {author} {\bibfnamefont {R.~M.}\ \bibnamefont
  {Wald}},\ }\href {https://doi.org/10.7208/chicago/9780226870373.001.0001}
  {\emph {\bibinfo {title} {{General Relativity}}}}\ (\bibinfo  {publisher}
  {Chicago Univ. Pr.},\ \bibinfo {address} {Chicago, USA},\ \bibinfo {year}
  {1984})\BibitemShut {NoStop}%
\bibitem [{\citenamefont {Kunduri}\ \emph {et~al.}(2006)\citenamefont
  {Kunduri}, \citenamefont {Lucietti},\ and\ \citenamefont
  {Reall}}]{Kunduri:2006qa}%
  \BibitemOpen
  \bibfield  {author} {\bibinfo {author} {\bibfnamefont {H.~K.}\ \bibnamefont
  {Kunduri}}, \bibinfo {author} {\bibfnamefont {J.}~\bibnamefont {Lucietti}},\
  and\ \bibinfo {author} {\bibfnamefont {H.~S.}\ \bibnamefont {Reall}},\ }\href
  {https://doi.org/10.1103/PhysRevD.74.084021} {\bibfield  {journal} {\bibinfo
  {journal} {Phys. Rev. D}\ }\textbf {\bibinfo {volume} {74}},\ \bibinfo
  {pages} {084021} (\bibinfo {year} {2006})},\ \Eprint
  {https://arxiv.org/abs/hep-th/0606076} {arXiv:hep-th/0606076} \BibitemShut
  {NoStop}%
\bibitem [{\citenamefont {Nakahara}(2003)}]{nakahara2003}%
  \BibitemOpen
  \bibfield  {author} {\bibinfo {author} {\bibfnamefont {M.}~\bibnamefont
  {Nakahara}},\ }\href@noop {} {\emph {\bibinfo {title} {Geometry, Topology and
  Physics}}},\ \bibinfo {edition} {2nd}\ ed.\ (\bibinfo  {publisher} {CRC
  Press},\ \bibinfo {year} {2003})\BibitemShut {NoStop}%
\bibitem [{\citenamefont {Chevalley}(1946)}]{chevalley1946}%
  \BibitemOpen
  \bibfield  {author} {\bibinfo {author} {\bibfnamefont {C.}~\bibnamefont
  {Chevalley}},\ }\href {http://www.jstor.org/stable/j.ctt1bpm9z7} {\emph
  {\bibinfo {title} {Theory of Lie Groups (PMS-8)}}}\ (\bibinfo  {publisher}
  {Princeton University Press},\ \bibinfo {year} {1946})\BibitemShut {NoStop}%
\bibitem [{\citenamefont {Fecko}(2006)}]{fecko_2006}%
  \BibitemOpen
  \bibfield  {author} {\bibinfo {author} {\bibfnamefont {M.}~\bibnamefont
  {Fecko}},\ }\href {https://doi.org/10.1017/CBO9780511755590} {\emph {\bibinfo
  {title} {Differential Geometry and Lie Groups for Physicists}}}\ (\bibinfo
  {publisher} {Cambridge University Press},\ \bibinfo {year}
  {2006})\BibitemShut {NoStop}%
\bibitem [{\citenamefont {Varshalovich}\ \emph {et~al.}(1988)\citenamefont
  {Varshalovich}, \citenamefont {Moskalev},\ and\ \citenamefont
  {Khersonskii}}]{varshalovich_1988}%
  \BibitemOpen
  \bibfield  {author} {\bibinfo {author} {\bibfnamefont {D.~A.}\ \bibnamefont
  {Varshalovich}}, \bibinfo {author} {\bibfnamefont {A.~N.}\ \bibnamefont
  {Moskalev}},\ and\ \bibinfo {author} {\bibfnamefont {V.~K.}\ \bibnamefont
  {Khersonskii}},\ }\href@noop {} {\emph {\bibinfo {title} {Quantum theory of
  angular momentum}}}\ (\bibinfo  {publisher} {World Scientific Pub.},\
  \bibinfo {year} {1988})\BibitemShut {NoStop}%
\bibitem [{\citenamefont {Sakurai}\ and\ \citenamefont
  {Napolitano}(2017)}]{sakurai_napolitano_2017}%
  \BibitemOpen
  \bibfield  {author} {\bibinfo {author} {\bibfnamefont {J.~J.}\ \bibnamefont
  {Sakurai}}\ and\ \bibinfo {author} {\bibfnamefont {J.}~\bibnamefont
  {Napolitano}},\ }\href {https://doi.org/10.1017/9781108499996} {\emph
  {\bibinfo {title} {Modern Quantum Mechanics}}},\ \bibinfo {edition} {2nd}\
  ed.\ (\bibinfo  {publisher} {Cambridge University Press},\ \bibinfo {year}
  {2017})\BibitemShut {NoStop}%
\bibitem [{\citenamefont {Hawking}\ and\ \citenamefont
  {Page}(1983)}]{Hawking:1982dh}%
  \BibitemOpen
  \bibfield  {author} {\bibinfo {author} {\bibfnamefont {S.}~\bibnamefont
  {Hawking}}\ and\ \bibinfo {author} {\bibfnamefont {D.~N.}\ \bibnamefont
  {Page}},\ }\href {https://doi.org/10.1007/BF01208266} {\bibfield  {journal}
  {\bibinfo  {journal} {Commun. Math. Phys.}\ }\textbf {\bibinfo {volume}
  {87}},\ \bibinfo {pages} {577} (\bibinfo {year} {1983})}\BibitemShut
  {NoStop}%
\bibitem [{\citenamefont {Policastro}\ \emph {et~al.}(2002)\citenamefont
  {Policastro}, \citenamefont {Son},\ and\ \citenamefont
  {Starinets}}]{Policastro:2002se}%
  \BibitemOpen
  \bibfield  {author} {\bibinfo {author} {\bibfnamefont {G.}~\bibnamefont
  {Policastro}}, \bibinfo {author} {\bibfnamefont {D.~T.}\ \bibnamefont
  {Son}},\ and\ \bibinfo {author} {\bibfnamefont {A.~O.}\ \bibnamefont
  {Starinets}},\ }\href {https://doi.org/10.1088/1126-6708/2002/09/043}
  {\bibfield  {journal} {\bibinfo  {journal} {JHEP}\ }\textbf {\bibinfo
  {volume} {09}},\ \bibinfo {pages} {043}},\ \Eprint
  {https://arxiv.org/abs/hep-th/0205052} {arXiv:hep-th/0205052} \BibitemShut
  {NoStop}%
\bibitem [{\citenamefont {Kovtun}(2019)}]{Kovtun:2019hdm}%
  \BibitemOpen
  \bibfield  {author} {\bibinfo {author} {\bibfnamefont {P.}~\bibnamefont
  {Kovtun}},\ }\href {https://doi.org/10.1007/JHEP10(2019)034} {\bibfield
  {journal} {\bibinfo  {journal} {JHEP}\ }\textbf {\bibinfo {volume} {10}},\
  \bibinfo {pages} {034}},\ \Eprint {https://arxiv.org/abs/1907.08191}
  {arXiv:1907.08191 [hep-th]} \BibitemShut {NoStop}%
\bibitem [{\citenamefont {Hoult}\ and\ \citenamefont
  {Kovtun}(2020)}]{Hoult:2020eho}%
  \BibitemOpen
  \bibfield  {author} {\bibinfo {author} {\bibfnamefont {R.~E.}\ \bibnamefont
  {Hoult}}\ and\ \bibinfo {author} {\bibfnamefont {P.}~\bibnamefont {Kovtun}},\
  }\href {https://doi.org/10.1007/JHEP06(2020)067} {\bibfield  {journal}
  {\bibinfo  {journal} {JHEP}\ }\textbf {\bibinfo {volume} {06}},\ \bibinfo
  {pages} {067}},\ \Eprint {https://arxiv.org/abs/2004.04102} {arXiv:2004.04102
  [hep-th]} \BibitemShut {NoStop}%
\bibitem [{\citenamefont {Bender}\ and\ \citenamefont
  {Orszag}(1999)}]{OrszagBender}%
  \BibitemOpen
  \bibfield  {author} {\bibinfo {author} {\bibfnamefont {C.}~\bibnamefont
  {Bender}}\ and\ \bibinfo {author} {\bibfnamefont {S.}~\bibnamefont
  {Orszag}},\ }\href@noop {} {\emph {\bibinfo {title} {Advanced Mathematical
  Methods for Scientists and Engineers I: Asymptotic Methods and Perturbation
  Theory}}},\ Advanced Mathematical Methods for Scientists and Engineers\
  (\bibinfo  {publisher} {Springer},\ \bibinfo {year} {1999})\BibitemShut
  {NoStop}%
\bibitem [{\citenamefont {Abbasi}(2022)}]{Abbasi:2021fcz}%
  \BibitemOpen
  \bibfield  {author} {\bibinfo {author} {\bibfnamefont {N.}~\bibnamefont
  {Abbasi}},\ }\href {https://doi.org/10.1007/JHEP04(2022)181} {\bibfield
  {journal} {\bibinfo  {journal} {JHEP}\ }\textbf {\bibinfo {volume} {04}},\
  \bibinfo {pages} {181}},\ \Eprint {https://arxiv.org/abs/2112.12751}
  {arXiv:2112.12751 [hep-th]} \BibitemShut {NoStop}%
\bibitem [{\citenamefont {Abbasi}\ \emph {et~al.}(2022)\citenamefont {Abbasi},
  \citenamefont {Kaminski},\ and\ \citenamefont {Tavakol}}]{Abbasi:2022aao}%
  \BibitemOpen
  \bibfield  {author} {\bibinfo {author} {\bibfnamefont {N.}~\bibnamefont
  {Abbasi}}, \bibinfo {author} {\bibfnamefont {M.}~\bibnamefont {Kaminski}},\
  and\ \bibinfo {author} {\bibfnamefont {O.}~\bibnamefont {Tavakol}},\
  }\href@noop {} {\  (\bibinfo {year} {2022})},\ \Eprint
  {https://arxiv.org/abs/2212.11499} {arXiv:2212.11499 [hep-th]} \BibitemShut
  {NoStop}%
\bibitem [{\citenamefont {Ishii}(2021)}]{Ishii:2021ncf}%
  \BibitemOpen
  \bibfield  {author} {\bibinfo {author} {\bibfnamefont {T.}~\bibnamefont
  {Ishii}},\ }\href {https://doi.org/10.21468/SciPostPhysProc.4.008} {\bibfield
   {journal} {\bibinfo  {journal} {SciPost Phys. Proc.}\ }\textbf {\bibinfo
  {volume} {4}},\ \bibinfo {pages} {008} (\bibinfo {year} {2021})}\BibitemShut
  {NoStop}%
\bibitem [{\citenamefont {Ishii}\ and\ \citenamefont
  {Murata}(2019)}]{Ishii:2018oms}%
  \BibitemOpen
  \bibfield  {author} {\bibinfo {author} {\bibfnamefont {T.}~\bibnamefont
  {Ishii}}\ and\ \bibinfo {author} {\bibfnamefont {K.}~\bibnamefont {Murata}},\
  }\href {https://doi.org/10.1088/1361-6382/ab1d76} {\bibfield  {journal}
  {\bibinfo  {journal} {Class. Quant. Grav.}\ }\textbf {\bibinfo {volume}
  {36}},\ \bibinfo {pages} {125011} (\bibinfo {year} {2019})},\ \Eprint
  {https://arxiv.org/abs/1810.11089} {arXiv:1810.11089 [hep-th]} \BibitemShut
  {NoStop}%
\bibitem [{\citenamefont {Dias}\ and\ \citenamefont
  {Santos}(2018)}]{Dias:2017tjg}%
  \BibitemOpen
  \bibfield  {author} {\bibinfo {author} {\bibfnamefont {O.~J.~C.}\
  \bibnamefont {Dias}}\ and\ \bibinfo {author} {\bibfnamefont {J.~E.}\
  \bibnamefont {Santos}},\ }\href {https://doi.org/10.1088/1361-6382/aad514}
  {\bibfield  {journal} {\bibinfo  {journal} {Class. Quant. Grav.}\ }\textbf
  {\bibinfo {volume} {35}},\ \bibinfo {pages} {185006} (\bibinfo {year}
  {2018})},\ \Eprint {https://arxiv.org/abs/1705.03065} {arXiv:1705.03065
  [hep-th]} \BibitemShut {NoStop}%
\bibitem [{\citenamefont {Dias}\ \emph {et~al.}(2015)\citenamefont {Dias},
  \citenamefont {Santos},\ and\ \citenamefont {Way}}]{Dias:2015rxy}%
  \BibitemOpen
  \bibfield  {author} {\bibinfo {author} {\bibfnamefont {O.~J.~C.}\
  \bibnamefont {Dias}}, \bibinfo {author} {\bibfnamefont {J.~E.}\ \bibnamefont
  {Santos}},\ and\ \bibinfo {author} {\bibfnamefont {B.}~\bibnamefont {Way}},\
  }\href {https://doi.org/10.1007/JHEP12(2015)171} {\bibfield  {journal}
  {\bibinfo  {journal} {JHEP}\ }\textbf {\bibinfo {volume} {12}},\ \bibinfo
  {pages} {171}},\ \Eprint {https://arxiv.org/abs/1505.04793} {arXiv:1505.04793
  [hep-th]} \BibitemShut {NoStop}%
\bibitem [{\citenamefont {Walker}(1950)}]{walker:1950alg}%
  \BibitemOpen
  \bibfield  {author} {\bibinfo {author} {\bibfnamefont {R.~J.}\ \bibnamefont
  {Walker}},\ }\href@noop {} {\emph {\bibinfo {title} {Algebraic curves}}},\
  Vol.~\bibinfo {volume} {58}\ (\bibinfo  {publisher} {Springer},\ \bibinfo
  {year} {1950})\BibitemShut {NoStop}%
\bibitem [{\citenamefont {Wall}(2004)}]{wall_2004}%
  \BibitemOpen
  \bibfield  {author} {\bibinfo {author} {\bibfnamefont {C.~T.~C.}\
  \bibnamefont {Wall}},\ }\href {https://doi.org/10.1017/CBO9780511617560}
  {\emph {\bibinfo {title} {Singular Points of Plane Curves}}},\ London
  Mathematical Society Student Texts\ (\bibinfo  {publisher} {Cambridge
  University Press},\ \bibinfo {year} {2004})\BibitemShut {NoStop}%
\bibitem [{\citenamefont {Grozdanov}\ \emph {et~al.}(2018)\citenamefont
  {Grozdanov}, \citenamefont {Schalm},\ and\ \citenamefont
  {Scopelliti}}]{Schalm:2018pjh}%
  \BibitemOpen
  \bibfield  {author} {\bibinfo {author} {\bibfnamefont {S.}~\bibnamefont
  {Grozdanov}}, \bibinfo {author} {\bibfnamefont {K.}~\bibnamefont {Schalm}},\
  and\ \bibinfo {author} {\bibfnamefont {V.}~\bibnamefont {Scopelliti}},\
  }\href {https://doi.org/10.1103/PhysRevLett.120.231601} {\bibfield  {journal}
  {\bibinfo  {journal} {Phys. Rev. Lett.}\ }\textbf {\bibinfo {volume} {120}},\
  \bibinfo {pages} {231601} (\bibinfo {year} {2018})},\ \Eprint
  {https://arxiv.org/abs/1710.00921} {arXiv:1710.00921 [hep-th]} \BibitemShut
  {NoStop}%
\bibitem [{\citenamefont {Shenker}\ and\ \citenamefont
  {Stanford}(2014)}]{Shenker:2013pqa}%
  \BibitemOpen
  \bibfield  {author} {\bibinfo {author} {\bibfnamefont {S.~H.}\ \bibnamefont
  {Shenker}}\ and\ \bibinfo {author} {\bibfnamefont {D.}~\bibnamefont
  {Stanford}},\ }\href {https://doi.org/10.1007/JHEP03(2014)067} {\bibfield
  {journal} {\bibinfo  {journal} {JHEP}\ }\textbf {\bibinfo {volume} {03}},\
  \bibinfo {pages} {067}},\ \Eprint {https://arxiv.org/abs/1306.0622}
  {arXiv:1306.0622 [hep-th]} \BibitemShut {NoStop}%
\bibitem [{\citenamefont {Shenker}\ and\ \citenamefont
  {Stanford}(2015)}]{Shenker:2015keq}%
  \BibitemOpen
  \bibfield  {author} {\bibinfo {author} {\bibfnamefont {S.~H.}\ \bibnamefont
  {Shenker}}\ and\ \bibinfo {author} {\bibfnamefont {D.}~\bibnamefont
  {Stanford}},\ }\href {https://doi.org/10.1007/JHEP05(2015)132} {\bibfield
  {journal} {\bibinfo  {journal} {JHEP}\ }\textbf {\bibinfo {volume} {05}},\
  \bibinfo {pages} {132}},\ \Eprint {https://arxiv.org/abs/1412.6087}
  {arXiv:1412.6087 [hep-th]} \BibitemShut {NoStop}%
\bibitem [{\citenamefont {Liu}\ and\ \citenamefont {Raju}(2020)}]{Liu:2020poq}%
  \BibitemOpen
  \bibfield  {author} {\bibinfo {author} {\bibfnamefont {Y.}~\bibnamefont
  {Liu}}\ and\ \bibinfo {author} {\bibfnamefont {A.}~\bibnamefont {Raju}},\
  }\href {https://doi.org/10.1007/JHEP12(2020)027} {\bibfield  {journal}
  {\bibinfo  {journal} {JHEP}\ }\textbf {\bibinfo {volume} {12}},\ \bibinfo
  {pages} {027}},\ \Eprint {https://arxiv.org/abs/2005.08508} {arXiv:2005.08508
  [hep-th]} \BibitemShut {NoStop}%
\bibitem [{\citenamefont {Mezei}\ and\ \citenamefont
  {S\'arosi}(2020)}]{Mezei:2019dfv}%
  \BibitemOpen
  \bibfield  {author} {\bibinfo {author} {\bibfnamefont {M.}~\bibnamefont
  {Mezei}}\ and\ \bibinfo {author} {\bibfnamefont {G.}~\bibnamefont
  {S\'arosi}},\ }\href {https://doi.org/10.1007/JHEP01(2020)186} {\bibfield
  {journal} {\bibinfo  {journal} {JHEP}\ }\textbf {\bibinfo {volume} {01}},\
  \bibinfo {pages} {186}},\ \Eprint {https://arxiv.org/abs/1908.03574}
  {arXiv:1908.03574 [hep-th]} \BibitemShut {NoStop}%
\bibitem [{\citenamefont {Jahnke}\ \emph {et~al.}(2019)\citenamefont {Jahnke},
  \citenamefont {Kim},\ and\ \citenamefont {Yoon}}]{Jahnke:2019heq}%
  \BibitemOpen
  \bibfield  {author} {\bibinfo {author} {\bibfnamefont {V.}~\bibnamefont
  {Jahnke}}, \bibinfo {author} {\bibfnamefont {K.-Y.}\ \bibnamefont {Kim}},\
  and\ \bibinfo {author} {\bibfnamefont {J.}~\bibnamefont {Yoon}},\ }\href
  {https://doi.org/10.1007/JHEP05(2019)037} {\bibfield  {journal} {\bibinfo
  {journal} {JHEP}\ }\textbf {\bibinfo {volume} {05}},\ \bibinfo {pages}
  {037}},\ \Eprint {https://arxiv.org/abs/1903.09086} {arXiv:1903.09086
  [hep-th]} \BibitemShut {NoStop}%
\bibitem [{\citenamefont {Blake}\ and\ \citenamefont
  {Davison}(2022)}]{Blake:2021hjj}%
  \BibitemOpen
  \bibfield  {author} {\bibinfo {author} {\bibfnamefont {M.}~\bibnamefont
  {Blake}}\ and\ \bibinfo {author} {\bibfnamefont {R.~A.}\ \bibnamefont
  {Davison}},\ }\href {https://doi.org/10.1007/JHEP01(2022)013} {\bibfield
  {journal} {\bibinfo  {journal} {JHEP}\ }\textbf {\bibinfo {volume} {01}},\
  \bibinfo {pages} {013}},\ \Eprint {https://arxiv.org/abs/2111.11093}
  {arXiv:2111.11093 [hep-th]} \BibitemShut {NoStop}%
\bibitem [{\citenamefont {Kabat}\ and\ \citenamefont
  {Ortiz}(1992)}]{Kabat:1992tb}%
  \BibitemOpen
  \bibfield  {author} {\bibinfo {author} {\bibfnamefont {D.~N.}\ \bibnamefont
  {Kabat}}\ and\ \bibinfo {author} {\bibfnamefont {M.}~\bibnamefont {Ortiz}},\
  }\href {https://doi.org/10.1016/0550-3213(92)90627-N} {\bibfield  {journal}
  {\bibinfo  {journal} {Nucl. Phys. B}\ }\textbf {\bibinfo {volume} {388}},\
  \bibinfo {pages} {570} (\bibinfo {year} {1992})},\ \Eprint
  {https://arxiv.org/abs/hep-th/9203082} {arXiv:hep-th/9203082} \BibitemShut
  {NoStop}%
\bibitem [{\citenamefont {Aichelburg}\ and\ \citenamefont
  {Sexl}(1971)}]{Aichelburg:1970dh}%
  \BibitemOpen
  \bibfield  {author} {\bibinfo {author} {\bibfnamefont {P.~C.}\ \bibnamefont
  {Aichelburg}}\ and\ \bibinfo {author} {\bibfnamefont {R.~U.}\ \bibnamefont
  {Sexl}},\ }\href {https://doi.org/10.1007/BF00758149} {\bibfield  {journal}
  {\bibinfo  {journal} {Gen. Rel. Grav.}\ }\textbf {\bibinfo {volume} {2}},\
  \bibinfo {pages} {303} (\bibinfo {year} {1971})}\BibitemShut {NoStop}%
\bibitem [{\citenamefont {Dray}\ and\ \citenamefont
  {'t~Hooft}(1985)}]{Dray:1984ha}%
  \BibitemOpen
  \bibfield  {author} {\bibinfo {author} {\bibfnamefont {T.}~\bibnamefont
  {Dray}}\ and\ \bibinfo {author} {\bibfnamefont {G.}~\bibnamefont
  {'t~Hooft}},\ }\href {https://doi.org/10.1016/0550-3213(85)90525-5}
  {\bibfield  {journal} {\bibinfo  {journal} {Nucl. Phys. B}\ }\textbf
  {\bibinfo {volume} {253}},\ \bibinfo {pages} {173} (\bibinfo {year}
  {1985})}\BibitemShut {NoStop}%
\bibitem [{\citenamefont {Sfetsos}(1995)}]{Sfetsos:1994xa}%
  \BibitemOpen
  \bibfield  {author} {\bibinfo {author} {\bibfnamefont {K.}~\bibnamefont
  {Sfetsos}},\ }\href {https://doi.org/10.1016/0550-3213(94)00573-W} {\bibfield
   {journal} {\bibinfo  {journal} {Nucl. Phys. B}\ }\textbf {\bibinfo {volume}
  {436}},\ \bibinfo {pages} {721} (\bibinfo {year} {1995})},\ \Eprint
  {https://arxiv.org/abs/hep-th/9408169} {arXiv:hep-th/9408169} \BibitemShut
  {NoStop}%
\bibitem [{\citenamefont {Rocha}\ \emph {et~al.}(2014)\citenamefont {Rocha},
  \citenamefont {Santarelli},\ and\ \citenamefont {Delsate}}]{Rocha:2014gza}%
  \BibitemOpen
  \bibfield  {author} {\bibinfo {author} {\bibfnamefont {J.~V.}\ \bibnamefont
  {Rocha}}, \bibinfo {author} {\bibfnamefont {R.}~\bibnamefont {Santarelli}},\
  and\ \bibinfo {author} {\bibfnamefont {T.}~\bibnamefont {Delsate}},\ }\href
  {https://doi.org/10.1103/PhysRevD.89.104006} {\bibfield  {journal} {\bibinfo
  {journal} {Phys. Rev. D}\ }\textbf {\bibinfo {volume} {89}},\ \bibinfo
  {pages} {104006} (\bibinfo {year} {2014})},\ \Eprint
  {https://arxiv.org/abs/1402.4161} {arXiv:1402.4161 [gr-qc]} \BibitemShut
  {NoStop}%
\bibitem [{\citenamefont {Maldacena}\ \emph {et~al.}(2016)\citenamefont
  {Maldacena}, \citenamefont {Shenker},\ and\ \citenamefont
  {Stanford}}]{Maldacena:2015waa}%
  \BibitemOpen
  \bibfield  {author} {\bibinfo {author} {\bibfnamefont {J.}~\bibnamefont
  {Maldacena}}, \bibinfo {author} {\bibfnamefont {S.~H.}\ \bibnamefont
  {Shenker}},\ and\ \bibinfo {author} {\bibfnamefont {D.}~\bibnamefont
  {Stanford}},\ }\href {https://doi.org/10.1007/JHEP08(2016)106} {\bibfield
  {journal} {\bibinfo  {journal} {JHEP}\ }\textbf {\bibinfo {volume} {08}},\
  \bibinfo {pages} {106}},\ \Eprint {https://arxiv.org/abs/1503.01409}
  {arXiv:1503.01409 [hep-th]} \BibitemShut {NoStop}%
\bibitem [{\citenamefont {Abbasi}\ and\ \citenamefont
  {Tahery}(2020)}]{Abbasi:2020ykq}%
  \BibitemOpen
  \bibfield  {author} {\bibinfo {author} {\bibfnamefont {N.}~\bibnamefont
  {Abbasi}}\ and\ \bibinfo {author} {\bibfnamefont {S.}~\bibnamefont
  {Tahery}},\ }\href {https://doi.org/10.1007/JHEP10(2020)076} {\bibfield
  {journal} {\bibinfo  {journal} {JHEP}\ }\textbf {\bibinfo {volume} {10}},\
  \bibinfo {pages} {076}},\ \Eprint {https://arxiv.org/abs/2007.10024}
  {arXiv:2007.10024 [hep-th]} \BibitemShut {NoStop}%
\bibitem [{\citenamefont {Grozdanov}\ \emph
  {et~al.}(2019{\natexlab{b}})\citenamefont {Grozdanov}, \citenamefont
  {Kovtun}, \citenamefont {Starinets},\ and\ \citenamefont
  {Tadi\'c}}]{Grozdanov:2019kge}%
  \BibitemOpen
  \bibfield  {author} {\bibinfo {author} {\bibfnamefont {S.}~\bibnamefont
  {Grozdanov}}, \bibinfo {author} {\bibfnamefont {P.~K.}\ \bibnamefont
  {Kovtun}}, \bibinfo {author} {\bibfnamefont {A.~O.}\ \bibnamefont
  {Starinets}},\ and\ \bibinfo {author} {\bibfnamefont {P.}~\bibnamefont
  {Tadi\'c}},\ }\href {https://doi.org/10.1103/PhysRevLett.122.251601}
  {\bibfield  {journal} {\bibinfo  {journal} {Phys. Rev. Lett.}\ }\textbf
  {\bibinfo {volume} {122}},\ \bibinfo {pages} {251601} (\bibinfo {year}
  {2019}{\natexlab{b}})},\ \Eprint {https://arxiv.org/abs/1904.01018}
  {arXiv:1904.01018 [hep-th]} \BibitemShut {NoStop}%
\bibitem [{\citenamefont {Romatschke}\ and\ \citenamefont
  {Romatschke}(2019)}]{Romatschke:2017ejr}%
  \BibitemOpen
  \bibfield  {author} {\bibinfo {author} {\bibfnamefont {P.}~\bibnamefont
  {Romatschke}}\ and\ \bibinfo {author} {\bibfnamefont {U.}~\bibnamefont
  {Romatschke}},\ }\href {https://doi.org/10.1017/9781108651998} {\emph
  {\bibinfo {title} {Relativistic Fluid Dynamics In and Out of Equilibrium}}},\
  Cambridge Monographs on Mathematical Physics\ (\bibinfo  {publisher}
  {Cambridge University Press},\ \bibinfo {year} {2019})\ \Eprint
  {https://arxiv.org/abs/1712.05815} {arXiv:1712.05815 [nucl-th]} \BibitemShut
  {NoStop}%
\bibitem [{\citenamefont {Molenkamp}\ and\ \citenamefont
  {de~Jong}(1994)}]{Molenkamp:1994}%
  \BibitemOpen
  \bibfield  {author} {\bibinfo {author} {\bibfnamefont {L.~W.}\ \bibnamefont
  {Molenkamp}}\ and\ \bibinfo {author} {\bibfnamefont {M.~J.~M.}\ \bibnamefont
  {de~Jong}},\ }\href {https://doi.org/10.1103/PhysRevB.49.5038} {\bibfield
  {journal} {\bibinfo  {journal} {Phys. Rev. B}\ }\textbf {\bibinfo {volume}
  {49}},\ \bibinfo {pages} {5038} (\bibinfo {year} {1994})}\BibitemShut
  {NoStop}%
\bibitem [{\citenamefont {Moll}\ \emph {et~al.}(2016)\citenamefont {Moll},
  \citenamefont {Kushwaha}, \citenamefont {Nandi}, \citenamefont {Schmidt},\
  and\ \citenamefont {Mackenzie}}]{Moll:2016}%
  \BibitemOpen
  \bibfield  {author} {\bibinfo {author} {\bibfnamefont {P.~J.~W.}\
  \bibnamefont {Moll}}, \bibinfo {author} {\bibfnamefont {P.}~\bibnamefont
  {Kushwaha}}, \bibinfo {author} {\bibfnamefont {N.}~\bibnamefont {Nandi}},
  \bibinfo {author} {\bibfnamefont {B.}~\bibnamefont {Schmidt}},\ and\ \bibinfo
  {author} {\bibfnamefont {A.~P.}\ \bibnamefont {Mackenzie}},\ }\href
  {https://doi.org/10.1126/science.aac8385} {\bibfield  {journal} {\bibinfo
  {journal} {Science}\ }\textbf {\bibinfo {volume} {351}},\ \bibinfo {pages}
  {1061} (\bibinfo {year} {2016})},\ \Eprint
  {https://arxiv.org/abs/https://www.science.org/doi/pdf/10.1126/science.aac8385}
  {https://www.science.org/doi/pdf/10.1126/science.aac8385} \BibitemShut
  {NoStop}%
\bibitem [{\citenamefont {Bloch}\ \emph {et~al.}(2008)\citenamefont {Bloch},
  \citenamefont {Dalibard},\ and\ \citenamefont {Zwerger}}]{Bloch:2008}%
  \BibitemOpen
  \bibfield  {author} {\bibinfo {author} {\bibfnamefont {I.}~\bibnamefont
  {Bloch}}, \bibinfo {author} {\bibfnamefont {J.}~\bibnamefont {Dalibard}},\
  and\ \bibinfo {author} {\bibfnamefont {W.}~\bibnamefont {Zwerger}},\ }\href
  {https://doi.org/10.1103/RevModPhys.80.885} {\bibfield  {journal} {\bibinfo
  {journal} {Rev. Mod. Phys.}\ }\textbf {\bibinfo {volume} {80}},\ \bibinfo
  {pages} {885} (\bibinfo {year} {2008})}\BibitemShut {NoStop}%
\bibitem [{\citenamefont {Sinova}\ \emph {et~al.}(2015)\citenamefont {Sinova},
  \citenamefont {Valenzuela}, \citenamefont {Wunderlich}, \citenamefont
  {Back},\ and\ \citenamefont {Jungwirth}}]{Sinova:2015}%
  \BibitemOpen
  \bibfield  {author} {\bibinfo {author} {\bibfnamefont {J.}~\bibnamefont
  {Sinova}}, \bibinfo {author} {\bibfnamefont {S.~O.}\ \bibnamefont
  {Valenzuela}}, \bibinfo {author} {\bibfnamefont {J.}~\bibnamefont
  {Wunderlich}}, \bibinfo {author} {\bibfnamefont {C.~H.}\ \bibnamefont
  {Back}},\ and\ \bibinfo {author} {\bibfnamefont {T.}~\bibnamefont
  {Jungwirth}},\ }\href {https://doi.org/10.1103/RevModPhys.87.1213} {\bibfield
   {journal} {\bibinfo  {journal} {Rev. Mod. Phys.}\ }\textbf {\bibinfo
  {volume} {87}},\ \bibinfo {pages} {1213} (\bibinfo {year}
  {2015})}\BibitemShut {NoStop}%
\bibitem [{\citenamefont {Florkowski}\ and\ \citenamefont
  {Ryblewski}(2011)}]{Florkowski:2010cf}%
  \BibitemOpen
  \bibfield  {author} {\bibinfo {author} {\bibfnamefont {W.}~\bibnamefont
  {Florkowski}}\ and\ \bibinfo {author} {\bibfnamefont {R.}~\bibnamefont
  {Ryblewski}},\ }\href {https://doi.org/10.1103/PhysRevC.83.034907} {\bibfield
   {journal} {\bibinfo  {journal} {Phys. Rev. C}\ }\textbf {\bibinfo {volume}
  {83}},\ \bibinfo {pages} {034907} (\bibinfo {year} {2011})},\ \Eprint
  {https://arxiv.org/abs/1007.0130} {arXiv:1007.0130 [nucl-th]} \BibitemShut
  {NoStop}%
\bibitem [{\citenamefont {Martinez}\ and\ \citenamefont
  {Strickland}(2010)}]{Martinez:2010sc}%
  \BibitemOpen
  \bibfield  {author} {\bibinfo {author} {\bibfnamefont {M.}~\bibnamefont
  {Martinez}}\ and\ \bibinfo {author} {\bibfnamefont {M.}~\bibnamefont
  {Strickland}},\ }\href {https://doi.org/10.1016/j.nuclphysa.2010.08.011}
  {\bibfield  {journal} {\bibinfo  {journal} {Nucl. Phys. A}\ }\textbf
  {\bibinfo {volume} {848}},\ \bibinfo {pages} {183} (\bibinfo {year}
  {2010})},\ \Eprint {https://arxiv.org/abs/1007.0889} {arXiv:1007.0889
  [nucl-th]} \BibitemShut {NoStop}%
\bibitem [{\citenamefont {Strickland}(2014)}]{Strickland:2014pga}%
  \BibitemOpen
  \bibfield  {author} {\bibinfo {author} {\bibfnamefont {M.}~\bibnamefont
  {Strickland}},\ }\href {https://doi.org/10.5506/APhysPolB.45.2355} {\bibfield
   {journal} {\bibinfo  {journal} {Acta Phys. Polon. B}\ }\textbf {\bibinfo
  {volume} {45}},\ \bibinfo {pages} {2355} (\bibinfo {year} {2014})},\ \Eprint
  {https://arxiv.org/abs/1410.5786} {arXiv:1410.5786 [nucl-th]} \BibitemShut
  {NoStop}%
\bibitem [{\citenamefont {Bhattacharyya}\ \emph {et~al.}(2008)\citenamefont
  {Bhattacharyya}, \citenamefont {Hubeny}, \citenamefont {Minwalla},\ and\
  \citenamefont {Rangamani}}]{Bhattacharyya:2007vjd}%
  \BibitemOpen
  \bibfield  {author} {\bibinfo {author} {\bibfnamefont {S.}~\bibnamefont
  {Bhattacharyya}}, \bibinfo {author} {\bibfnamefont {V.~E.}\ \bibnamefont
  {Hubeny}}, \bibinfo {author} {\bibfnamefont {S.}~\bibnamefont {Minwalla}},\
  and\ \bibinfo {author} {\bibfnamefont {M.}~\bibnamefont {Rangamani}},\ }\href
  {https://doi.org/10.1088/1126-6708/2008/02/045} {\bibfield  {journal}
  {\bibinfo  {journal} {JHEP}\ }\textbf {\bibinfo {volume} {02}},\ \bibinfo
  {pages} {045}},\ \Eprint {https://arxiv.org/abs/0712.2456} {arXiv:0712.2456
  [hep-th]} \BibitemShut {NoStop}%
\bibitem [{\citenamefont {Waeber}\ and\ \citenamefont
  {Yaffe}(2023)}]{Waeber:2022vgf}%
  \BibitemOpen
  \bibfield  {author} {\bibinfo {author} {\bibfnamefont {S.}~\bibnamefont
  {Waeber}}\ and\ \bibinfo {author} {\bibfnamefont {L.~G.}\ \bibnamefont
  {Yaffe}},\ }\href {https://doi.org/10.1007/JHEP03(2023)208} {\bibfield
  {journal} {\bibinfo  {journal} {JHEP}\ }\textbf {\bibinfo {volume} {03}},\
  \bibinfo {pages} {208}},\ \Eprint {https://arxiv.org/abs/2211.09190}
  {arXiv:2211.09190 [hep-th]} \BibitemShut {NoStop}%
\bibitem [{\citenamefont {Kovtun}\ \emph {et~al.}(2005)\citenamefont {Kovtun},
  \citenamefont {Son},\ and\ \citenamefont {Starinets}}]{Kovtun:2004de}%
  \BibitemOpen
  \bibfield  {author} {\bibinfo {author} {\bibfnamefont {P.}~\bibnamefont
  {Kovtun}}, \bibinfo {author} {\bibfnamefont {D.~T.}\ \bibnamefont {Son}},\
  and\ \bibinfo {author} {\bibfnamefont {A.~O.}\ \bibnamefont {Starinets}},\
  }\href {https://doi.org/10.1103/PhysRevLett.94.111601} {\bibfield  {journal}
  {\bibinfo  {journal} {Phys. Rev. Lett.}\ }\textbf {\bibinfo {volume} {94}},\
  \bibinfo {pages} {111601} (\bibinfo {year} {2005})},\ \Eprint
  {https://arxiv.org/abs/hep-th/0405231} {arXiv:hep-th/0405231} \BibitemShut
  {NoStop}%
\bibitem [{\citenamefont {Schwinger}(1952)}]{Schwinger:1952}%
  \BibitemOpen
  \bibfield  {author} {\bibinfo {author} {\bibfnamefont {J.}~\bibnamefont
  {Schwinger}}\ }\href {https://doi.org/10.2172/4389568} {10.2172/4389568}
  (\bibinfo {year} {1952})\BibitemShut {NoStop}%
\bibitem [{\citenamefont {Tajima}(2015)}]{Tajima:2015owa}%
  \BibitemOpen
  \bibfield  {author} {\bibinfo {author} {\bibfnamefont {N.}~\bibnamefont
  {Tajima}},\ }\href {https://doi.org/10.1103/PhysRevC.91.014320} {\bibfield
  {journal} {\bibinfo  {journal} {Phys. Rev. C}\ }\textbf {\bibinfo {volume}
  {91}},\ \bibinfo {pages} {014320} (\bibinfo {year} {2015})},\ \Eprint
  {https://arxiv.org/abs/1501.06347} {arXiv:1501.06347 [nucl-th]} \BibitemShut
  {NoStop}%
\end{thebibliography}%





\end{document}